\documentclass[11pt]{article}

\usepackage{epsfig,amssymb,amsmath,psfrag,hyperref}
\usepackage{color}
\usepackage{xcolor}
\usepackage{slashed} %Slash notation
\usepackage{float}
\usepackage[utf8]{inputenc}
\numberwithin{equation}{section}
\usepackage{amsfonts}
\usepackage{caption}
\usepackage{multirow}
\usepackage{hyperref}
\usepackage{cite}
\usepackage[a4paper, total={6in, 8.3in}]{geometry}
\usepackage{sidecap}
\usepackage{latexsym}
\usepackage{epsfig}
\graphicspath{{projectplots/}}
\renewcommand\thefootnote{\textcolor{red}{\fnsymbol{footnote}}}

\definecolor{nicered}{rgb}{0.7,0.1,0.1}
\definecolor{nicegreen}{rgb}{0.1,0.5,0.1}
\definecolor{rosso}{cmyk}{0,1,1,0.4}
\definecolor{babypink}{rgb}{0.96, 0.76, 0.76}
\definecolor{babyblueeyes}{rgb}{0.63, 0.79, 0.95}
\definecolor{azure(colorwheel)}{rgb}{0.0, 0.5, 1.0}
\definecolor{amethyst}{rgb}{0.6, 0.4, 0.8}
\definecolor{MyDarkBlue}{rgb}{0,0.1,0.7}
\definecolor{PetuniaColor}{RGB}{150,23,147}
\definecolor{secnum}{RGB}{13,151,225}
\definecolor{ptcbackground}{RGB}{212,237,252}
\definecolor{myblue}{RGB}{10,10,200}
\definecolor{ptctitle}{RGB}{0,177,235}
\definecolor{blus}{cmyk}{1,1,0,0.1}
\definecolor{verdes}{cmyk}{0.99,0,0.59,0.65}
\definecolor{rossos}{cmyk}{0,1,1,0.55}
\definecolor{redy}{cmyk}{0,1,1,0.7}
\definecolor{greeny}{cmyk}{0.99,0,0.59,0.98}
\definecolor{green-go}{cmyk}{0.79,0,0.59,0.5}
\definecolor{lime}{RGB}{0,255,0}
\definecolor{ForestGreen}{RGB}{34,139,34}
\definecolor{Purple}{RGB}{128,0,128}

\hypersetup{colorlinks,bookmarksopen,bookmarksnumbered,
citecolor={nicered},
linkcolor={MyDarkBlue},pdfstartview=FitH,
urlcolor={nicegreen}}

\usepackage{subcaption}
\def\beq{\begin{equation}}
\def\eeq{\end{equation}}
\def\bea{\begin{eqnarray}}
\def\eea{\end{eqnarray}}

\def\eq$#1${\begin{equation}#1\end{equation}}
\def\gat$#1${\begin{gather}#1\end{gather}}
\def\bal$#1${\begin{align}#1\end{align}}
\def\eqarr$#1${\begin{eqnarray}#1\end{eqnarray}}

\newcommand{\myref}[2]{{\color{myblue}\ref{#1}(\subref{#2})}}

\usepackage{relsize}
\usepackage{indentfirst}
\usepackage{amssymb}
\usepackage{bookmark}
\usepackage{amsthm}
\usepackage{amsmath}

\newcommand{\newc}{\newcommand}
\def\eq$#1${\begin{equation}#1\end{equation}}
\def\gat$#1${\begin{gather}#1\end{gather}}
\def\bal$#1${\begin{align}#1\end{align}}
\def\eqarr$#1${\begin{eqnarray}#1\end{eqnarray}}
\newc{\pa}{\partial}
\newc{\alp}{\alpha}
\newc{\gam}{\gamma}
\newc{\Gam}{\Gamma}
\newc{\del}{\delta}
\newc{\eps}{\epsilon}
\newc{\lam}{\lambda}
\newc{\sig}{\sigma}
\newc{\ups}{\upsilon}
\newc{\ome}{\omega}
\newc{\pphi}{\varphi}
\newc{\nonum}{\nonumber}
\newc{\hami}{\text{\textbf{\lat{H}}}}
\newc{\gren}{\mathcal{G}}
\newc{\lagr}{\mathcal{L}}
\newc{\timor}{\mathcal{T}}
\newc{\prop}{\mathcal{K}}
\newc{\zcal}{\mathcal{Z}}
\newc{\cinf}{\mathcal{C}_\infty}
\newc{\operx}{\text{\textbf{\lat{x}}}}
\newc{\opera}{\text{\textbf{\lat{a}}}}
\newc{\operp}{\text{\textbf{\lat{p}}}}
\newc{\operl}{\text{\textbf{\lat{L}}}}
\newc{\gfv}{g^{(5)}}
\newc{\kfv}{\kappa_{(5)}}
\newc{\tf}{\tilde{f}}
\newc{\tlam}{\tilde{\Lambda}}
\newc{\tl}{\tilde{\lam}}
\newc{\dist}{\displaystyle}
\newc{\ra}{\rightarrow}
\newc{\Ra}{\Rightarrow}

\newc{\teo}[1]{\textcolor{PetuniaColor}{#1}}
\newc{\corr}[1]{\textcolor{red}{#1}}

\begin{document}

\begin{titlepage}

\vspace*{1cm}
\begin{center}
{\bf \Large{Incorporating Physical Constraints in Braneworld Black-String\\[2mm]
Solutions for a Minkowski Brane in Scalar-Tensor Gravity}}

\bigskip \bigskip \medskip

{\bf Theodoros Nakas},$^{a,}$\footnote[1]{\,t.nakas@uoi.gr}
{\bf Panagiota Kanti},$^{a,}$\footnote[2]{\,pkanti@uoi.gr} and
{\bf Nikolaos Pappas}$^{b,}$\footnote[3]{\,npappas@uoi.gr}

\bigskip
$^a${\it Division of Theoretical Physics, Department of Physics,\\
University of Ioannina, Ioannina GR-45110, Greece}

\bigskip %\medskip
$^b${\it Nuclear and Particle Physics Section, Physics Department,\\
National and Kapodistrian University of Athens, Athens GR-15771, Greece}

\bigskip \medskip
{\bf Abstract}
\end{center}

In the framework of a general scalar-tensor theory, where the scalar field is non-minimally
coupled to the five-dimensional Ricci scalar curvature, we investigate the emergence of
complete brane-world solutions. By assuming a variety of forms for the coupling function,
we solve the field equations in the bulk, and determine in an analytic way the form of the
gravitational background and scalar field in each case. The solutions are always characterised
by a regular scalar field, a finite energy-momentum tensor, and an exponentially decaying
warp factor even in the absence of a negative bulk cosmological constant. The space-time on
the brane is described by the Schwarzschild solution leading to either a non-homogeneous
black-string solution in the bulk, when the mass parameter $M$ is non-zero, or a regular
anti-de Sitter space-time, when $M=0$. We construct physically-acceptable solutions by
demanding in addition a positive effective gravitational constant on our brane, a positive 
total energy-density for our brane and the validity of the weak energy condition in the bulk.
We find that, although the theory does not allow for all three conditions to be simultaneously
satisfied, a plethora of solutions emerge which satisfy the first two, and most fundamental,
conditions.

\end{titlepage}

\renewcommand*{\thefootnote}{\arabic{footnote}}

\setcounter{page}{1}
%%%%%%%%%%%%%%%%%%%%%%%%%%%%%%%%%%%%%%%%%%%%%%%%%%%%%%%%%%%%%%%%%%%%%%

\section{Introduction}
The first higher-dimensional formulation of the General Theory of Relativity
\cite{einstein1, einstein2, einstein3} by Kaluza \cite{Kaluza:1984ws} and Klein 
\cite{Klein1926} is almost as old as the original theory itself. In the 80's,
the postulation of the existence of extra space-like dimensions in nature was
combined with the string-inspired notion of the brane, which plays the role of 
our four-dimensional world \cite{Misha, Akama}. At the turn of the last century, 
the modern brane-world theories were proposed \cite{ADD1, ADD2, ADD3, RS1, RS2} 
in which the extra spatial dimensions may be large compared to the Planck scale or, 
even, infinite. This
radical change in the structure and topology of space-time has significantly
affected the properties of all gravitational solutions which emerge in the framework
of the new theories. In addition, the phase-space of solutions of a higher-dimensional
gravitational theory now contains a variety of black objects, namely black holes, black
strings, black branes, black rings, or black saturns \cite{Emparan-review}. 

The warped brane-world model \cite{RS1, RS2} admits an infinitely long extra dimension
which is nevertheless accompanied by the localisation of the graviton close to our brane. 
This is realised with the help of an exponentially decreasing warp factor in the expression
of the line-element which describes the higher-dimensional gravitational background. 
However, the presence of this factor has proven to be an insurmountable 
obstacle in the derivation of an analytical, non-approximate solution describing a
regular, localized-close-to-our-brane black hole (see Refs. \cite{CHR, EHM1, EHM2,
tidal, KT, Papanto, CasadioNew, Frolov, KOT,
Kofinas, Karasik, GGI, CGKM, Cuadros, AS, Heydari, Ovalle1, KPZ, Ovalle2,
Harko, Ovalle3, daRocha1, daRocha2, Ovalle4, KPP2, review1, review2, review3,
review4, review5, review6, review7, Nakas} for an impartial list of works on this topic;
for a number of numerical solutions describing in principle regular brane-world black
holes, see \cite{KTN, Kudoh, TT, Kleihaus, FW, Page1, Page2}).

The aforementioned attempts to derive an analytical solution of a localised brane-world
black hole have in fact proven that solutions describing a different type of a black
object, namely a black string, are much easier to construct. Although the first such
solution \cite{CHR} in the context of the warped brane-world models \cite{RS1,RS2}
was proven to be unstable \cite{GL, RuthGL}, a variety of higher-dimensional 
black strings have since been derived in the context of different theories in the 
literature---see, for example Refs.  \cite{Gubser, Wiseman, Kudoh2, Sorkin1, Sorkin2, Kleihaus2,
Headrick, Charmousis1, Charmousis2, Figueras, Kalisch, Emparan2, Cisterna1, KNP,
Cisterna2, Cisterna3, CFLO, KNP2, Rezvanjou}. In \cite{KPZ, KPP2}, a brane-world model, that
contained a bulk scalar field with an arbitrary potential and a non-minimal coupling
to gravity, was studied. Scalar-tensor theories of this type are very popular and
have been extensively studied in the context of four-dimensional gravity, while brane-world
generalizations have been studied in the literature before, both in static and non-static
backgrounds \cite{Farakos1, Bogdanos1, Farakos2, Farakos3, Bhatta1, Bhatta2}. 
The objective of the analyses in \cite{KPZ, KPP2} was to derive an analytical solution
describing a regular, localised black hole; although no such solution was found, these
studies hinted that black-string solutions were in fact much easier to emerge
in the context of a non-minimally coupled scalar-tensor brane-world model.

To demonstrate this, in \cite{KNP}, we launched a comprehensive study of the types of
black-string solutions that emerge in the context of this theory. Solving analytically the
complete set of gravitational and scalar-field equations in the bulk, we determined
novel black-string solutions which reduced to a Schwarzschild-(anti-)de Sitter space-time
on the brane. The sign of the effective cosmological constant on the brane was shown
to determine not only the topology of our brane---leading to a de Sitter, anti-de Sitter
or Minkowski four-dimensional background---but also the properties of the coupling function
between the bulk scalar field and the five-dimensional Ricci scalar curvature. In \cite{KNP},
we focused on the case of a positive cosmological constant on the brane, and showed
that, in order for the scalar field to be real-valued, the coupling function had to be
negative over a particular regime in the bulk. Nevertheless, we were able to derive
solutions which had a robust four-dimensional effective theory on the brane and a 
number of interesting, yet provocative, features in the bulk. In a follow-up work
\cite{KNP2}, we considered the case of a negative cosmological constant on the
brane, which allowed positive-definite coupling functions; by employing two particular
forms of the latter, we produced two complete analytical solutions that were characterised
by a regular scalar field and a localised-close-to-our brane energy-momentum tensor.
In addition, the solutions featured a negative-definite bulk potential which supported by
itself the warping of the space-time even in the absence of the traditional, negative,
bulk cosmological constant. 

Having covered the cases of a de Sitter and anti-de Sitter space-time on our brane
in \cite{KNP, KNP2}, in this third instalment we turn our attention to the case of a
Minkowski brane, i.e. with a vanishing effective cosmological constant. The objective
would be the same, namely to perform a comprehensive study of the complete set
of field equations and derive analytical solutions for the gravitational background and
scalar field in the bulk. As we will demonstrate, this case is the least restrictive and
most flexible of the three, and allows for a variety of profiles for the coupling function
and scalar field along the extra coordinate. In order to construct physically-acceptable
solutions, we will demand the finiteness of both the coupling function and scalar field
everywhere in the bulk; in fact, we will consider forms of the coupling function that
become trivial at large distances from our brane thus leading to a minimally-coupled
scalar-tensor theory in that limit. Even under the above assumptions, we will present
a large number of solutions; they will all be characterised by a regular scalar field
and a finite energy-momentum tensor localised near our brane. In addition, the
bulk potential of the scalar field may take a variety of forms at our will, while
supporting  in all cases an exponentially decaying warp factor even in the absence of
a negative bulk cosmological constant.  Negative values of the coupling function in the
bulk will not be necessary in our analysis, nevertheless, they will be allowed. The form
of the effective theory on the brane will thus be of primary importance and a necessary
ingredient of our analysis in the study of each solution presented. We will naturally
demand a positive effective gravitational constant on our brane, and investigate whether
this demand may be simultaneously satisfied with the condition of a positive total
energy of our brane and the validity of the weak energy conditions in the bulk. 
The gravitational background on the brane will be described by the Schwarzschild solution
leading to either a non-homogeneous black-string solution in the bulk, when the mass
parameter $M$ is non-zero, or a regular anti-de Sitter space-time, when $M=0$. 

Our paper has the following outline: in Sec. \ref{th-frame}, we present our theory, the field
equations and impose a number of physical constraints on the scalar field and its
coupling function. In Sec. \ref{linear} to \ref{hyper-tang}, we present a large number of complete brane-world
solutions, and discuss in detail their physical properties in the bulk, the junction conditions,
the effective theory on the brane and the parameter space where the
optimum solutions---from the physical point of view---emerge in each case.
We present our conclusions in Sec. \ref{conc}.

%%%%%%%%%%%%%%%%%%%%%%%%%%%%%%%%%%%%%%%%%%%%%%%%%%%%%%%%%%%%%%%%%%%%
%
%
%%%%%%%%%%%%%%%%%%%%%%%%%%%%%%%%%%%%%%%%%%%%%%%%%%%%%%%%%%%%%%%%%%%%

\section{The Theoretical Framework}
\label{th-frame}

We consider the following action functional which describes a five-dimensional
scalar-tensor theory of gravity 
%%%%%%%%%%%%%%%%
\beq
\label{action}
S_B=\int d^4x\int dy \,\sqrt{-g^{(5)}}\left[\frac{f(\Phi)}{2\kappa_5^2}R
-\Lambda_5-\frac{1}{2}\,\pa_L\Phi\,\pa^L\Phi-V_B(\Phi)\right].
\eeq
%%%%%%%%%%%%%
The theory contains the five-dimensional scalar curvature $R$, a bulk
cosmological constant $\Lambda_5$, and a five-dimensional scalar field
$\Phi$. The latter is characterised by a self-interacting potential $V_B(\Phi)$
and a non-minimal coupling to $R$ via a coupling function $f(\Phi)$. As in our
previous works \cite{KNP, KNP2}, we will initially keep this function arbitrary
so that our formalism is applicable to a large class of theories. 
In the above, $\kappa_5^2=8\pi G_5$, where $G_5$ is the five-dimensional gravitational
constant $G_5$, and $g^{(5)}_{MN}$ is the metric tensor of the five-dimensional
space-time. 

Embedded in this five-dimensional space-time is a 3-brane, our four-dimensional
world, located at $y=0$ along the extra spatial dimension. The energy content of
our brane is described by the following action 
%%%%%%%%%%%%
\beq
\label{action_br}
S_{br}=\int d^4x\sqrt{-g^{(br)}}(\lagr_{br}-\sigma)=
-\int d^4x\int dy\sqrt{-g^{(br)}}\,[V_b(\Phi)+\sigma]\,\delta(y)\,,
\eeq
%%%%%%%%%%%
which should be added to the bulk action (\ref{action}) to complete the theory. 
The brane Lagrangian $\lagr_{br}$ is assumed, for simplicity, to contain only an
interaction term $V_b(\Phi)$ of the bulk scalar field with the brane, while $\sigma$ is the
constant brane self-energy. Also, $g^{(br)}_{\mu\nu}=g_{\mu\nu}^{(5)}(x^\lam,y=0)$ is the
induced-on-the-brane metric tensor. In what follows, we will denote five-dimensional
indices with capital Latin letters $M,N,L,...$ and four-dimensional indices with
lower-case Greek letters $\mu,\nu,\lambda,...$ as usual.

The field equations of the theory follow if we vary the complete action $S=S_B+S_{br}$
with respect to the metric tensor $g^{(5)}_{MN}$ and scalar field $\Phi$. Then, we
obtain the gravitational field equations
%%%%%%%%%%%%
\eq$\label{grav_eqs}
f(\Phi)\,G_{MN}\sqrt{-g^{(5)}}=\kappa_5^2\left[(T^{(\Phi)}_{MN}-g_{MN}\Lambda_5)
\sqrt{-g^{(5)}}-[V_b(\Phi)+\sigma]\,g^{(br)}_{\mu\nu} \delta^\mu_M\delta^\nu_N
\delta(y)\sqrt{-g^{(br)}}\right],$
%%%%%%%%%%%%%%
with the energy-momentum tensor of the theory given by the expression
\eq$\label{Tmn}
T^{(\Phi)}_{MN}=\pa_M\Phi\,\pa_N\Phi+g_{MN}\left[-\frac{\pa_L\Phi\pa^L\Phi}{2}-
V_B(\Phi)\right]+\frac{1}{\kappa_5^2}\left[\nabla_M\nabla_Nf(\Phi)-g_{MN}\Box 
f(\Phi)\right],$
%%%%%%%%%%%%%%
and the scalar-field equation
%%%%%%%%%%%%%%%
\beq
-\frac{1}{\sqrt{-g^{(5)}}}\,\pa_M\left(\sqrt{-\gfv}g^{MN}\pa_N\Phi\right)=
\frac{\pa_\Phi f}{2 \kappa^2_5} R-\pa_\Phi V_B 
-\frac{\sqrt{-g^{(br)}}}{\sqrt{-\gfv}}\,\partial_\Phi V_b\,\delta(y)\,,
\label{phi-eq-0}
\eeq
%%%%%%%%%%%%%%%%%%
respectively. 

The form of the five-dimensional gravitational background needs to be specified
next. As in \cite{KNP, KNP2}, we consider the following line-element
%%%%%%%%%%%%
\eq$\label{metric}
ds^2=e^{2A(y)}\left\{-\left[1-\frac{2m(r)}{r}\right]dv^2+2dvdr+r^2(d\theta^2+
\sin^2\theta d\varphi^2)\right\}+dy^2\,,$
%\vspace*{2em}
%%%%%%%%%%%%%%%
which describes a five-dimensional space-time warped along the fifth dimension
due to the presence of the warp factor $e^{2A(y)}$. Its four-dimensional part
has the form of a generalised Vaidya line-element: if $m(r)$ is a constant $M$,
this reduces, after a coordinate transformation, to the Schwarzschild solution.
The four-dimensional observer at $y=0$ would then see a black-hole line-element
on the brane, however, its embedding in the extra dimension as in Eq. (\ref{metric})
results, in the context of the original Randall-Sundrum model \cite{RS1,RS2}, in a
black-string solution \cite{CHR} with an infinitely-long singularity plagued by
instabilities \cite{GL, RuthGL}. 

By introducing a dependence of the mass function on the extra coordinate $y$,
it is possible to localise the black hole close to the brane but this demands
a form of bulk matter
that can not be supported by ordinary fields \cite{KT, KOT}. A more general
ansatz for the mass function of the form $m=m(v,r,y)$, that was employed in
subsequent works \cite{KPZ, KPP2}, increased the flexibility of the
model but failed, too, to lead to localised black-hole solutions in the
context of a variety of scalar-field theories. Up to today, the analytical
determination---in a closed form---of regular, localised black holes in
warped brane-world models remains an open problem.

However, the five-dimensional scalar-tensor theory of gravity described by
Eq. (\ref{action}) was shown \cite{KPZ, KPP2} to admit novel black-string
solutions that may be constructed analytically. In our previous works
\cite{KNP, KNP2}, we performed a comprehensive study of the types of
black-string solutions that emerge in the context of this theory when the
cosmological constant on the brane is positive or negative, respectively.
Here, we complete our study by considering the case of a Minkowski brane.
As we will demonstrate, this case is the most flexible of all that allows
for a larger variety of profiles for the scalar field and its coupling
function while retaining all the attractive characteristics of the previous
two cases. 

We will employ again the line-element (\ref{metric}), and proceed to derive
the explicit form of the field equations (\ref{grav_eqs})-(\ref{phi-eq-0}).
We will focus on solving this set of equations first in the bulk, and thus
ignore for now all $\delta(y)$-terms. The explicit form of the gravitational
equations follows by combining the non-vanishing components of the Einstein
$G^{M}{}_N$ and energy-momentum $T^{(\Phi)M}{}_N$ tensors. In mixed form,
these are:
%%%%%%%%%%%%
\bea
&~&G^0{}_0=G^1{}_1=6A'^2+3A''-\frac{2e^{-2A}\pa_rm}{r^2},\nonumber \\[1mm] 
&~&G^2{}_2=G^3{}_3=6A'^2+3A''-\frac{e^{-2A}\pa_r^2m}{r}, \label{GMN} \\[2mm] 
&~&G^4{}_4=6A'^2-\frac{e^{-2A}\left(2\pa_rm+r\pa_r^2m\right)}{r^2}, \nonumber
\eea
%%%%%%%%%%%%%%%%%
and  
%%%%%%%%%%%%%
\begin{gather}
T^{(\Phi)0}{}_0=T^{(\Phi)1}{}_1=T^{(\Phi)2}{}_2=T^{(\Phi)3}{}_3=
A' \Phi'\,\pa_\Phi f+\lagr_\Phi-\Box f, \nonumber \\[2mm]
T^{(\Phi)4}{}_4=(1+\pa_\Phi^2 f)\Phi'^2+\Phi''\,\pa_\Phi f+\lagr_\Phi-\Box f,
\label{TMN-mixed}
\end{gather}
%%%%%%%%%%%%%
respectively, where a prime ($'$) denotes the derivative with respect to the $y$-coordinate.
Above, we have made the assumption that the scalar field depends only
on the coordinate along the fifth dimension, i.e. $\Phi=\Phi(y)$, and we have defined
the quantities
%%%%%%%%%%%%%%%%%%
\beq\label{Lagr}\lagr_{\Phi}=-\frac{1}{2}\,\pa_L\Phi\,\pa^L\Phi-V_B(\Phi)
=-\frac{1}{2}\,\Phi'^2-V_B(\Phi),
\eeq
%%%%%%%%%%%%%%
and
%%%%%%%%%%%%%%%
\beq
\label{Box_f}
\Box f=4A' \Phi'\,\pa_\Phi f+\Phi'^2\,\pa_\Phi^2 f+\Phi''\,\pa_\Phi f.
\eeq
%%%%%%%%%%%%%%%%%%
Employing the above, and upon some simple manipulation \cite{KNP}, we obtain
three equations having the following form
%%%%%%%%%%
\beq
\label{eq-mass}
r\,\pa_r^2m-2\pa_rm=0\,,
\eeq
%%%%%%%%%%%
\beq \label{grav-1}
f\left(3A''+e^{-2A}\frac{\pa_r^2m}{r}\right)=\pa_\Phi f \left(A'\Phi'-\Phi''\right)
-(1+\pa_\Phi^2 f)\Phi'^2\,,
\eeq
%%%%%%%%%%%%%
\beq
\label{grav-2}
f\left(6A'^2+3A''-\frac{2e^{-2A}\pa_rm}{r^2}\right)=A'\Phi'\,\pa_\Phi f+
\lagr_\Phi -\Box f-\Lambda_5\,.
\eeq
%%%%%%%%%%%%%%
Note that, for notational simplicity, we have absorbed the gravitational constant
$\kappa_5^2$ in the expression of the general coupling function $f(\Phi)$. 
Turning next to the scalar-field equation in the bulk (\ref{phi-eq-0}), this takes 
the explicit form
\beq \label{phi-eq}
\Phi'' + 4A' \Phi' =\pa_\Phi f \left(10A'^2+4A''-e^{-2A}\frac{2\pa_rm+
r\,\pa_r^2m}{r^2}\right) +\pa_\Phi V_B\,.
\eeq
%%%%%%%%%%%%%%

In order to increase the flexibility of the theory, the form of the mass function
$m=m(r)$ in the gravitational background (\ref{metric}) was left arbitrary.
Nevertheless, this will be duly determined via Eq. (\ref{eq-mass}); by direct
integration, we obtain the unique solution 
%%%%%%%%%%%%%%
\beq
m(r)=M+ \Lambda r^3/6\,, \label{mass-sol}
\eeq
%%%%%%%%%%%%%%%
where $M$ and $\Lambda$ are arbitrary integration constants. The
projected-on-the-brane gravitational background follows by setting $y=0$
in the line-element (\ref{metric}) and using the above result for the mass
function; then, we find the expression
%%%%%%%%%%%%%%%%%
\eq$\label{metric-brane}
ds^2_4=-\left(1-\frac{2M}{r}-\frac{\Lambda r^2}{3}\right)dv^2+2dv dr+
r^2(d\theta^2+\sin^2\theta\ d\varphi^2)\,.$
%%%%%%%%%%%%%%%
By employing an appropriate coordinate transformation, the above Vaidya form
of the four-dimensional line-element may be transformed to the usual 
Schwarzschild-(anti-)de Sitter solution \cite{KNP}. As a result, we may 
interpret the two arbitrary parameters $M$ and $\Lambda$ as the mass of
the black-hole on the brane and the cosmological constant on the brane. 
The cases of positive and negative cosmological constant on the brane
(i.e. $\Lambda>0$ and $\Lambda<0$) were studied respectively in our previous
two works \cite{KNP, KNP2}; in the context of the present analysis, we will
focus on the case of a zero four-dimensional cosmological constant
($\Lambda = 0$).

Returning to the remaining field equations (\ref{grav-1})-(\ref{phi-eq}),
one may demonstrate that only two of them are independent \cite{KNP}. 
We may therefore ignore altogether the scalar field equation (\ref{phi-eq})
and work only with the gravitational equations (\ref{grav-1})-(\ref{grav-2}).
The former equation will provide the solution for the scalar field $\Phi$
while the latter will help us to determine the scalar potential in the
bulk $V_B(\Phi)$. To this end, we need also the expression of the warp function
$A(y)$ for which we will use the well-known form $A(y)=-k |y|$ \cite{RS1, RS2},
with $k$ a positive constant, as this ensures the localization of gravity
near the brane. Setting also the mass function to be $m(r)=M$ (since $\Lambda=0$),
Eq. (\ref{grav-1}) takes the form\footnote{We assume a
${\bf Z}_2$-symmetry in the bulk under the change $y \rightarrow -y$ therefore,
henceforth, we focus on the positive $y$-regime.}
\eq$\label{grav-1-1}
(1+\pa_\Phi^2f)\Phi'^2+\pa_\Phi f(\Phi''+k \Phi')=0\,,$
or
\eq$\label{grav-1-2}
\Phi'^2+\pa_y^2f+k\,\pa_yf=0\, ,$
while Eq. \eqref{grav-2}, with the use of Eq. \eqref{grav-1-2}, can be solved for $V_B(y)$:
\eq$\label{V-B}
V_B(y)=-\Lambda_5-6k^2 f(y)+\frac{7}{2}\,k\,\pa_y f-\frac{1}{2}\,\pa_y^2f\, .$
In the above, we have also used the relations
\eq$\label{dif-f}
\pa_yf=\Phi' \,\pa_\Phi f, \hspace{1.5em}\pa_y^2f=
\Phi'^2\,\pa_\Phi^2 f+\Phi''\,\pa_\Phi f\,.$

The topology of the 5-dimensional spacetime in the bulk may be inferred from the
form of the curvature invariant quantities. Using the 5-dimensional line-element
(\ref{metric}), together
with the relations $m(r)=M$ and $A=-k |y|$, we find the following expressions
%%%%%%%%%%%%%%%%
\beq
R=-20 k^2\,, \quad R_{MN} R^{MN}=80 k^4\,, \quad 
R_{MNRS} R^{MNRS}= 40 k^4 +\frac{48 M^2\,e^{4k|y|}}{r^6}\,.
\eeq
%%%%%%%%%%%%%%%%%
For $M=0$, the bulk spacetime is characterised by a constant
negative curvature at every point, and is therefore an AdS spacetime. This holds
despite the presence of a non-trivial distribution of energy in the bulk, i.e. that
of a non-minimally coupled scalar field with a potential, and is ensured through
the field equations which, like Eqs. (\ref{grav-1-2}) and (\ref{V-B}), relate the
different bulk quantities among themselves. It is for this reason that, as we will
see, the exponentially decaying warp factor will be supported even in the absence
of the negative bulk cosmological constant $\Lambda_5$. In the case where
$M \neq 0$, the above invariants describe a 5-dimensional black-string solution
with an infinitely-long spacetime singularity extending throughout the extra
dimension. The black-string  singularity reaches the boundary of spacetime
which is by itself a singular hypersurface.

The solution for both the scalar field and the bulk potential depends, through 
Eqs. (\ref{grav-1-2})-(\ref{V-B}), on the form of the non-minimal coupling function $f(\Phi)$. 
In our previous work \cite{KNP2}, we assigned the following constraints to the
scalar field $\Phi(y)$ and its coupling function $f[\Phi(y)]$:
\begin{enumerate}
\item[\bf(i)] Both functions should be real and finite in their whole domain 
and of class $C^{\infty}$.
\item[\bf(ii)] At $y\ra \pm\infty$, both functions should satisfy the following relations, otherwise
the finiteness of the theory at infinity cannot be ascertained,
\eq$\label{con.1}
\lim_{y\ra\pm\infty}\frac{d^n[f(y)]}{dy^n}=0,\hspace{1.5em}\forall n\geq 1,$
\eq$\label{con.2}
\lim_{y\ra\pm\infty}\frac{d^n[\Phi(y)]}{dy^n}=0,\hspace{1.5em}\forall n\geq 1.$
\end{enumerate} 
The second constraint amounts to considering profiles of the scalar field and
forms of the coupling function that both reduce to a constant value far away
from the brane. Together with the first constraint, they ensure a physically
acceptable behaviour for our scalar-tensor theory. The sign, however, of the
coupling function $f(y)$ will not be fixed. In \cite{KNP}, where the case of
a positive cosmological constant on the brane was studied, i.e. $\Lambda>0$, 
the coupling
function had to be negative-definite away from our brane for the reality of
the scalar field to be ensured; nevertheless, the effective theory on the brane
could still be well-defined. In the case of $\Lambda <0$ \cite{KNP2},
%a negative cosmological constant on the brane 
no such requirement was necessary and the coupling 
function was assumed to be everywhere positive-definite in terms of the
$y$-coordinate; then, gravity was normal over the entire five-dimensional
space-time leading to a well-defined effective field theory on the brane.

In the context of the present analysis, where $\Lambda=0$, we may consider
coupling functions that are either positive or negative-definite for particular
regions of the $y$-coordinate. As we will demonstrate, it is possible to obtain
a positive effective four-dimensional gravitational constant in every case. This
will hold even when five-dimensional gravity behaves in an anti-gravitating way
at particular regimes of space-time---as it turns out, such a behaviour is not 
physically forbidden as long as the effective theory on our brane is well-defined.
To this end, the derivation of the effective theory on the brane is going to play
an important role in our forthcoming analysis, and will thus supplement every
bulk solution we derive. 

%%%%%%%%%%%%%%%%%%%%%%%%%%%%%%%%%%%%%%%%%%%%%%%%%%%%%%%%%%%%%%%%%%%%%%%%%%%%
%
%
%%%%%%%%%%%%%%%%%%%%%%%%%%%%%%%%%%%%%%%%%%%%%%%%%%%%%%%%%%%%%%%%%%%%%%%%%%%%
 
\section{A Linear Coupling Function}
\label{linear}

Choosing $\Lambda=0$ on our brane simplifies the set of field equations of
the theory, but more importantly, relaxes constraints that had to be imposed
on the coupling function. As a result, the latter is now allowed to adopt a
variety of physically-acceptable forms, all obeying the criteria (i) and (ii)
of the previous section. These forms lead to viable brane-world models (for
$M=0$) or black-string solutions (for $M \neq 0$). In an effort to construct
the most realistic solutions, we will also study, in every case, the energy
conditions both in the bulk and on the brane.

We start our analysis with the case of the linear coupling function:
\eq$\label{linear-f}
f(\Phi)=f_0+\Phi_0\Phi\, ,$
where $f_0$ and $\Phi_0$ are arbitrary parameters of the theory. In what
follows, we will first solve the system of field equations 
(\ref{grav-1-1}) and (\ref{V-B}) in the bulk and then consider the
effective theory on the brane as well as the energy conditions.

\subsection{The bulk solution}
\label{linear-bulk}

\par Substituting the aforementioned coupling function in Eq. \eqref{grav-1-1}
and solving the resulting second-order differential equation, we obtain the solution:
\eq$
\Phi(y)=\Phi_0\left[-ky+\ln(e^{ky}+\xi)\right], \label{linear-Phi}$
where $\xi$ is an integration constant.
Note that the gravitational field equation \eqref{grav-1-1} possesses a
translational symmetry with respect to the scalar field $\Phi(y)$.
Hence, we are free to fix the value of a second integration constant,
that should in principle appear additively on the right-hand-side of
Eq. (\ref{linear-Phi}), to zero without loss of generality. Then,
using Eq. \eqref{linear-Phi} in \eqref{linear-f}, we find
\eq$\label{linear-f-y}
f(y)=f_0+\Phi_0^2\left[-ky+\ln(e^{ky}+\xi)\right] .$
As we mentioned earlier, both functions $f(y)$ and $\Phi(y)$ should be real and finite;
therefore $\xi\in(-1,0)\cup(0,\infty)$, and $\Phi_0\in\mathbb{R}\setminus\{0\}$.
It is clear from Eqs. \eqref{linear-Phi} and \eqref{linear-f-y} that if we allow
$\xi$ to become equal to zero, then we nullify the scalar field everywhere in the
bulk and reduce the coupling function to a constant, which makes our model trivial.
The allowed range of values for the parameter $f_0$ will be determined shortly.

\begin{figure}[t]
\begin{center}
 \begin{subfigure}[b]{0.47\textwidth}
        \includegraphics[width=\textwidth]{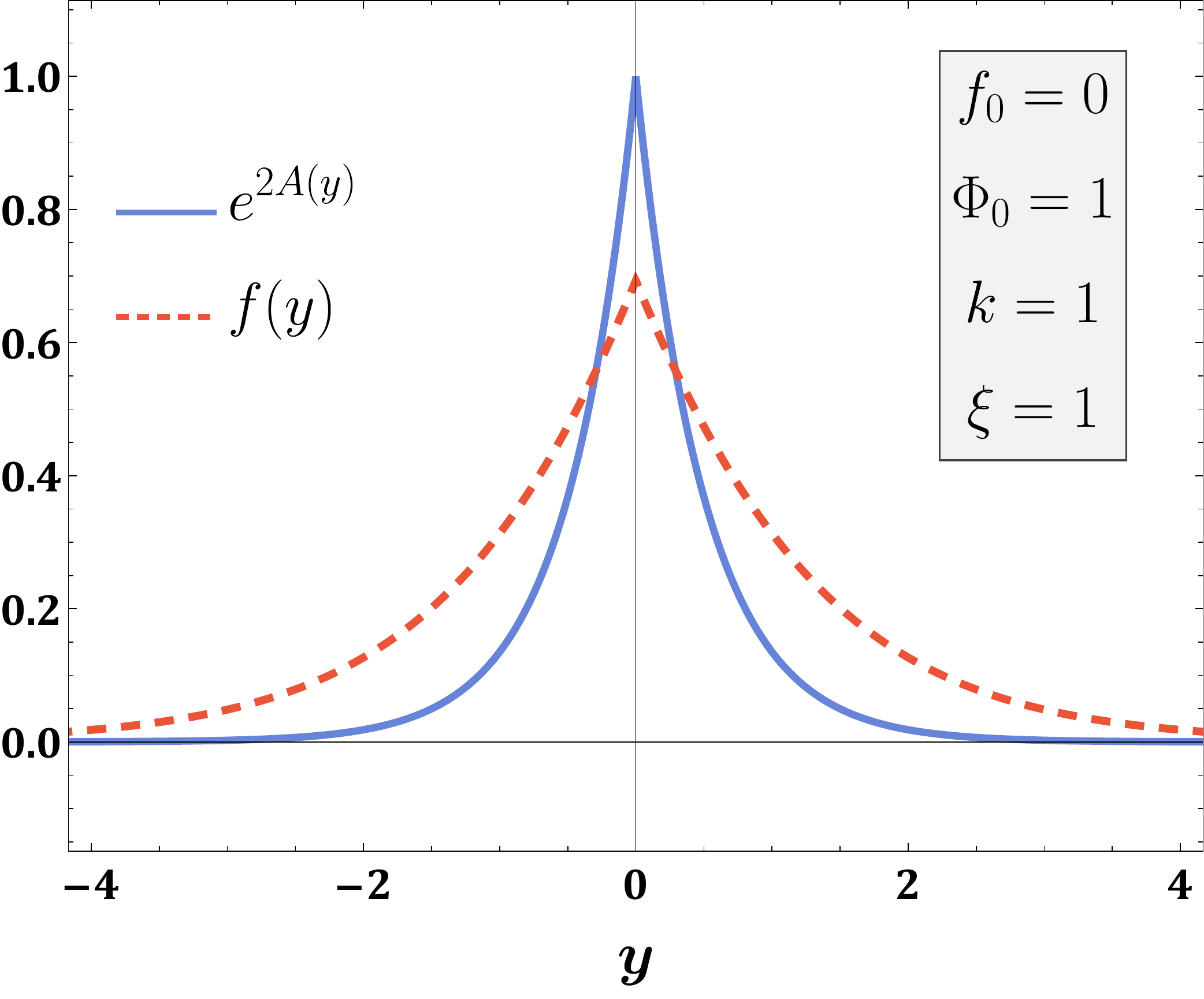}
        \caption{\hspace*{-1.5em}}
        \label{linear-plot1}
    \end{subfigure}
    ~ %add desired spacing between images, e. g. ~, \quad, \qquad, \hfill etc. 
      %(or a blank line to force the subfigure onto a new line)
 \begin{subfigure}[b]{0.44\textwidth}
        \includegraphics[width=\textwidth]{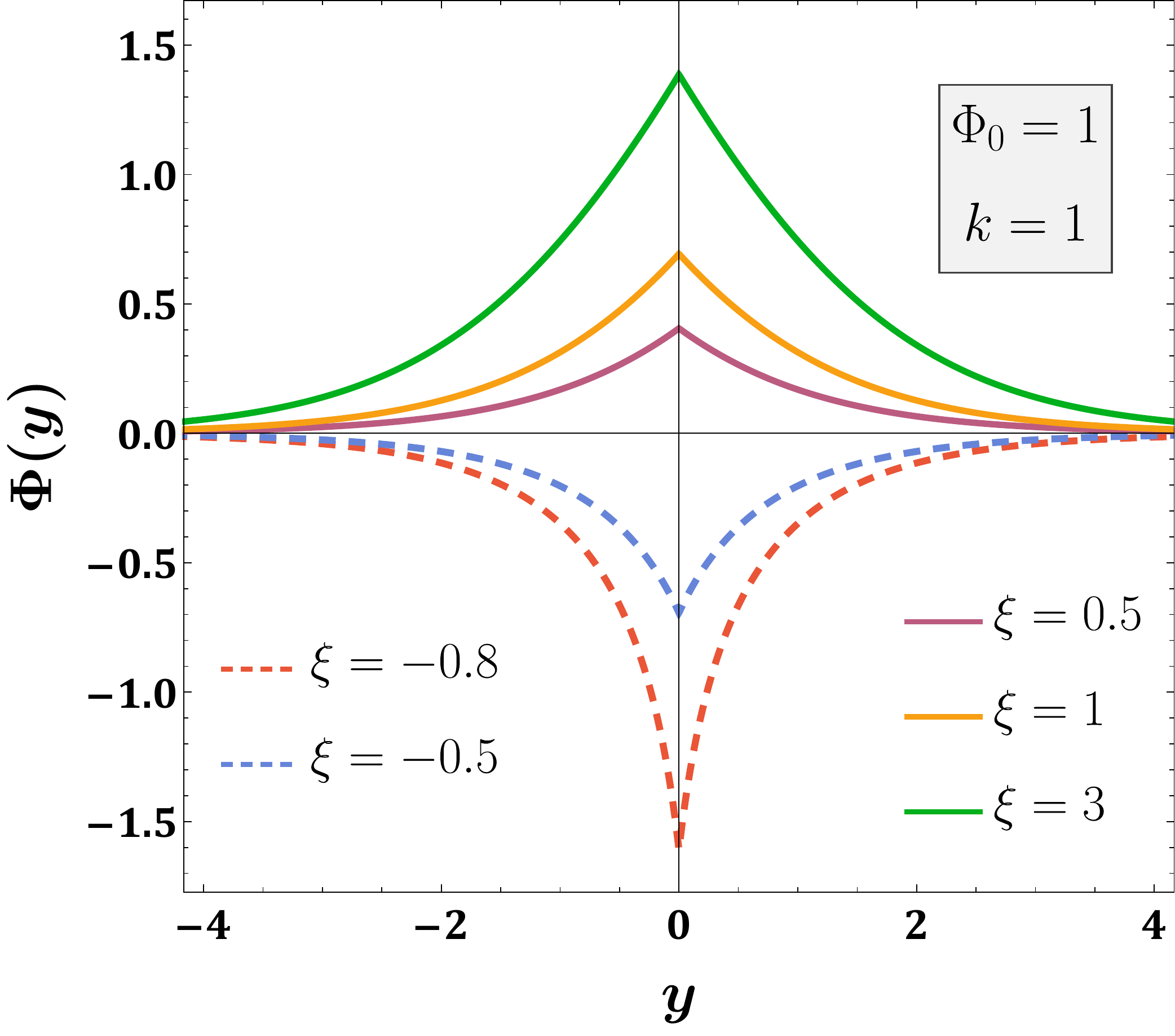}
        \caption{\hspace*{-3.7em}}
        \label{linear-plot2}
    \end{subfigure}
    ~ %add desired spacing between images, e. g. ~, \quad, \qquad, \hfill etc. 
    %(or a blank line to force the subfigure onto a new line)
    \vspace{-0.5em}
     \caption{(a) The warp factor $e^{2A(y)}=e^{-2k|y|}$ and coupling
    function $f(y)$ in terms of the coordinate $y$ for $f_0=0,\, \Phi_0=1,\,
    k=1,\,\xi=1$, and (b) the scalar field $\Phi(y)$ for
    different values of the parameter $\xi=-0.8,-0.5,0.5,\, 1,\, 3$ (from bottom 
    to top).}
    \vspace{-1em}
   \label{linear-plot-1-2}
  \end{center}
\end{figure}

In Fig. \myref{linear-plot-1-2}{linear-plot1}, we depict the warp factor $e^{2A(y)}=e^{-2k|y|}$
and coupling function $f(y)$ in terms of the coordinate $y$ for $f_0=0,\, \Phi_0=1,\,
k=1$, and $\xi=1$. We observe that, similarly to the warp factor, the coupling function
remains localised close to our brane and reduces to zero at large distances although
with a smaller rate. According to this behaviour, the non-minimal coupling of the
scalar field to the five-dimensional Ricci scalar takes its maximum value at the
location of the brane whereas, for large values of $y$, this coupling vanishes leading
to a minimally-coupled scalar-tensor theory of gravity. The profile of the scalar
field $\Phi(y)$ itself is presented in Fig. \myref{linear-plot-1-2}{linear-plot2} for
$\Phi_0=1$ and $k=1$. We also display the dependence of this profile on the
value of the parameter $\xi=-0.8,-0.5,0.5,\, 1,\, 3$ (from bottom to top).
It is clear that also the scalar field exhibits a localised behaviour with
the value of $\xi$ determining the overall sign and maximum value of $\Phi$
on our brane. The dependence of the coupling function $f(y)$ on the value of $\xi$
is similar to that of the scalar field, as one can easily deduce from the relation
\eqref{linear-f}. 

\par
\begin{figure}[t]
\begin{center}
\begin{subfigure}[b]{0.47\textwidth}
        \includegraphics[width=\textwidth]{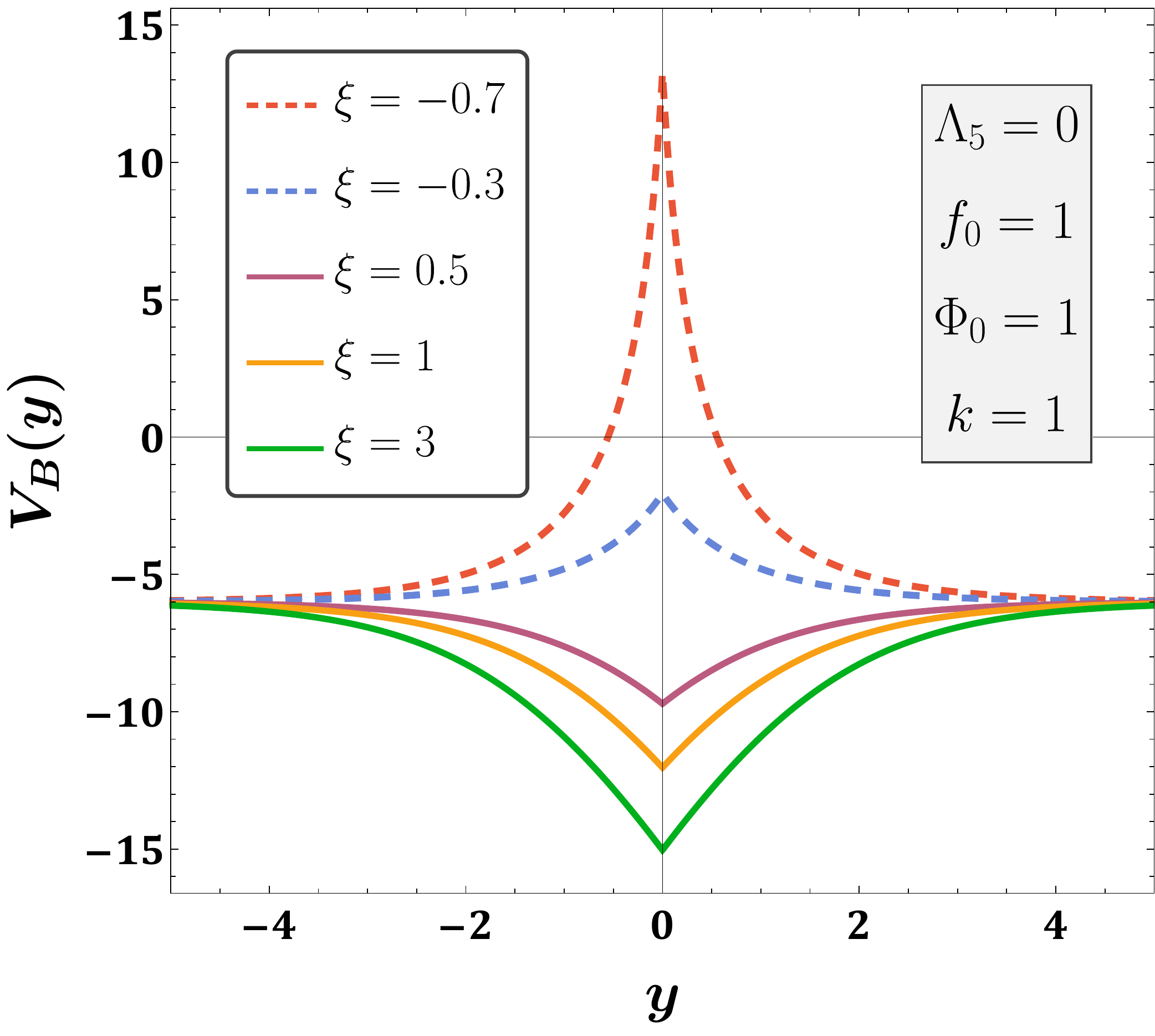}
        \caption{\hspace*{-3.7em}}
        \label{linear-plot3}
    \end{subfigure}
    \qquad %add desired spacing between images, e. g. ~, \quad, \qquad, \hfill etc. 
      %(or a blank line to force the subfigure onto a new line)
 \begin{subfigure}[b]{0.44\textwidth}
        \includegraphics[width=\textwidth]{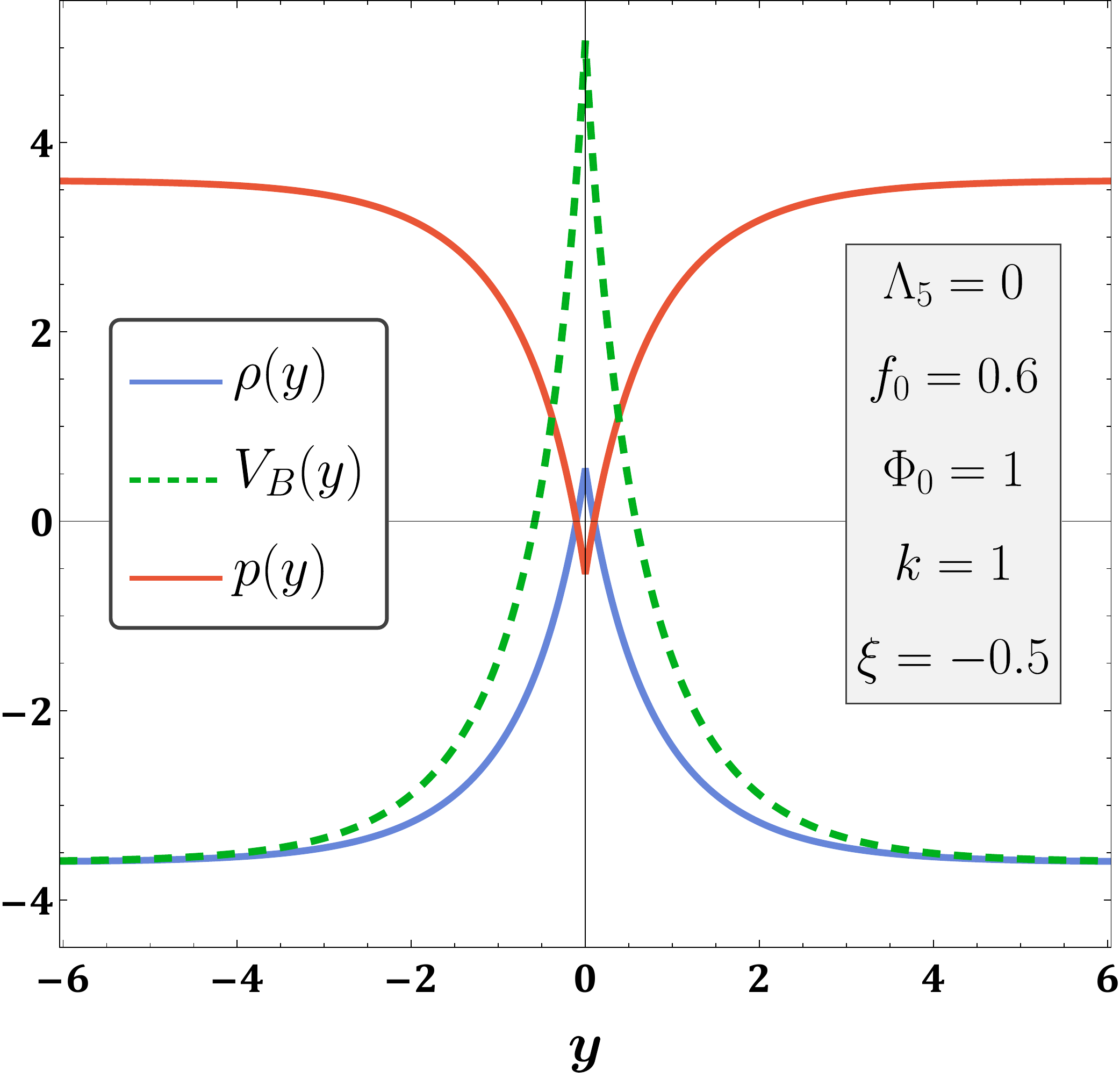}
        \caption{\hspace*{-1.3em}}
        \label{linear-plot4}
    \end{subfigure}
    ~ %add desired spacing between images, e. g. ~, \quad, \qquad, \hfill etc. 
    %(or a blank line to force the subfigure onto a new line)
    \vspace{-0.5em}
     \caption{(a) The scalar potential $V_B$ in terms of the coordinate $y$ for different
     values of the parameter $\xi=-0.7,\,-0.3,\,0.5,\, 1,\, 3$ (from top to bottom),
     (b) the energy density $\rho$, pressure components 
    $p^i=p^y=p$ and scalar potential $V_B$ in terms of the coordinate $y$ for
    the case $f_0=0.6$ and $\xi=-0.5$.}
    \vspace{-1em}
   \label{linear-plot3-4}
  \end{center}
\end{figure}

 The potential of the scalar field $V_B(y)$ in the bulk can be determined from
Eq. \eqref{V-B} using the expression of the coupling function $f(y)$ (\ref{linear-f-y}).
Thus, we obtain
\eq$\label{linear-V-y}
V_B(y)=-\Lambda_5-6k^2f_0+\frac{k^2\Phi_0^2}{2}\left[12ky-\frac{\xi(8e^{ky}+7\xi)}{(\xi+e^{ky})^2}\right]
-6k^2\Phi_0^2\ln(\xi+e^{ky})\, .$
Using Eq. \eqref{linear-Phi}, we can express the potential in terms of the scalar
field in a closed form, as follows
\eq$\label{linear-V}
V_B(\Phi)=-\Lambda_5-6k^2f_0-6k^2\Phi_0\Phi-4k^2\Phi_0^2\left(1-e^{-\Phi/\Phi_0}\right)
+\frac{k^2\Phi_0^2}{2}\left(1-e^{-\Phi/\Phi_0}\right)^2\,.$
%%%%%%%%%%
We observe that the parameter $f_0$ appearing in the expression of the coupling
function (\ref{linear-f-y}) gives a constant contribution to the scalar bulk
potential. Depending on the value of $f_0$, the asymptotic value of $V_B$ in the
bulk (when $\Phi$ vanishes) can be either positive, zero or negative. In the latter
case, this contribution may be considered to play the role of the negative bulk
cosmological constant $\Lambda_5$, which is usually introduced in an ad hoc way.
Therefore, such a quantity is not necessary any more in order to support
the exponentially decreasing warp factor \'a la Randall-Sundrum \cite{RS1, RS2}.
As mentioned earlier, it is the non-minimal coupling of the scalar field combined
with the form of the bulk potential that supports the AdS bulk spacetime and the
chosen form of the warp factor. To this end, we will henceforth choose a vanishing
value for $\Lambda_5$ in any numerical evaluation, however, for completeness,
we will retain it in our equations. The profile of the bulk potential $V_B$ is presented
in Fig. \myref{linear-plot3-4}{linear-plot3} for $f_0=1$, which leads to a negative
asymptotic value of $V_B$. The figure depicts the dependence of $V_B$ on the
parameter $\xi$: the scalar potential may be negative everywhere in the bulk or
assume a positive value on our brane depending on the value of $\xi$. 

%%%%%%%%%%%%%%%%%%%%%
%%%%%%%%%%%%%%%%%

We may also compute the components of the energy-momentum tensor of the theory in the bulk.
Using the relations $\rho=-T^0{}_0$, $p^i=T^i{}_i$, $p^y=T^y{}_y$, we obtain the following
expressions:
\gat$\label{linear-rho}
\rho(y)=-\left(T^{(\Phi)0}{}_0-\Lambda_5\right)=-6k^2f(y)\, ,\\[2mm]
\label{linear-p}
p^i(y)=T^{(\Phi)i}{}_i-\Lambda_5=6k^2f(y)\, ,\\[3mm]
\label{linear-py}
p^y(y)=T^{(\Phi)y}{}_y-\Lambda_5=6k^2f(y)\, .$
The above relations hold in general, for arbitrary form of the coupling function and profile
of the scalar field. From the above expressions, we can immediately observe that the 
energy-momentum tensor in the bulk is isotropic ($p^y=p^i\equiv p$) and satisfies an
equation of state of the form $p=-\rho$. The sign of all energy-momentum tensor
components depends on that of the coupling function: at bulk regimes where $f(y)$
is negative-definite, the energy density $\rho(y)$ will be positive while the pressure
$p(y)$ would have the opposite sign. At these regimes, the weak energy conditions\footnote{The
weak energy conditions postulate that $\rho\geq 0,\ \rho+p\geq 0$.} will be satisfied. 
We are primarily interested in satisfying these on and close to our brane.
Thus, if we impose the condition that $f(0)<0$ and combine this inequality with the form of
Eq. \eqref{linear-f-y}, we may obtain the range of values for the parameter $f_0$, with respect
to $\xi$  and $\Phi_0$, for which the weak energy conditions on our brane are satisfied.
Hence, we get:
\eq$\label{linear-weak-con}
\frac{f_0}{\Phi_0^2}<-\ln(1+\xi)\, .$
A particular, indicative case where the weak energy conditions are satisfied on our brane
is depicted in Fig. \myref{linear-plot3-4}{linear-plot4}: it corresponds to the set of values $\Phi_0=1$,
$\xi=-0.5$ and $f_0=0.6$, which satisfy the above inequality. Both the bulk potential and
energy density are positive on our brane while the pressure components assume a negative
value of equal magnitude to that of $\rho$.

%%%%%%%%%%%%%%%%%%%%%%%%%%%%%%%%%%%%%%%%%%%%%%%%%%%%%%%%%%%%%%%%%%%%%%%

\subsection{Junction conditions and effective theory}

Let us now address the junction conditions that should be imposed on our bulk solution
due to the presence of the brane at $y=0$. The energy content of the brane will be
given by the combination $\sigma + V_b(\Phi)$, where $\sigma$ is the constant self-energy
of the brane and $V_b(\Phi)$ an interaction term of the bulk scalar field with the brane. 
Since this distribution of energy is located at a single point along the extra dimension,
i.e. at $y=0$, it creates a discontinuity in the second derivatives of the warp factor,
the coupling function and the scalar field at the location of the brane. We may then
write 
$A''=\hat A'' + [A']\,\delta(y)$,  $f''=\hat f''+[f']\ \del(y)$ and $\Phi''=\hat \Phi''
+ [\Phi']\,\delta(y)$, where the hat quantities denote the distributional (i.e. regular)
parts of the second derivatives and $[\cdots]$ stand for the discontinuities of the
corresponding first derivatives across the brane \cite{BDL}. 
Then, in the complete field equations (\ref{grav-1}) and (\ref{phi-eq}), we match the
coefficients of the $\delta$-function terms\footnote{We note that the line-element
(\ref{metric}) satisfies the relation $\sqrt{-g^{(5)}}=\sqrt{-g^{(4)}}$.} and obtain
the following two conditions
%%%%%%%%%%%%%%%%%%%5
\gat$\label{jun_con1}
3f(y)[A']=-[\Phi']\,\pa_\Phi f-(\sig+V_b)\, , \\[4mm]
\label{jun_con2}
[\Phi']=4[A']\,\pa_\Phi f+\pa_\Phi V_b\, ,$
respectively, where all quantities are evaluated at $y=0$. The above expressions also
hold in general for arbitrary forms of the coupling function $f(\Phi)$. In the case
of a linear $f(\Phi)$, employing the form of the warp function $A(y)=-k |y|$ and the
solution (\ref{linear-Phi}) for the scalar field $\Phi(y)$, we obtain the constraints
\gat$\label{linear-jc1}
\sig+V_b(\Phi)\Big|_{y=0}=\frac{2k\xi\Phi_0^2}{1+\xi}+6kf_0+6k\Phi_0^2\ln(1+\xi)\,, \\[4mm] 
\label{linear-jc2}
\pa_\Phi V_b\Big|_{y=0}=\frac{2k\Phi_0(4+3\xi)}{1+\xi}\,.$
In the above relations, we have used the assumed $\mathbf{Z}_2$ symmetry in the bulk.

The first constraint (\ref{linear-jc1}) relates the total energy density of the brane
with bulk parameters. It may be used to fix one of the bulk parameters of our solution,
for example, the warping constant $k$; then, the warping of space-time is naturally
determined by the distribution of energy in the bulk and on the brane. The second constraint
(\ref{linear-jc2}) may in turn be used to fix one parameter of the brane interaction term
$V_b$ of the scalar field. Going further, we may demand that, for physically interesting
situations, the total energy density of the brane should be positive; then, the
r.h.s. of Eq. (\ref{linear-jc1})
leads to 
\eq$\label{linear-brane-ene-con}
\frac{f_0}{\Phi_0^2}>-\ln(1+\xi)-\frac{\xi}{3(1+\xi)}\, .$
The above is therefore an additional constraint that the bulk parameters ($f_0, \Phi_0, \xi$)
should satisfy which, as the one of Eq. (\ref{linear-weak-con}), follows not from the
mathematical consistency of the solution but from strictly physical arguments.

We now turn to the effective theory on the brane that follows by integrating the complete
five-dimensional theory, given by $S=S_B+S_{br}$, over the fifth coordinate $y$. We would
like to derive first the effective four-dimensional gravitational constant that governs
all gravitational interactions on our brane. For this, it is of key importance to express
the five-dimensional Ricci scalar $R$ in terms of the four-dimensional projected-on-the-brane
Ricci scalar $R^{(4)}$. One can easily prove that the five-dimensional Ricci scalar $R$ of
the following line-element
\eq$\label{gen-ds2}
ds^2=e^{-2k|y|}g^{(br)}_{\mu\nu}(x)\, dx^\mu dx^\nu+dy^2\,$
can be written in the form
\eq$\label{Ricci-5}
R=-20k^2+8k\frac{d^2|y|}{dy^2}+e^{2k|y|}R^{(4)}\, .$
Equation \eqref{Ricci-5} holds even if the projected-on-the-brane four-dimensional metric
$g^{(br)}_{\mu\nu}$ leads to a zero four-dimensional Ricci scalar $R^{(4)}$ when the latter
is evaluated for particular solutions (as is the case for our Vaidya induced metric).
The part of the complete action $S=S_B+S_{br}$ that is relevant for the evaluation of the
effective gravitational constant is the following:
\gat$ \label{action_eff}
S\supset\int d^4 x\,dy\,\sqrt{-g^{(5)}}\ \frac{f(\Phi)}{2}e^{2k|y|}R^{(4)}\,.$
%%%%%%%%%%%%%%%%%%
Then, using also that $\sqrt{-g^{(5)}}=e^{-4k|y|} \sqrt{-g^{(br)}}$, where
$g^{(br)}_{\mu\nu}$ is the metric tensor of the projected on the brane space-time,
the four-dimensional, effective gravitational constant is given by the integral
%%%%%%%%%%%%%%%%%%%
\eq$\frac{1}{\kappa_4^2}\equiv 2\,\int_{0}^{\infty} dy\, e^{-2 k y}\,f(y)
=2\,\int_{0}^{\infty} dy\ e^{-2ky}\left[f_0-\Phi_0^2\, ky+\Phi_0^2\ln(e^{ky}+\xi)\right]\,.
\label{linear_effG}$
Using the relation $1/\kappa_4^2=M_{Pl}^2/8\pi$ and calculating the above integral,
we obtain the following expression for the effective Planck scale:
\eq$
M_{Pl}^2=\frac{8\pi\Phi_0^2}{k}\left\{\frac{f_0}{\Phi_0^2}-\frac{1}{2}+\frac{1}{\xi^2}\left[\xi+
\left(\xi^2-1\right)\ln(1+\xi)\right]\right\}\,.$
Note, that, due to the localisation of both the coupling function and scalar
field close to our brane, no need arises for the introduction of a second brane in
the model. The above value for $M_{Pl}^2$ is therefore finite as demanded, however,
it is not sign-definite. We should therefore demand that the aforementioned expression
is positive-definite which leads to the third, and most, important constraint
on the values of ($f_0, \Phi_0, \xi$), namely 
\eq$\label{linear-eff-con}
\frac{f_0}{\Phi_0^2}>\frac{\xi-2}{2\xi}+\frac{1-\xi^2}{\xi^2}\ln(1+\xi)\, .$

The integral of all the remaining terms of the five-dimensional action $S=S_B+S_{br}$,
apart from the one appearing in Eq. (\ref{action_eff}), will yield the effective
cosmological constant on the brane. This is due to the fact that the scalar field
$\Phi$ is only $y$-dependent; therefore, when the integration over the extra 
coordinate $y$ is performed, no dynamical degree of freedom remains in the
four-dimensional effective theory. The effective cosmological constant is
thus given by the expression
%%%%%%%%%%%%%%%%
\bal$\label{cosm_eff}
-\Lambda_4&=\int_{-\infty}^{\infty} dy\,e^{-4k|y|}\Bigl[-10 k^2 f(y)-\Lambda_5-
\frac{1}{2}\,\Phi'^2 -V_B(y)+f(y)(-4A'')|_{y=0}-[\sigma +V_b(\Phi)]\,\delta(y)\Bigr]
\nonum\\[1mm]
&=2\int_0^\infty dy\ e^{-4ky}\left[-10 k^2 f(y)-\Lambda_5-
\frac{1}{2}\,\Phi'^2 -V_B(y)\right]+8kf(0)-[\sigma +V_b(\Phi)]_{y=0}\,.$
In the above, we have also added the Gibbons-Hawking term \cite{Gibbons-terms}
due to the presence of the brane, that acts as a boundary for the five-dimensional
space-time.
Substituting the expressions for the coupling function and the bulk potential of the
scalar field, and employing the junction condition \eqref{linear-jc1}, we finally
obtain the result 
\eq$\Lambda_4=0\, .$
As in our previous works for positive \cite{KNP} and negative cosmological constant
\cite{KNP2} on the brane, it is clear that the parameter $\Lambda$ appearing in the expression
of the mass function \eqref{mass-sol}, and in the projected-on-the-brane line-element
\eqref{metric-brane} is indeed related to the four-dimensional cosmological constant
$\Lambda_4$. Therefore, in the context of the present analysis where we have set
$\Lambda=0$, we derived a vanishing $\Lambda_4$ as anticipated.

\subsection{The energy conditions in the parameter space}

\par
\begin{SCfigure}
    \centering
    \includegraphics[width=0.5\textwidth]{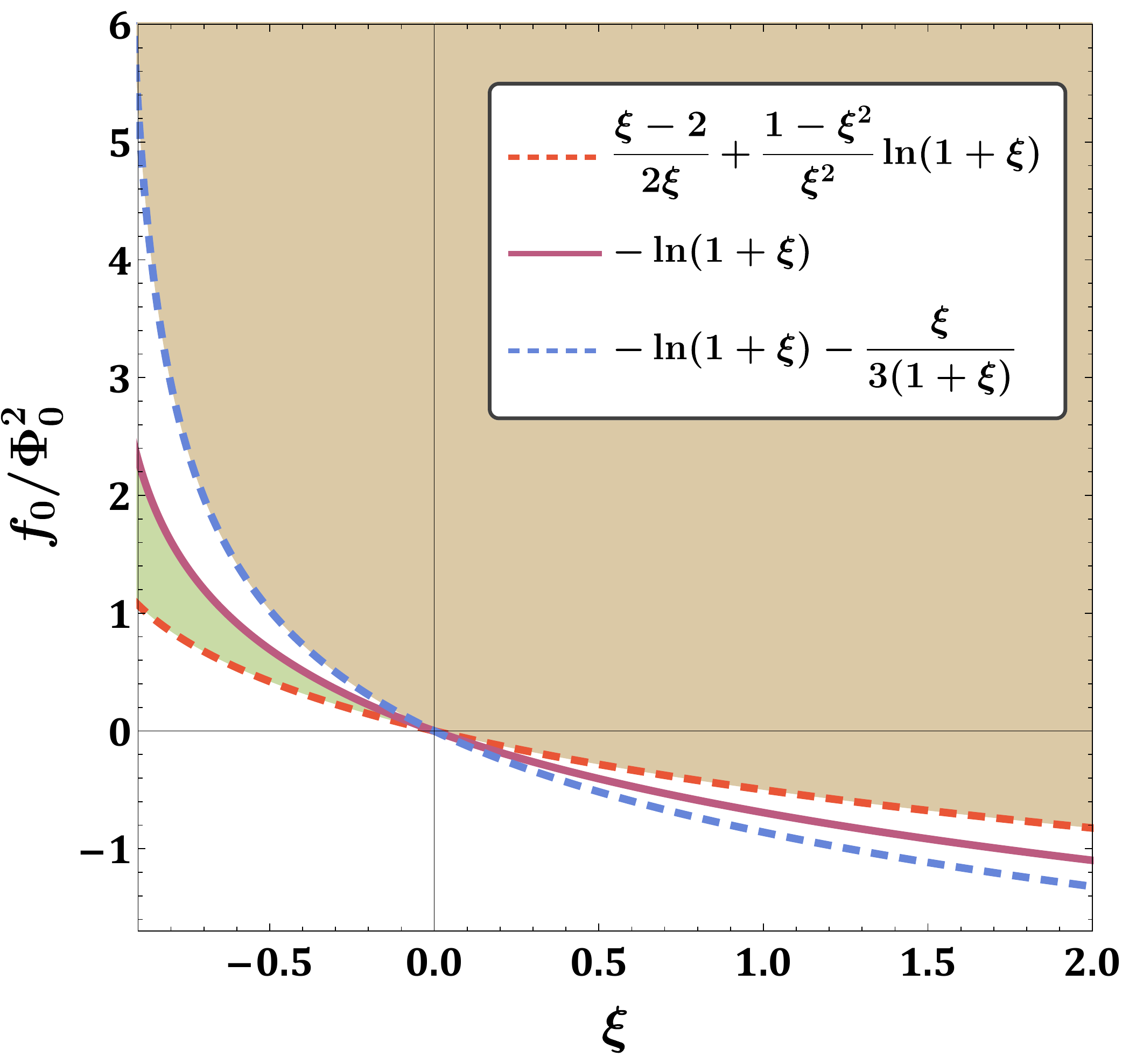}
    \caption{The parameter space between the ratio $f_0/\Phi_0^2$ 
    and the parameter $\xi$. The graphs of the expressions of the r.h.s.
    of the inequalities \eqref{linear-weak-con}, \eqref{linear-brane-ene-con}, 
    \eqref{linear-eff-con} are depicted as well.\\
    \vspace{2em}}
    \label{linear-plot-par}  
\end{SCfigure}

 We will now focus on the inequalities \eqref{linear-weak-con}, 
\eqref{linear-brane-ene-con}, and \eqref{linear-eff-con}, and in particular 
investigate whether it is possible to simultaneously satisfy all three of them.
To this end, we study the parameter space defined by the ratio $f_0/\Phi_0^2$ 
and the parameter $\xi$: this is depicted in Fig. \ref{linear-plot-par}, where we
have plotted the expressions of the r.h.s.'s of the inequalities \eqref{linear-weak-con},
\eqref{linear-brane-ene-con}, \eqref{linear-eff-con} with respect to the parameter $\xi$.
From a physical point of view, the most important inequality to satisfy is \eqref{linear-eff-con},
which ensures that the four-dimensional effective gravitational constant on our brane
is positive: this demands that $f_0/\Phi_0^2$ should be always greater than
$\frac{\xi-2}{2\xi}+\frac{1-\xi^2}{\xi^2}\ln(1+\xi)$ and corresponds to
the area above the red dashed curve in Fig. \ref{linear-plot-par}.
The inequality \eqref{linear-weak-con} ensures that the bulk energy-momentum
tensor satisfies the weak energy conditions at the location of our brane, and demands
that $f_0/\Phi_0^2$ should be smaller than $-\ln(1+\xi)$, this corresponds to the
area below the purple continuous line in Fig. \ref{linear-plot-par}. Finally,  inequality
\eqref{linear-brane-ene-con} expresses the demand that the total energy-density of our brane
is positive; this is satisfied if $f_0/\Phi_0^2$ is greater than $-\ln(1+\xi)-\frac{\xi}{3(1+\xi)}$, this
is the area above the blue dashed curve in Fig.  \ref{linear-plot-par}.

It is straightforward to see that it is impossible to satisfy all three inequalities simultaneously.
However, it is always possible to satisfy two out of these three at a time---in
Fig. \ref{linear-plot-par}, we have highlighted the regions where the most important
inequality  \eqref{linear-eff-con} is one of the two satisfied conditions---we observe that
this area covers a very large part of the parameter space. Which one of the two remaining
inequalities is the second satisfied condition depends on the value of the parameter $\xi$;
therefore, we distinguish the following cases:
\begin{enumerate}
\item[\bf(i)] For $\xi\in(-1,0)$, it is easy to see that the following sequence of
inequalities holds
\eq$\frac{\xi-2}{2\xi}+\frac{1-\xi^2}{\xi^2}\ln(1+\xi)<-\ln(1+\xi)<-\ln(1+\xi)-\frac{\xi}
{3(1+\xi)}\,.$
Thus, we can simultaneously satisfy either the inequalities 
\eqref{linear-eff-con} and \eqref{linear-weak-con} (green region in Fig. 
\ref{linear-plot-par}) or \eqref{linear-eff-con} and \eqref{linear-brane-ene-con} (brown region
in Fig. \ref{linear-plot-par}). In the former case, we have a physically acceptable four-dimensional
effective theory on the brane and the weak energy-conditions are satisfied on and close to our 
brane; the total energy density of the brane $\sig+V_b(\Phi)|_{y=0}$, however, is negative.
In the latter case, we still have a physically acceptable effective theory and
the total energy density of our brane is now positive; the weak energy conditions though
are not satisfied  by the bulk matter close to our brane. 

\item[\bf(ii)] For $\xi>0$, it now holds
\eq$-\ln(1+\xi)-\frac{\xi}{3(1+\xi)}<-\ln(1+\xi)<\frac{\xi-2}{2\xi}+\frac{1-\xi^2}{\xi^2}
\ln(1+\xi)\, ,$
In this case, we are able to simultaneously satisfy only the inequalities
\eqref{linear-eff-con} and \eqref{linear-brane-ene-con} (brown region in Fig. \ref{linear-plot-par}).
Then, we can have a regular four-dimensional effective theory and a positive total 
energy density on our brane. However, in this range of values for the parameter
$\xi$, it is impossible to satisfy the weak energy condition close to our brane
and have a well-behaved effective theory. 

\end{enumerate}

Going back to Figs. \myref{linear-plot3-4}{linear-plot3} and \myref{linear-plot3-4}{linear-plot4}, we 
observe that the solution depicted in Fig. \myref{linear-plot3-4}{linear-plot4} as well as the solution 
for $\xi=-0.7$ in Fig. \myref{linear-plot3-4}{linear-plot3} fall in the green area of Fig. \ref{linear-plot-par} 
and thus respect the energy conditions---indeed $V_B$ and $\rho$ are positive on and close to the brane. In contrast, the
remaining solutions of Fig. \myref{linear-plot3-4}{linear-plot3} belong to the brown area of
Fig. \ref{linear-plot-par}, and thus violate the weak energy conditions; they have,
however, a positive total energy-density through the junction condition \eqref{linear-jc1}.
We stress that all depicted solutions have a well-defined four-dimensional effective theory,
i.e. a positive effective gravitational constant. 

%%%%%%%%%%%%%%%%%%%%%%%%%%%%%%%%%%%%%%%%%%%%%%%%%%%%%%%%%%%%%%%%%%%%%%%%%%%%%
%
%
%%%%%%%%%%%%%%%%%%%%%%%%%%%%%%%%%%%%%%%%%%%%%%%%%%%%%%%%%%%%%%%%%%%%%%%%%%%%%

\section{A Quadratic Coupling Function}
\label{quad}

In this section, we proceed to consider the case of the quadratic coupling function, and
thus we write
\eq$\label{quad-f}
f(\Phi)=f_0+\Phi_0 \Phi+\lam \Phi^2\,,$
where again ($f_0$, $\Phi_0$, $\lambda$) are arbitrary parameters. Throughout this section,
it will be assumed that $\lambda \neq 0$
otherwise the analysis reduces to the one of the linear case studied in the previous section.
As before, we start with the derivation of the bulk solution and then turn to the effective
theory on the brane. 

\subsection{The bulk solution and the effective theory on the brane}

\par Substituting the aforementioned form of the coupling function in Eq. \eqref{grav-1-1} we obtain
the equation:
\eq$(1+2\lam)\Phi'^2+(2\lam\Phi+\Phi_0)(\Phi''+k\Phi')=0\, .$
Integrating, we find the following solution for the scalar field
\eq$\label{quad-Phi-gen}
\Phi(y)=\left\{\begin{array}{ll}
\displaystyle{\frac{1}{2\lam}\left[\Phi_1(\mu+e^{-ky})^\frac{2\lam}{1+4\lam}-\Phi_0\right]}, & 
\lam\in\mathbb{R}\setminus\{-\frac{1}{4},0\}\\[4mm]
2\Phi_0+\Phi_1\,e^{\,\mu\, e^{-ky}},& \lam=-\frac{1}{4}\end{array}\right\}\,,$
%%%%%%%%%%%%%%%%%%
where $\mu$ and $\Phi_1$ are integration constants. We note that the case with $\Phi_0=0$
was studied in \cite{Bogdanos1}; here, we generalise the aforementioned analysis by assuming
that $\Phi_0 \neq 0$. We also perform a more comprehensive analysis of the ensuing solutions
by studying the different profiles of the coupling function, scalar field and bulk potential, which
emerge as the values of the parameters of the model vary. In addition, we supplement our
analysis with the study of the effective theory on the brane and of the physical constraints
imposed on the solutions. In order to simplify our notation, we set $\Phi_1=\xi\Phi_0$, where
$\xi$ is a new integration constant. Then, Eq. \eqref{quad-Phi-gen} is written as
%%%%%%%%%%%%%%%%%%%%%
\eq$\label{quad-Phi}
\Phi(y)=\left\{\begin{array}{ll}
\displaystyle{\frac{\Phi_0}{2\lam}\left[\xi(\mu+e^{-ky})^\frac{2\lam}{1+4\lam}-1\right]}, &
\lam\in\mathbb{R}\setminus\{-\frac{1}{4},0\}\\[4mm]
\Phi_0\left(2+\xi\,e^{\,\mu\, e^{-ky}}\right),& \lam=-\frac{1}{4}\end{array}\right\}\,.$
%%%%%%%%%%%%%%
Substituting the above expression in Eq. \eqref{quad-f}, we obtain the following profile for
the coupling function in terms of the extra coordinate
%%%%%%%%%%%%%%%%%%%
\gat$\label{quad-f-y}
f(y)=\left\{\begin{array}{ll}
\displaystyle{f_0+\frac{\Phi_0^2}{4\lam}\left[\xi^2(\mu+e^{-ky})^\frac{4\lam}{1+4\lam}-1\right]}, &
\lam\in\mathbb{R}\setminus\{-\frac{1}{4},0\}\\[4mm]
\displaystyle{f_0+\Phi_0^2\left(1-\frac{\xi^2}{4}\,e^{2\mu\, e^{-ky}}\right)},& \lam=-\frac{1}{4}
\end{array}\right\}\,.$

\par The theory seems to contain five independent parameters: $f_0$, $\Phi_0$, $\lambda$,
$\mu$ and $\xi$. However, the range of values for two of these will be constrained by the
physical demands imposed on the model. To start with, both the scalar field $\Phi(y)$ and the coupling
function $f(y)$ must be real and finite in their whole domain, according to the discussion in Sec. \ref{th-frame}.
From Eq. \eqref{quad-Phi}, we observe that the allowed range of values of the parameter
$\mu$ depends on the values that the parameter $\lambda$ assumes. In Appendix
\ref{App-mu}, we consider in detail all possible values for $\lambda$ and the ensuing
allowed ranges of values for $\mu$ -- the different cases and corresponding results are
summarised in Table \ref{quad-par-val}\footnote{The symbol $\land$ that was used in Table \ref{quad-par-val}
simply denotes the \textit{logical and}. For example, the statement $A\land B$ is true if $A$ and $B$ are both
true; else it is false.}.
%
%%%%%%%%%%%%%%%%%%%%%%%%%%%%%%%%%%%%%%%%%%%%%%%%%%%%%%
\begin{center} 
\begin{table}[t]
\hspace*{0.8cm}
\begin{tabular}{ |c|c|c| } 
\hline
\multicolumn{3}{|c|}{\textbf{Range of values for all parameters}}\\
\hline
\multirow{20}{10em}{\begin{center}\vspace{-3.5em}$\xi\in\mathbb{R}\setminus\{0\}$,\\ \vspace{1em}$\Phi_0
\in\mathbb{R}\setminus\{0\}$,\\ \vspace{1em}$f_0$ is given by Eq. \eqref{quad-eff-con}.
\end{center}}
& \multirow{2}{7em}{$\hspace{2.5em}\lam>0$} & \multirow{2}{5em}{\hspace{1em}$\mu\geq 0$}\\
&  & \\ \cline{2-3}
& \multirow{2}{15em}{\hspace{0.5em}\small{$\displaystyle{\lam\in\left(-\frac{1}{4},0
\right)\ \land\ \frac{2\lam}{1+4\lam}\neq n,\  n\in\mathbb{Z}^<}$}}
& \multirow{4}{5em}{\hspace{1em}$\mu> 0$} \\
& & \\
& & \\
& & \\ \cline{2-3}
& \multirow{2}{15em}{\hspace{2.9em}\small{$\displaystyle{\lam\in\left(-\frac{1}{4},0\right)\ \land\ 
\frac{2\lam}{1+4\lam}=n,\  n\in\mathbb{Z}^<}$}}
& \multirow{4}{11em}{\hspace{0.2em}$\mu\in(-\infty,-1)\cup(0,+\infty)$} \\
& &\\
& & \\
& & \\ \cline{2-3}
& \multirow{2}{7em}{$\hspace{1.7em}\displaystyle{\lam=-1/4}$} 
& \multirow{2}{10em}{\vspace{0em}$\mu\in(-\infty,0)\cup(0,+\infty)$}\\
&  & \\ \cline{2-3}
& \multirow{3}{17em}{\hspace{2.9em}\small{$\displaystyle{\lam<-\frac{1}{4}\ \land\ \frac{2\lam}
{1+4\lam}\neq n,\  n\in\mathbb{Z}^>}$}}
& \multirow{3}{3em}{\hspace{0.2em}$\mu\geq 0$} \\
& &\\
& & \\ \cline{2-3}
& \multirow{3}{17em}{\hspace{2.9em}\small{$\displaystyle{\lam<-\frac{1}{4}\ \land\ \frac{2\lam}
{1+4\lam}= n,\  n\in\mathbb{Z}^>}$}}
& \multirow{3}{3em}{\hspace{0.2em}$\mu\in\mathbb{R}$} \\
%& \multirow{3}{20em}{\small{\hspace{3em}$\displaystyle{\lam<-\frac{1}{4}\
%\&\ \frac{2\lam}{1+4\lam}=2n},\ n\in\mathbb{Z}^>$}}
%& \multirow{3}{3em}{$\mu\in\mathbb{R}$}\\
& &\\
& & \\ \cline{2-3}
\hline
\end{tabular}
\caption{Range of values for all parameters of the model.}
\label{quad-par-val}
\end{table}
\end{center}
%%%%%%%%%%%%%%%%%%%%%%%%%%%%%%%%%%%%%%%%%%%%

\vskip -0.5cm
In addition, from the analysis of the previous section, it became clear that the theory is not
robust unless a positive effective gravitational constant is obtained on the brane. This
demand will impose a constraint on one of the remaining parameters of the theory: we
choose this parameter to be $f_0$. Thus, in order to appropriately choose the values
of $f_0$ to study the profile of the scalar field and coupling function, at this point we
turn to the effective theory and compute the effective gravitational constant. We will
employ Eq. \eqref{action_eff}, and consider separately the cases with $\lam\neq-1/4$
and $\lam=-1/4$. In the first case, using also Eq. \eqref{quad-f-y}, we obtain
%%%%%%%%%%%%%%%%%%%
\bal$\frac{1}{\kappa_4^2}=2 \int_{0}^{\infty} dy\, e^{-2 k y}\,f(y)=
\frac{1}{k}\left(f_0-\frac{\Phi_0^2}{4\lam}\right)+\frac{\Phi_0^2\xi^2}{2\lam}
\int_{0}^{\infty} dy\ e^{-2ky}\left(\mu+e^{-ky}\right)^\frac{4\lam}{1+4\lam}\, .$
%%%%%%%%%%%%
In order to evaluate the integral on the r.h.s. of the above equation, we perform the change
of variable $t=e^{-ky}$. 
%Thus, we have:
%\bal$\label{quad-int}
%\int_{0}^{\infty} dy\,e^{-2ky}\left(\mu+e^{-ky}\right)^\frac{4\lam}{1+4\lam}&=
%\frac{\mu^{\frac{4\lam}{1+4\lam}}}{k}\int_0^1 dt\ t\left[1-\left(-\frac{1}{\mu}\right)t
%\right]^{\frac{4\lam}{1+4\lam}}%\nonum\\[2mm]
%=\frac{\mu^{\frac{4\lam}{1+4\lam}}}{2k}\,_2F_1\left(-\frac{4\lam}{1+4\lam},2;3;
%-\frac{1}{\mu}\right)\, ,$
If we also use the integral representation of the hypergeometric function \cite{Abramowitz}
\eq$\label{int-rep-hyper}
\,_2F_1\left(a,b;c;z\right)=\frac{\Gamma(c)}{\Gamma(b)\Gamma(c-b)}\int_0^1
dt\ t^{b-1}(1-t)^{c-b-1}(1-zt)^{-a},\hspace{1.5em}Re(c)>Re(b)>0\, ,$
we finally obtain the result:
%%%%%%%%%%%%%%
\eq$\frac{1}{\kappa_4^2}=\frac{M_{Pl}^2}{8\pi}=\frac{\Phi_0^2}{k}
\left[\frac{f_0}{\Phi_0^2}-\frac{1}{4\lam}+\frac{\xi^2\,\mu^{
\frac{4\lam}{1+4\lam}}}{4\lam}\,_2F_1\left(-\frac{4\lam}{1+4\lam},2;3;-\frac{1}{\mu}
\right)\right]\, .$
%%%%%%%%%%%%%%%%%%%

We can further simplify the above expression using the following relations
%%%%%%%%%%%%%%%%%%%%%
\eq$\,_2F_1\left(a,2;3;z\right)=\left\{\begin{array}{ll}
\displaystyle{\frac{2(1-z)^{-a}\left[z(a+z-az)+(1-z)^a-1\right]}{(a-2)(a-1)z^2}}\, , & a\in
\mathbb{R}\setminus\{1,2\}\\[3mm]
\displaystyle{\frac{2}{z^2}\left[-z-\ln(1-z)\right]}\, , & a=1\\[3mm]
\displaystyle{\frac{2}{z^2(1-z)}\left[z+\ln(1-z)-z\ln(1-z)\right]}\, , &a=2
\end{array}\right\}\, .$
Then, for $\lam\neq-1/4$, the four-dimensional effective Planck scale may be written in
terms of elementary functions as follows
%%%%%%%%%%%%%%%%%%%%%%%%
\eq$
M_{Pl}^2=\left\{\begin{array}{ll}
\frac{8\pi\Phi_0^2}{k}\left\{\frac{f_0}{\Phi_0^2}-\frac{1}{4\lam}+\frac{(1+
4\lam)^2\xi^2}{4\lam(1+6\lam)(1+8\lam)}\left[\mu^{\frac{2+12\lam}{1+4\lam}}-
(1+\mu)^{\frac{1+8\lam}{1+4\lam}}\left(\mu-\frac{1+8\lam}{1+4\lam}\right)
\right]\right\}, &\lam\in\mathbb{R}\setminus\{-\frac{1}{4},-\frac{1}{6},-
\frac{1}{8},0\}\\[4mm]
\frac{8\pi\Phi_0^2}{k}\left\{\frac{f_0}{\Phi_0^2}+2-4\xi^2
\left[1-\mu\ln\left(\frac{1+\mu}{\mu}\right)
\right]\right\}, & \lam=-\frac{1}{8}\\[4mm]
\frac{8\pi\Phi_0^2}{k}\left\{\frac{f_0}{\Phi_0^2}+\frac{3}{2}-\frac{3\xi^2
}{1+\mu}\left[-1+(1+\mu)\ln\left(\frac{1+\mu}
{\mu}\right)\right]\right\}, & \lam=-\frac{1}{6}\end{array}\right\}. \label{MPl-quad}$
%%%%%%%%%%%%%%%%%%
On the other hand, for $\lam=-1/4$, we readily obtain
\gat$M_{Pl}^2=\frac{8\pi}{\kappa_4^2}=\frac{8\pi\Phi_0^2}{k}\left\{\frac{f_0}{\Phi_0^2}+1-
\frac{\xi^2}{8\mu^2}\left[1+e^{2\mu}(2\mu-1)
\right]\right\}\,. \label{Mpl-1/4}$
%%%%%%%%%%%%%%%%%%%%%

Since the effective four-dimensional gravitational scale $M_{Pl}^2$ should be
a positive number, Eqs. \eqref{MPl-quad} and \eqref{Mpl-1/4} impose the following constraints
on the values of the ratio $f_0/\Phi_0^2$:
%%%%%%%%%%%%%%%%%%%
\eq$\label{quad-eff-con}
\begin{array}{ll}
\frac{f_0}{\Phi_0^2}>\frac{1}{4\lam}\left\{1-\frac{(1+
4\lam)^2\xi^2}{(1+6\lam)(1+8\lam)}\left[\mu^{\frac{2+12\lam}{1+4\lam}}-
(1+\mu)^{\frac{1+8\lam}{1+4\lam}}\left(\mu-\frac{1+8\lam}{1+4\lam}\right)
\right]\right\}, &\lam\in\mathbb{R}\setminus\{-\frac{1}{4},-\frac{1}{6},-
\frac{1}{8},0\}\\[5mm]
\frac{f_0}{\Phi_0^2}>-2\left\{1-2\xi^2
\left[1-\mu\ln\left(\frac{1+\mu}{\mu}\right)
\right]\right\}, & \lam=-\frac{1}{8}\\[5mm]
\frac{f_0}{\Phi_0^2}>-\frac{3}{2}\left\{1-\frac{2\xi^2
}{1+\mu}\left[-1+(1+\mu)\ln\left(\frac{1+\mu}
{\mu}\right)\right]\right\}, & \lam=-\frac{1}{6}\\[5mm]
\frac{f_0}{\Phi_0^2}>-1+\frac{\xi^2}{8\mu^2}\left[
1+e^{2\mu}(2\mu-1)\right], & \lam=-\frac{1}{4}\end{array}$
%%%%%%%%%%%%%%%%%%%
We choose to use the above constraints in order to limit the range of values of the
parameter $f_0$. The remaining parameters $\Phi_0$, $\lam$ and $\xi$ may then take
values in almost the entire set of real numbers, specifically $\Phi_0\in\mathbb{R}\setminus\{0\}$, 
$\xi\in\mathbb{R}\setminus\{0\}$ and $\lam\in\mathbb{R}\setminus\{0\}$.  These ranges of
values are also summarised in Table \ref{quad-par-val}. We finally note that the above constraints
for the positivity of the effective four-dimensional gravitational constant allow for both positive
and negative values of the parameter $f_0$.

\begin{figure}[t!]
    \centering
    \begin{subfigure}[b]{0.47\textwidth}
        \includegraphics[width=\textwidth]{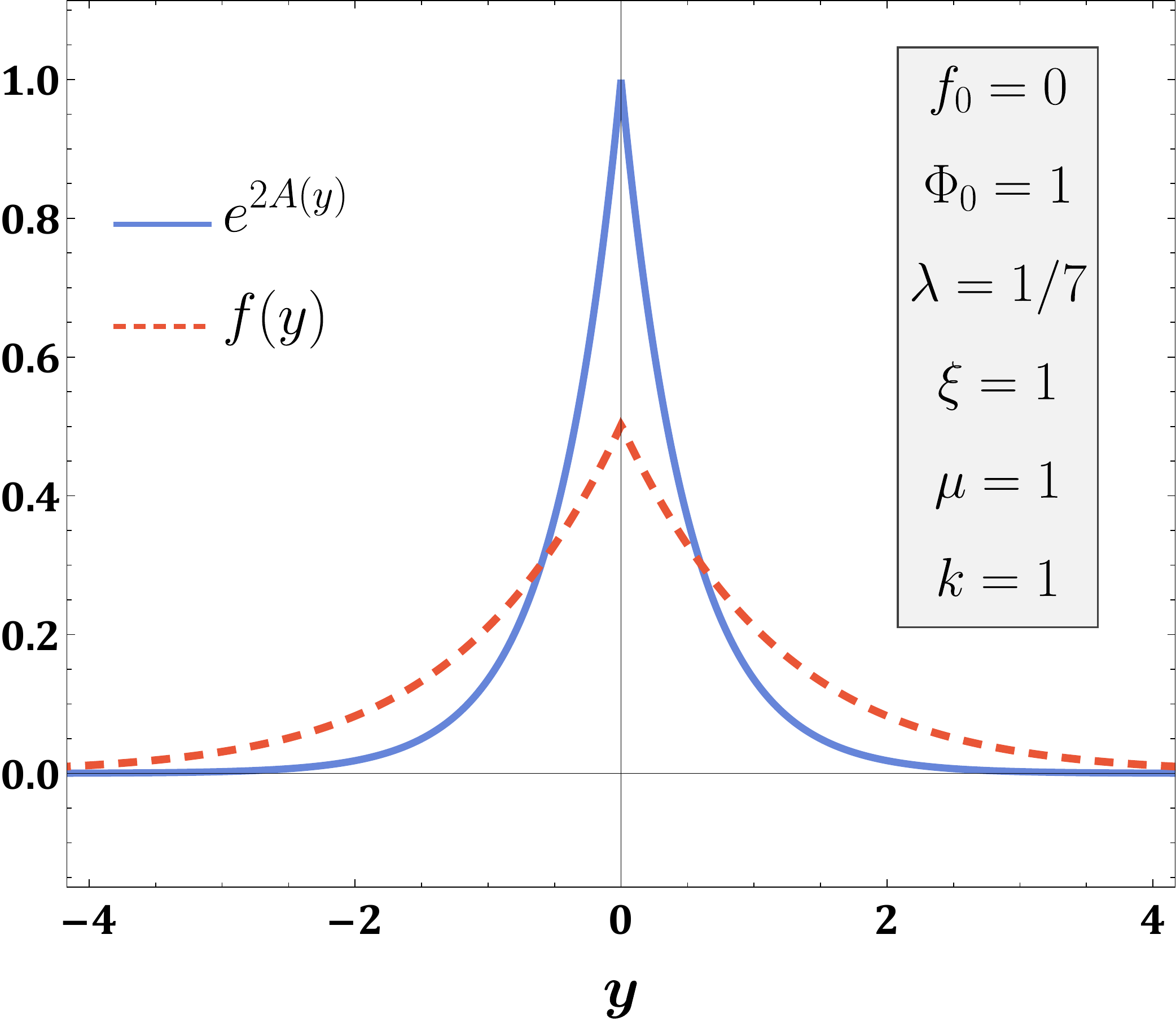}
        \caption{\hspace*{-1.6em}}
        \label{quad-plot1}
    \end{subfigure}
    ~ %add desired spacing between images, e. g. ~, \quad, \qquad, \hfill etc. 
      %(or a blank line to force the subfigure onto a new line)
    \begin{subfigure}[b]{0.44\textwidth}
        \includegraphics[width=\textwidth]{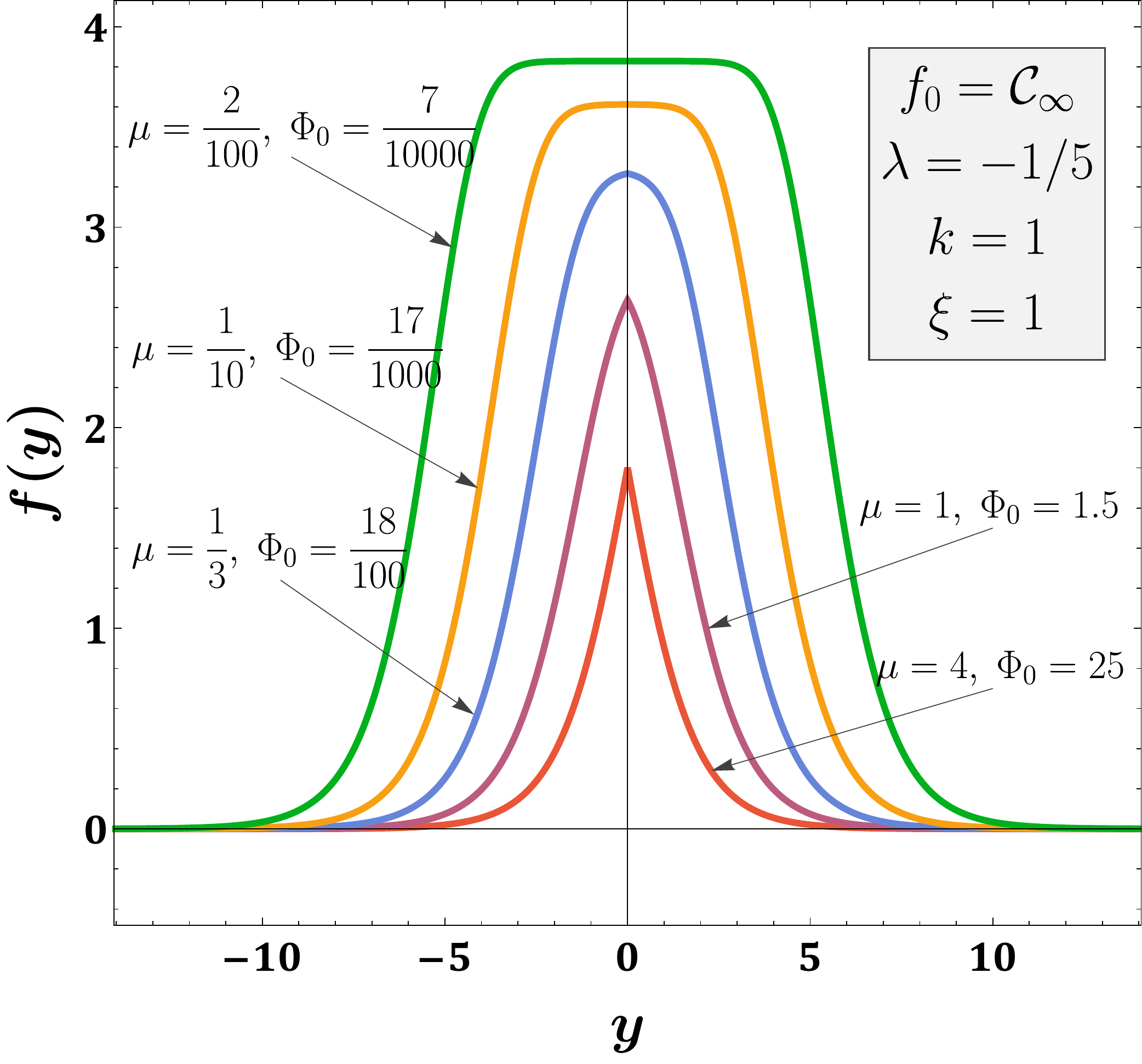}
        \caption{\hspace*{-2.5em}}
        \label{quad-plot2}
    \end{subfigure}
    ~ %add desired spacing between images, e. g. ~, \quad, \qquad, \hfill etc. 
    %(or a blank line to force the subfigure onto a new line)
    \vspace{-0.5em}
    \caption{(a) The warp factor $e^{2A(y)}=e^{-2k|y|}$ and coupling function $f(y)$ in terms 
    of the coordinate $y$ for $\lam=1/7$, and (b) the coupling function for $\lam=-1/5$ 
    in the regime $(-1/4,0)$ which satisfies $\frac{2\lam}{1+4\lam}=-2$ and different values 
    of parameters $\mu$ and $\Phi_0$. In
    figure (b), $\cinf$ indicates the value that the parameter $f_0$ should have in
    order to get a vanishing coupling function $f(y)$ at $y\ra\infty$.}
   \label{quad-plot-1-2}
  %\end{center}
\end{figure}
%%%%%%%%%%%%%%%%%%%%%%%%%%%%%%%%%%%%

\par We now proceed to study the profile of our solution.
In Fig. \myref{quad-plot-1-2}{quad-plot1}, we depict the form of the warp factor 
$e^{-2k|y|}$ and the coupling function $f(y)$ in terms of the coordinate $y$ along 
the fifth dimension, for $\Phi_0=1$, $\lam=1/7$, $\xi=1$, $\mu=1$, $k=1$ and $f_0=0$, 
which, as one can verify, is allowed by Eq. \eqref{quad-eff-con}. The warp factor is
always localised close to the brane and vanishes at the boundary of space-time 
independently of the values of the parameters. The behaviour of the coupling function 
though depends strongly on the values of the parameters of the model. For
$\lam>0$ and $\mu\geq 0$, the qualitative behaviour of the coupling function is the
same as the one that is  illustrated in Fig. \myref{quad-plot-1-2}{quad-plot1}. In Fig. 
\myref{quad-plot-1-2}{quad-plot2}, we present the behaviour of the coupling function for 
various values of the parameters $\mu$ and $\Phi_0$ while $\lam$ is now in the regime
$(-1/4,0)$. We note that, for generic values of the parameters, the
asymptotic value of $f(y)$, as $y \rightarrow \infty$, is not zero: if so desired, one may
choose $f_0$ to be equal to $\cinf$, which indicates the value that $f_0$ should have
in order to get a vanishing coupling function at infinity; from Eq. \eqref{quad-f-y}, we can
immediately calculate that $\mathcal{C}_\infty=\frac{\Phi_0^2}{4\lam}\left(1-\xi^2\,
\mu^{\frac{4\lam}{1+4\lam}}\right)$. In Fig. \myref{quad-plot-1-2}{quad-plot2},
one can clearly see the strong dependence of the profile of the coupling function also
on the value of the parameter $\mu$. As $\mu$ approaches zero, the coupling function is 
characterized by a plateau around our brane; the closer the value of $\mu$ is to zero, the 
wider the plateau. On the contrary, $\Phi_0$ does not significantly affect the behaviour of
$f(y)$; it just scales the function as a whole.  The behaviour depicted in Fig. 
\myref{quad-plot-1-2}{quad-plot2} holds for all values of $\lam$ in the regime 
$(-\frac{1}{4},0)$ as long as $\mu>0$. A different behaviour appears in the case where
$\frac{2 \lam}{1+4\lam}=-2n$, $n\in\mathbb{Z}^>$ and $\mu<-1$; in this case the behaviour 
of the coupling function is exactly the same as the one for $\lam<-\frac{1}{4}$ and $\frac{2\lam}
{1+4\lam}\neq n$, with $ n\in\mathbb{Z}^>$, which will be discussed next.

\par
\begin{figure}[t!]
    \centering
    \begin{subfigure}[b]{0.487\textwidth}
        \centering
        \includegraphics[height=0.289\textheight]{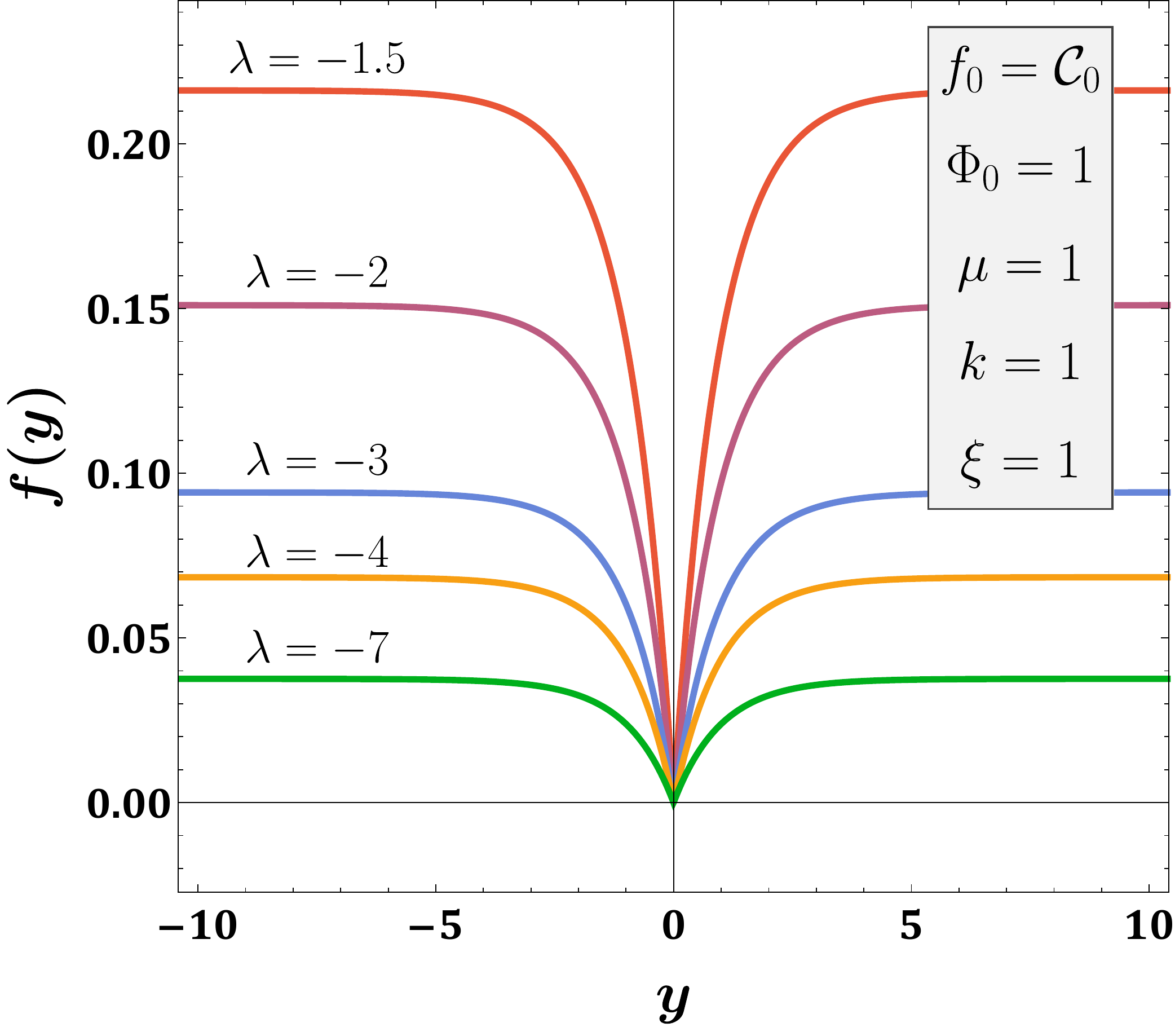}
        \caption{\hspace*{-3.7em}}
        \label{quad-plot3}
    \end{subfigure}
    ~ %add desired spacing between images, e. g. ~, \quad, \qquad, \hfill etc. 
      %(or a blank line to force the subfigure onto a new line)
    \begin{subfigure}[b]{0.487\textwidth}
        \includegraphics[height=0.289\textheight]{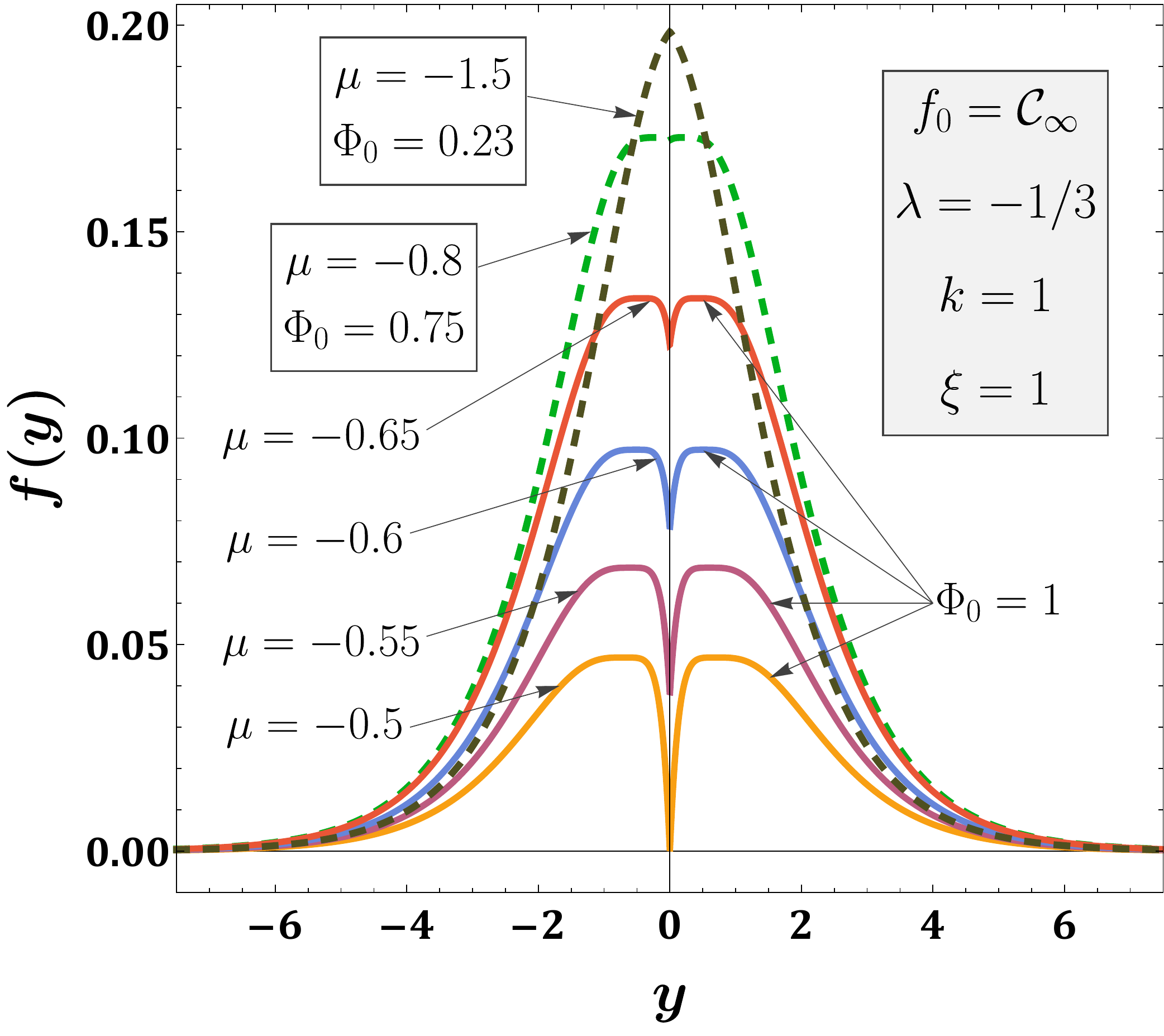}
        \caption{\hspace*{-3em}}
        \label{quad-plot5}
    \end{subfigure}
    ~ %add desired spacing between images, e. g. ~, \quad, \qquad, \hfill etc. 
    %(or a blank line to force the subfigure onto a new line)
    \vspace{-1.8em}
    \caption{(a) The coupling function $f(y)$ in terms of the coordinate $y$ for $\Phi_0=1$, 
    $\mu=1$, $k=1$, $\xi=1$, and values of $\lam$ smaller than $-\frac{1}{4}$ with
    $\frac{2\lam}{1+4\lam}\neq n$, $n\in\mathbb{Z}^>$. (b) The coupling function $f(y)$
    for $\lam=-1/3$, which satisfies  $\frac{2\lam}{1+4\lam}=2$, $k=1$, $\xi=1$ and
    $f_0=\cinf$, while $\mu$ takes values equal or lower than  $-1/2$.\\}
     \label{quad-plot-3-5}
  %\end{center}
\end{figure}

 In Fig. \myref{quad-plot-3-5}{quad-plot3}, we display the behaviour of the coupling function
$f(y)$ for $\Phi_0=1$, $\mu=1$, $k=1$, $\xi=1$ and values of $\lam$ in the regime $\lam < -1/4$. 
Since it holds that $\frac{2\lam}{1+4\lam} \neq n$, with $n\in\mathbb{Z}^>$, the parameter $\mu$
is constrained to values greater than or equal to zero. For easy comparison, the parameter $f_0$ has
been taken to be equal to $\mathcal{C}_0$, which is the value that leads to $f(0)=0$; again, from
Eq. \eqref{quad-f-y}, we find that 
$\mathcal{C}_0=\frac{\Phi_0^2}{4\lam}\left[1-\xi^2(\mu+1)^{\frac{4\lam}{1+4\lam}}\right]$.
In such a model, the non-minimal coupling of the scalar field to the five-dimensional scalar
curvature is non-vanishing in the bulk but disappears at the location of the brane.
In this range of values for the parameter $\lam$, the behaviour of the coupling function, as
depicted in Fig. \myref{quad-plot-3-5}{quad-plot3}, does not change regardless of the values
of all the other parameters. In contrast, when $\lam$ satisfies the condition $\frac{2\lam}{1+4\lam}=n$,
the profile of the coupling function is extremely sensitive to changes in the parameter $\mu$. 
Indeed, Fig. \myref{quad-plot-3-5}{quad-plot5} shows the behaviour of the coupling function $f(y)$
for $k=1$, $\xi=1$ and $f_0=\mathcal{C}_\infty$, while $\lam=-1/3$ or $\frac{2\lam}{1+
4\lam}=2$. In this figure, we focus on values of $\mu$ that are smaller than or equal to $-1/2$. 
We observe that, as $\mu$ approaches and exceeds $-1$, the behaviour of the coupling function
becomes similar to the one in Figs. \myref{quad-plot-1-2}{quad-plot1} 
and \myref{quad-plot-1-2}{quad-plot2}.  Here, we have chosen again $f_0=\mathcal{C}_\infty$,
therefore, the non-minimal coupling takes its maximum value on or close to our brane while it
vanishes at infinity. On the other hand, as $\mu$ approaches zero and takes on positive values,
the profile of $f(y)$ resembles more the one depicted in Fig. \myref{quad-plot-3-5}{quad-plot3}.

%%%%%%%%%%%%%%%%%%%%%%%%%%%%%%%%%%%%%%%%%%%%%%%%%%%
%%%%%%%%%%%%%%%%%%%%%%%%%%%%%%%%%%%%%%%%%%%%%%%%%%%%%%

\par Finally, in the case where $\lam=-1/4$, the coupling function, as presented in 
Eq. \eqref{quad-f-y}, is given by a double exponential expression. It is not hard 
to realize that the qualitative behaviour of $f(y)$ in this case is similar to 
the one in Fig. \myref{quad-plot-1-2}{quad-plot2} when $\mu<0$ and similar to Fig. 
\myref{quad-plot-3-5}{quad-plot3} when $\mu>0$. It is also necessary to stress that the 
behaviour of the scalar field $\Phi(y)$ is similar to that of the coupling function $f(y)$, 
as one may easily conclude by observing Eqs. \eqref{quad-Phi} and \eqref{quad-f-y}.
Therefore, it is redundant to present any graphs of the scalar field as a function of the 
$y$-coordinate. 

%%%%%%%%%%%%%%%%%%%%%%%%%%%%%%%%%%%%%%%%%%%%%%%%
\begin{figure}[t]
    \centering
    \begin{subfigure}[b]{0.49\textwidth}
        \includegraphics[width=\textwidth]{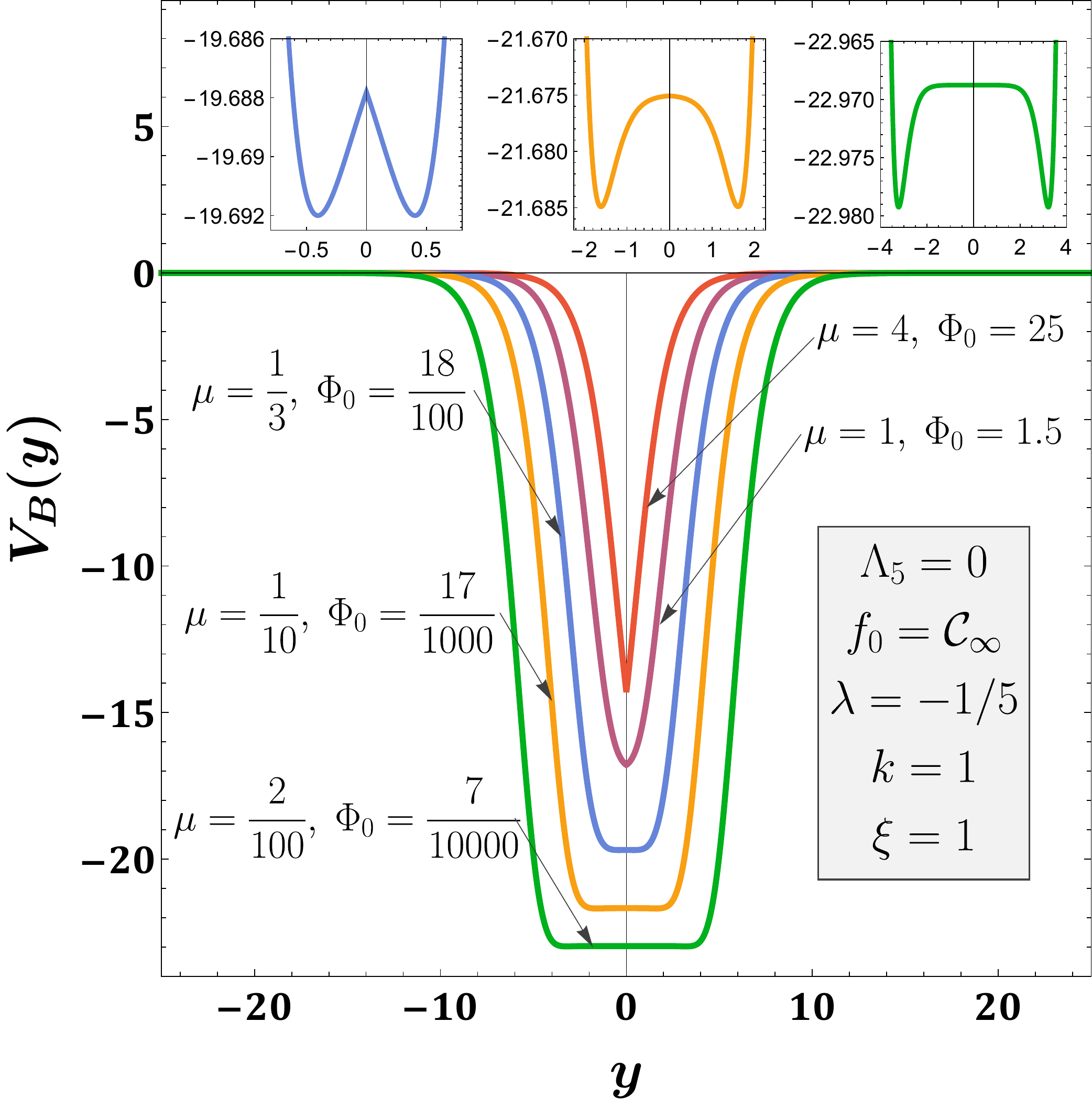}
        \caption{\hspace*{-3.8em}}
        \label{quad-plot6}
    \end{subfigure}
    \hfill
    ~ %add desired spacing between images, e. g. ~, \quad, \qquad, \hfill etc. 
      %(or a blank line to force the subfigure onto a new line)
    \begin{subfigure}[b]{0.49\textwidth}
        \includegraphics[width=\textwidth]{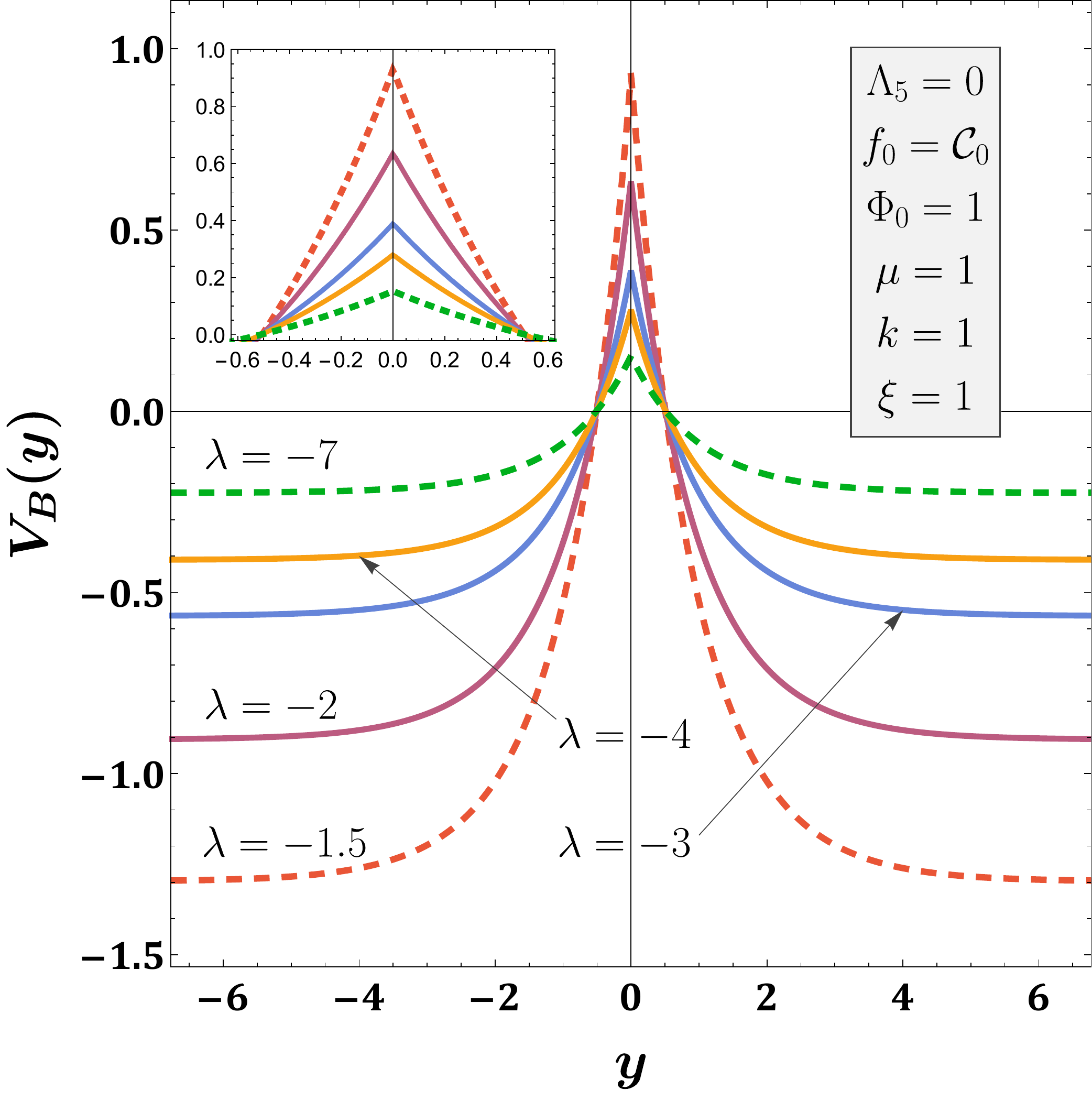}
        \caption{\hspace*{-4em}}
        \label{quad-plot7}
    \end{subfigure}
    
    \vspace{-0.5em}  
    \caption{The scalar potential $V_B$ in terms of the extra dimension $y$ for 
    $\Lambda_5=0$, $k=1$, $\xi=1$, and: (a) $\lambda=-1/5$, $f_0=\cinf$, and
    variable $\mu$ and $\Phi_0$, while, in (b) $\Phi_0=1$, $\mu=1$, $f_0=\mathcal{C}_0$
    and $\lam=-1.5, -2, -3, -4, -7$. In 
    each case, $\cinf$ and $\mathcal{C}_0$ should be evaluated separately.}
    \label{quad-plot-6-7}
\end{figure}
%%%%%%%%%%%%%%%%%%%%%%%%%%%%%%%%%%%%%%%%%%%%%%%%%%%%%%%%%%%%%%%%

\par
\begin{figure}[t]
    \centering
    \begin{subfigure}[b]{0.48\textwidth}
        \includegraphics[width=\textwidth]{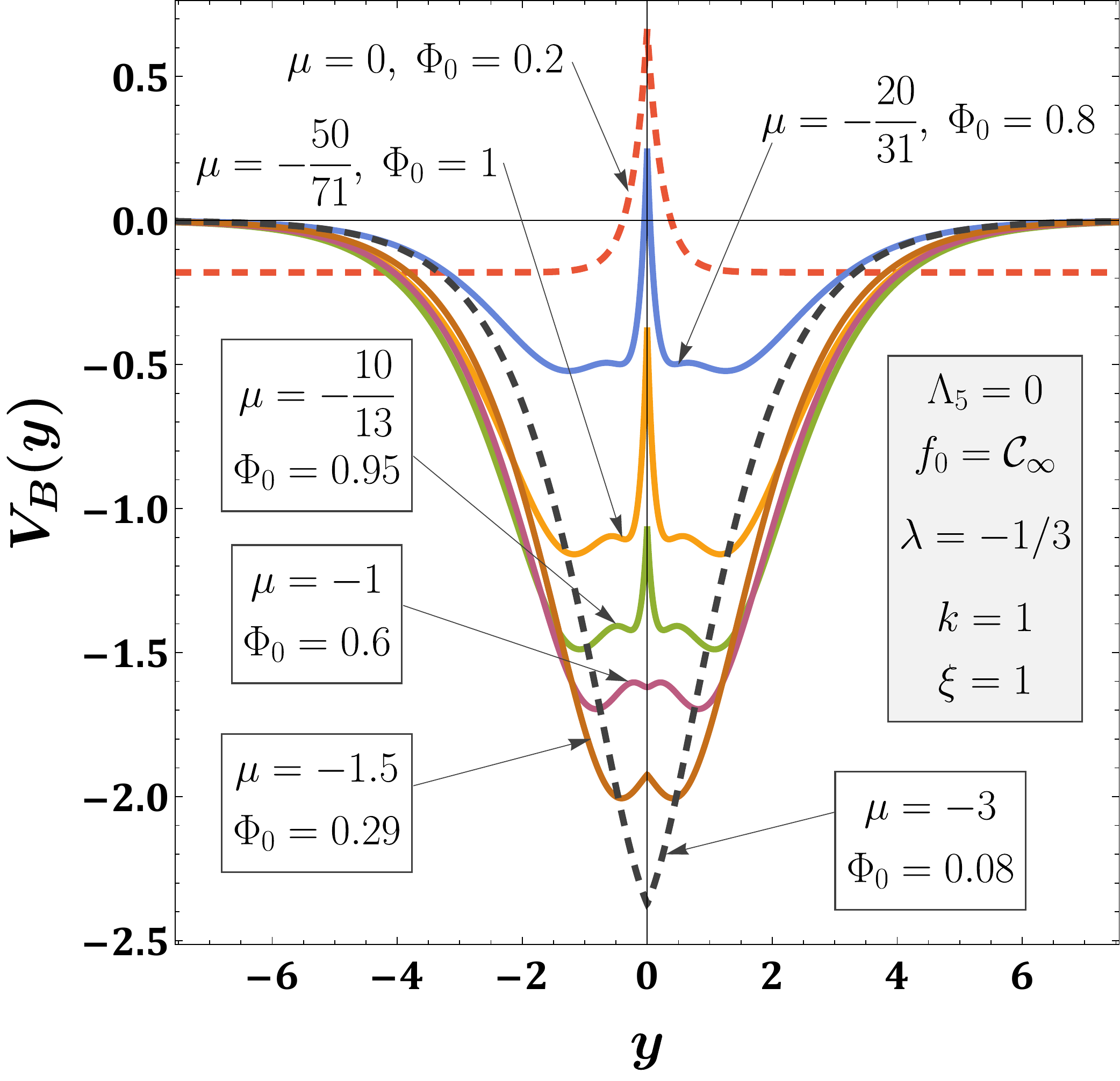}
        \caption{\hspace*{-3.9em}}
        \label{quad-plot8a}
    \end{subfigure}
    \hfill
    ~ %add desired spacing between images, e. g. ~, \quad, \qquad, \hfill etc. 
      %(or a blank line to force the subfigure onto a new line)
    \begin{subfigure}[b]{0.48\textwidth}
        \includegraphics[width=\textwidth]{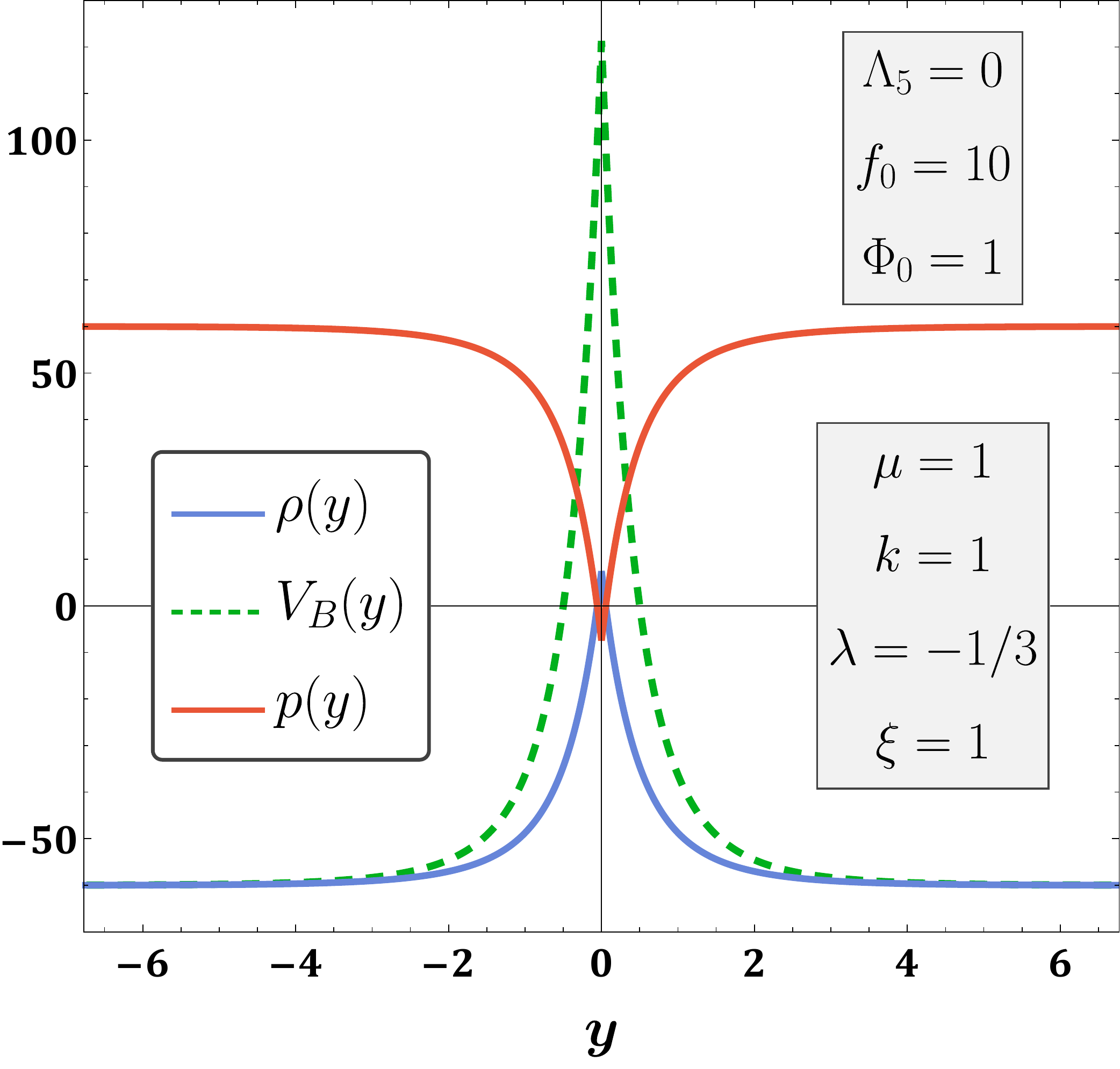}
        \caption{\hspace*{-2em}}
        \label{quad-plot8b}
    \end{subfigure}  
    \vspace{-0.5em}  
    \caption{(a) The scalar potential $V_B$ in terms of the extra dimension $y$ for 
    $\Lambda_5=0$, $\lam=-1/3$, $k=1$, $\xi=1$ and $f_0=\cinf$. 
    The varying parameters are $\Phi_0$ and $\mu$.
    (b) The energy density $\rho$ and pressure $p$ of the system together with
    the scalar potential $V_B$ in terms of the coordinate $y$ for $\Lambda_5=0$, $f_0=10$,
    $\Phi_0=1$, $\mu=1$, $k=1$, $\xi=1$, and $\lam=-1/3$.}
    \label{quad-plot-8}
\end{figure}

 The scalar potential $V_B$ can be determined in terms of the extra dimension $y$ from Eq.
\eqref{V-B} by substituting the function $f(y)$ given in Eq. \eqref{quad-f-y}. Consequently, we obtain
%%%%%%%%%%%%%%%%%%%%
\eq$\label{quad-V-y}
V_B(y)=\left\{\begin{array}{ll}
\displaystyle{-\Lambda_5-6k^2f_0+\frac{3k^2\Phi_0^2}{2\lam}-\frac{\xi^2 k^2\Phi_0^2\left(
\mu+e^{-ky}\right)^{-\frac{2+4\lam}{1+4\lam}}}{6\lam(1+4\lam)^2}\times} & \\[4mm]
\hspace{1cm}\times\left\{\left[e^{-ky}(3+16\lam)+3(1+4\lam)\mu\right]^2-\lam(3+16\lam)
e^{-2ky}\right\}\, , & \lam\in\mathbb{R}\setminus\{-\frac{1}{4},0\} \\[2em]
\displaystyle{-\Lambda_5-6k^2(f_0+\Phi_0^2)+\frac{\xi^2\Phi_0^2k^2}{2}\, e^{2\mu\,e^{-ky}}
\left(3+4\mu e^{-ky}+\mu^2e^{-2ky}\right)},&\lam=-\frac{1}{4}\end{array}\right\}.$
%%%%%%%%%%%%%%%%%%%%%
As in the linear case, it is possible to express the potential in terms of the scalar field $\Phi$
in closed form and obtain:
\eq$\label{quad-V-Phi}
V_B(\Phi)=\left\{\begin{array}{ll}
-\Lambda_5-6k^2f_0+\frac{3k^2\Phi_0^2}{2\lam}+\frac{\xi^2 k^2\Phi_0^2
}{6\lam(1+4\lam)^2}\left(\frac{2\lam\Phi}{\xi\Phi_0}+\frac{1}{\xi}\right)^{-\frac{1+2
\lam}{\lam}}\left\{3\mu^2\lam\right.  & \\[3mm]
\hspace{1em}\left.+6\lam(3+16\lam)\mu\left(\frac{2\lam\Phi}{\xi\Phi_0}+\frac{1}{\xi}
\right)^{\frac{1+4\lam}{2\lam}}-(3+16\lam)(3+15\lam)\left(\frac{2\lam\Phi}{\xi\Phi_0}+
\frac{1}{\xi}\right)^{\frac{1+4\lam}{\lam}}\right\}, 
&\lam\in\mathbb{R}\setminus\{-\frac{1}{4},0\} \\[2em]
\displaystyle{-\Lambda_5-6k^2(f_0+\Phi_0^2)+\frac{\Phi_0^2k^2}{2}\left(\frac{\Phi}{\Phi_0}-2
\right)^2\times} & \\[4mm]
\hspace{4em}\displaystyle{\times\left\{3+4\ln\left[\frac{1}{\xi}\left(\frac{\Phi}{\Phi_0}-2
\right)\right]+\ln^2\left[\frac{1}{\xi}\left(\frac{\Phi}{\Phi_0}-2\right)\right]\right\}}, 
& \lam=-1/4\end{array}\right\}.$
%%%%%%%%%%%%%%%%%%%%%%%%%%%%%
In Figs. \myref{quad-plot-6-7}{quad-plot6}, \myref{quad-plot-6-7}{quad-plot7} and
\myref{quad-plot-8}{quad-plot8a}, we display the behaviour of the scalar potential $V_B$ as a function
of the extra coordinate $y$ using the same values for the parameters as in Figs.
\myref{quad-plot-1-2}{quad-plot2}, \myref{quad-plot-3-5}{quad-plot3} and
 \myref{quad-plot-3-5}{quad-plot5}, respectively. It is worth observing the variety of
forms that one may achieve for $V_B$ by varying the values of the parameters of the model. 
In Fig. \myref{quad-plot-6-7}{quad-plot6}, constructed for $\lam \in (-1/4, 0)$, the
scalar potential adopts a negative value around the location of our brane, thus mimicking
locally a negative bulk cosmological constant $\Lambda_5$, while it vanishes away 
from our brane. In Fig.  \myref{quad-plot-6-7}{quad-plot7}, constructed for values
of $\lam$ in the regime $(-\infty, -1/4)$, the scalar potential has a positive value on
and close to our brane and then decreases rapidly to a constant negative value; this
asymptotic value depends on the values of the parameters and is preserved until the
space-time boundaries. Finally, Fig. \myref{quad-plot-8}{quad-plot8a}, constructed
for $\lam=-1/3$ and $\frac{2\lam}{1+4\lam}=2$, shows the sensitivity of the scalar
potential to the value of parameter $\mu$ with local
minima and maxima appearing in its profile. It should be however stressed that the warp
factor adopts its exponentially decaying form for all aforementioned profiles of the bulk
potential and independently of whether $\Lambda_5=0$ or not.

%%%%%%%%%%%%%%%%%%%%%%%%%%%%%%%%%%%%%%%%%%%%%%%%%%%

%%%%%%%%%%%%%%%%%%%%%%%%%%%%%%%%%%%%%%%%%%%%%%%%%%%%%

\par The components of the energy-momentum tensor of the theory may be finally computed by 
employing Eqs. \eqref{linear-rho}-\eqref{linear-p}. As in the linear case, we obtain
\eq$\label{quad-rho-pi-py}
\begin{gathered}
\rho(y)=-p(y)=-6k^2f(y)\,,\\[1mm]
p(y)=p^i(y)=p^y(y)\,.
\end{gathered}$
We discussed thoroughly in the previous section, that in order to satisfy the 
weak energy conditions on and close to the brane, we should allow the coupling 
function $f(y)$ to take negative values at these regimes. Thus, demanding that $f(0)<0$ 
and using Eq. \eqref{quad-f-y}, we obtain the constraints:
\eq$\label{quad-weak-con}
\left\{\begin{array}{ll}
\displaystyle{\frac{f_0}{\Phi_0^2}<\frac{1}{4\lam}\left[1-\xi^2(\mu+1)^{\frac{4\lam}
{1+4\lam}}\right]}, &\lam\in\mathbb{R}\setminus\{-\frac{1}{4},0\}\\[4mm]
\displaystyle{\frac{f_0}{\Phi_0^2}<\frac{\xi^2}{4}e^{2\mu}-1}, & \lam=-\frac{1}{4}
\end{array}\right\}\,.$
%%%%%%%%%%%%%%%%%%%%
In Fig. \myref{quad-plot-8}{quad-plot8b}, we present the energy density $\rho(y)$, the pressure
$p(y)=p^i(y)=p^y(y)$ and the scalar potential $V_B(y)$ in terms of the coordinate $y$.
It is obvious that, for this particular set of parameters, chosen to satisfy the above
constraints, the weak energy conditions are satisfied by the bulk matter on and close
to the brane. 

We should complete our bulk solution with the junction conditions introduced in the
model due to the presence of the brane at $y=0$. As discussed in the previous section,
the energy content of the brane is given by the combination $\sigma + V_b(\Phi)$,
and it creates a discontinuity in the second derivatives of the warp factor, coupling
function and scalar field  at the location of the brane. Using Eqs. \eqref{jun_con1}
and \eqref{jun_con2}, for $\lam\neq\{-\frac{1}{4},0\}$ and $\lam=-1/4$, we obtain
\gat$\label{quad-jun1-1}
\sigma+V_b(\Phi)|_{y=0}=6k\left(f_0-\frac{\Phi_0^2}{4\lam}\right)+
\frac{\Phi_0^2\xi^2k}{2\lam(1+4\lam)}\left[3(1+\mu)(1+4\lam)+4\lam\right]
(1+\mu)^{-\frac{1}{1+4\lam}}\, ,\\[3mm]
\label{quad-jun1-2}
\pa_\Phi V_b |_{y=0}=\frac{2k\xi\Phi_0}{1+4\lam}(1+\mu)^{-\frac{1+2\lam}{1+4\lam}}
\left[4(1+\mu)(1+4\lam)-1\right]\, ,$
and 
\gat$\label{quad-jun2-1}
\sigma+V_b(\Phi)|_{y=0}=6k(f_0+\Phi_0^2)-\frac{\Phi_0^2\xi^2k\,e^{2\mu}}{2}(3+2\mu)\, 
,\\[3mm]
\label{quad-jun2-2}
\pa_\Phi V_b |_{y=0}=-2\Phi_0\,k\,\xi\,\mu\,e^{\mu}(\mu+2)\, ,$
respectively. Using the constraints \eqref{quad-jun1-1} and \eqref{quad-jun2-1}, it is easy
to deduce that in order to have a positive total energy  density on the brane, namely
$\sig+V_b(\Phi)|_{y=0}>0$, we should have, respectively
\eq$\label{quad-brane-ene-con}
\left\{\begin{array}{ll}
\displaystyle{\frac{f_0}{\Phi_0^2}>\frac{1}{4\lam}\left\{1-\xi^2\left[(1+\mu)^{\frac{4\lam}{1+4
\lam}}+\frac{4\lam}{3(1+4\lam)}(1+\mu)^{-\frac{1}{1+4\lam}}\right]\right\}}\,,
& \lam\in\mathbb{R}\setminus\{-\frac{1}{4},0\}\\[5mm]
\displaystyle{\frac{f_0}{\Phi_0^2}>-1+\frac{\xi^2}{12}\,e^{2\mu}(3+2\mu)}\,, & \lam=-1/4
\end{array}\right\}\,.$
%%%%%%%%%%%%%%%%%%%%%
Let us also note that, from the constraint \eqref{quad-jun2-2}, we see that the brane
interaction term $V_b$ can be a constant, and thus absorbed into the brane tension
$\sigma$, under the condition $\mu=-2$. A similar fixing of the parameter $\mu$ follows
from Eq. \eqref{quad-jun1-2}, which leads to the result $\mu=-\frac{3+4\lam}{4(1+4\lam)}$.
However, in this case, care should be taken so that the resulting values of $\mu$, in
terms of $\lambda$, are allowed by Table \ref{quad-par-val}.

\par The effective four-dimensional gravitational scale on the brane has already been calculated
and is given in Eqs. \eqref{MPl-quad} and \eqref{Mpl-1/4}. The effective cosmological constant
on the brane $\Lambda_4$ can be calculated from Eq. \eqref{cosm_eff}, and is found to be
zero also in this case, as anticipated. 

%%%%%%%%%%%%%%%%%%%%%%%%%%%%%%%%%%%%%%%%%%%%%%%%%%%%%%5

\subsection{The energy conditions in the parameter space}

\par We will now study the inequalities \eqref{quad-eff-con}, \eqref{quad-weak-con}
and \eqref{quad-brane-ene-con} and investigate again whether these may be simultaneously
satisfied. In particular, we will study the parameter space between the ratio $f_0/\Phi_0^2$
and the parameters $\lam$, $\mu$, and $\xi$. Given the large number of parameters,
we will present three-dimensional graphs of the parameter space of the ratio $f_0/\Phi_0^2$
with two of the three parameters $\lam,\, \mu,\, \xi$, while keeping the remaining one
fixed. Before we continue, we elucidate that, in the forthcoming analysis, we will denote
the r.h.s. of inequality \eqref{quad-eff-con} with $F_{eff}(\lam,\mu,\xi)$, since it is associated
with the effective gravitational constant, the r.h.s. of inequality \eqref{quad-weak-con},
which refers to the energy conditions in the bulk, with $F_{B}(\lam,\mu,\xi)$, and
finally, the r.h.s. of inequality \eqref{quad-brane-ene-con}, which involves the total
energy density on the brane, with $F_{br}(\lam,\mu,\xi)$. 

While pursuing to satisfy simultaneously all the aforementioned inequalities, we have performed
a comprehensive study of the parameter space of the quantities $f_0/\Phi_0^2$, $\lam$, $\mu$
and $\xi$ following the classification of cases, regarding the values of the free parameters,
presented in Table \ref{quad-par-val}. We present the corresponding results below:

\begin{enumerate}
\item[\bf(i)] For $\lam>0$, we have $\mu\geq 0$, while the parameter $\xi$ can take 
values in the whole set of real numbers except zero. In Fig. \myref{quad-lam-pos-neg}{quad-lam-pos},
we depict the parameter space of the quantities $f_0/\Phi_0^2$, $\lam$ and $\mu$, for
$\xi=1$ and $\lam>0$. Although the surfaces representing the functions $F_{eff}(\lam,\mu,\xi)$, 
$F_{br}(\lam,\mu,\xi)$ and $F_{B}(\lam,\mu,\xi)$ change significantly for different values
of the parameter $\xi$, their relative positions remain the same satisfying always the relation
$$F_{eff}(\lam,\mu,\xi)>F_{B}(\lam,\mu,\xi)>F_{br}(\lam,\mu,\xi)\,.$$
This means that there is no point in the parameter space for $\lam>0$ at which all three
inequalities are satisfied simultaneously. It is possible though to satisfy simultaneously the
inequalities \eqref{quad-eff-con} and \eqref{quad-brane-ene-con}. Particularly, for every 
value of the ratio $f_0/\Phi_0^2$ which is greater than the value of the function $F_{eff}
(\lam,\mu,\xi)$ at any given point in the parameter space the aforementioned two inequalities
will be satisfied. This means that the positivity of both the effective four-dimensional
gravitational constant and the total energy-density on the brane is ensured. In contrast,
there is no point in the parameter space at which we can satisfy the inequality \eqref{quad-weak-con}
because the surface of the function $F_{B}(\lam,\mu,\xi)$ lies always below the surface of the
function $F_{eff}(\lam,\mu,\xi)$; as a result, the weak energy conditions are always violated
by the bulk matter close to the brane. 

%%%%%%%%%%%%%%%%%%%%%%%%%%%%%%%%%%%%%%%%%%%%%%%%%%%%%%

\begin{figure}[t]
    \centering
    \begin{subfigure}[b]{0.49\textwidth}
        \includegraphics[width=\textwidth]{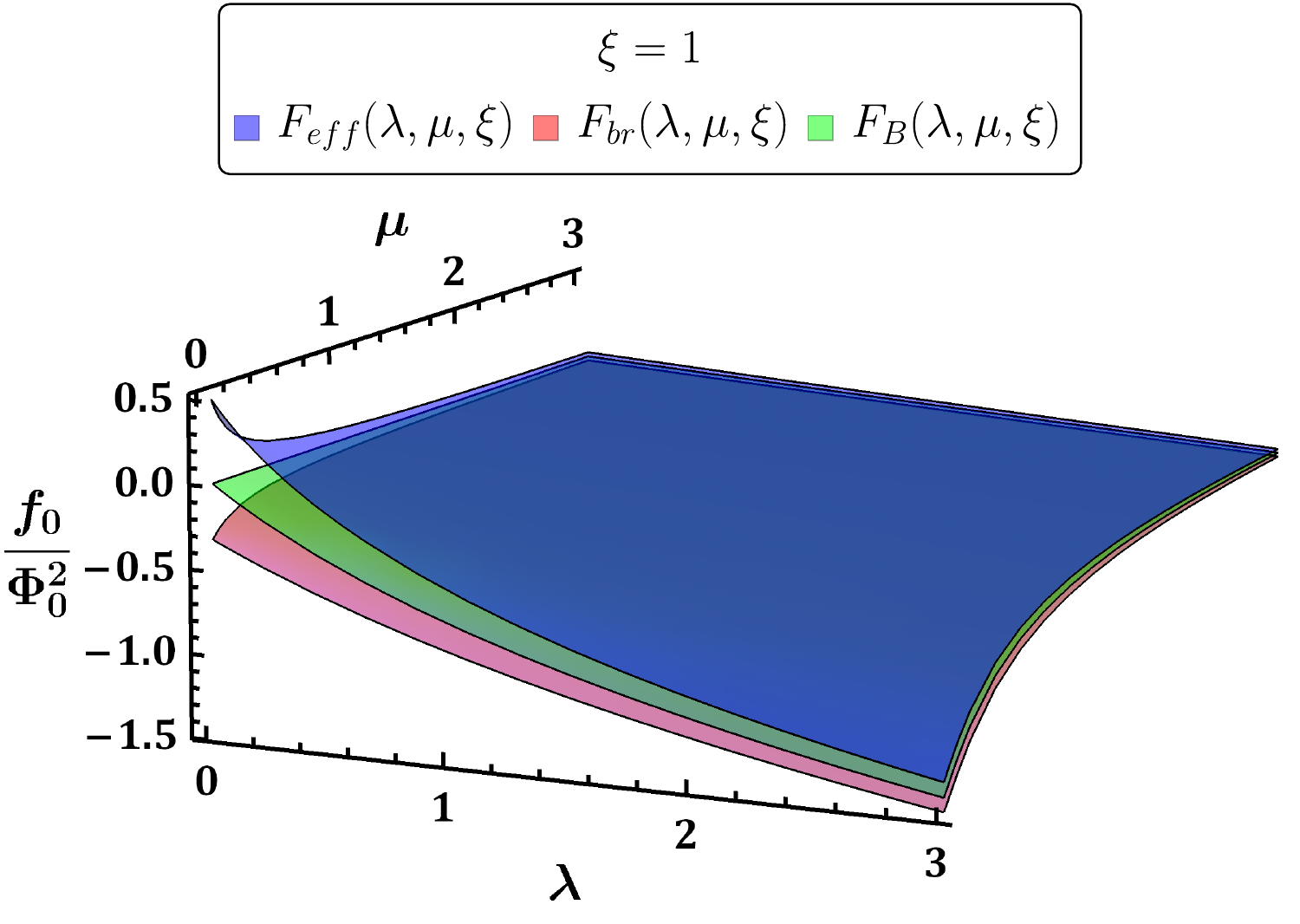}
        \caption{\hspace*{2.8em}}
        \label{quad-lam-pos}
    \end{subfigure}
    ~ %add desired spacing between images, e. g. ~, \quad, \qquad, \hfill etc. 
      %(or a blank line to force the subfigure onto a new line)
    \begin{subfigure}[b]{0.485\textwidth}
        \includegraphics[width=\textwidth]{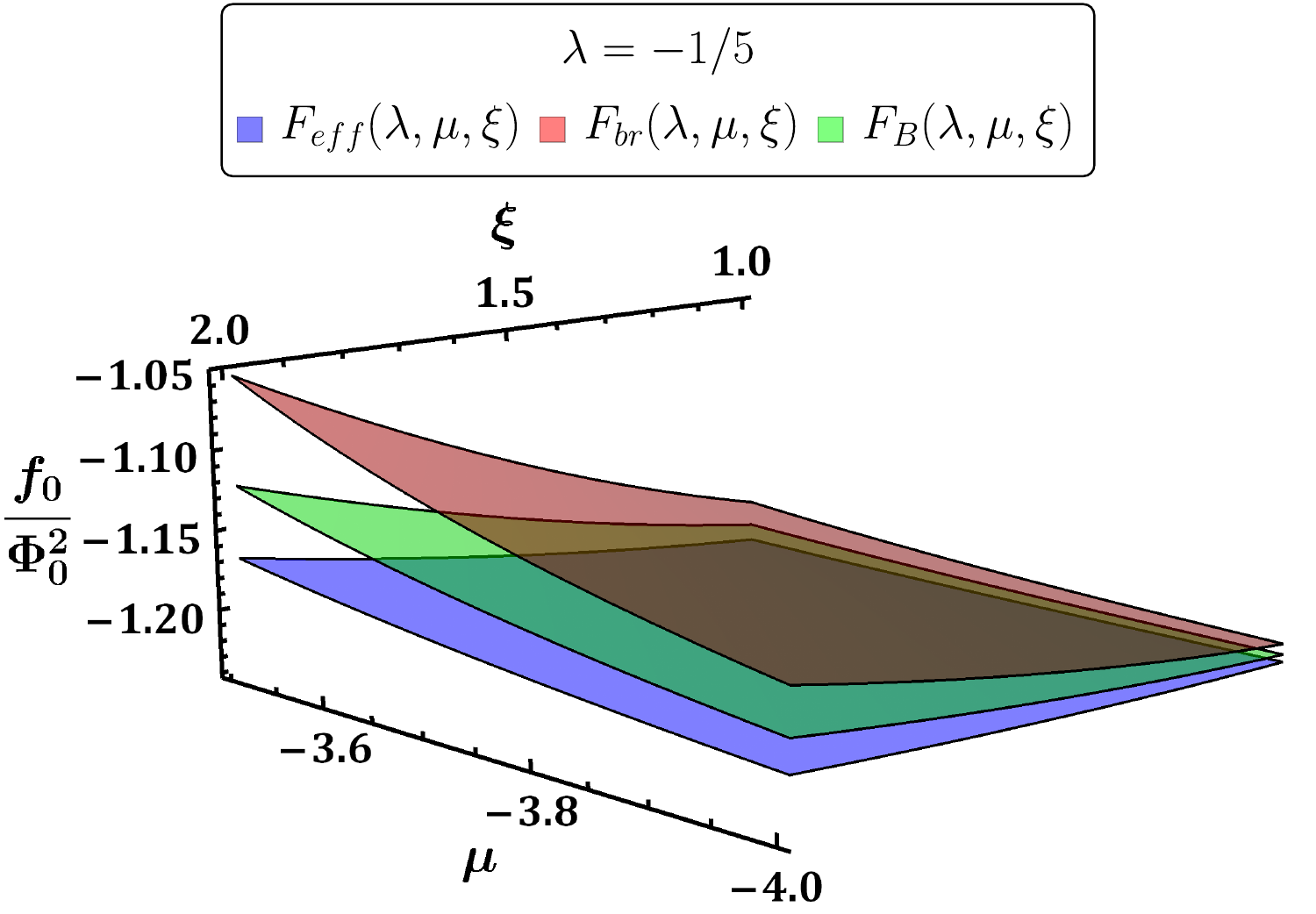}
        \caption{\hspace*{6.2em}}
        \label{quad-lam-neg-even1}
    \end{subfigure}
    ~ %add desired spacing between images, e. g. ~, \quad, \qquad, \hfill etc. 
    %(or a blank line to force the subfigure onto a new line)
    \vspace{-1.8em}    
    \caption{(a) The parameter space of the quantities $f_0/\Phi_0^2$, $\lam$ and $\mu$, for
    $\xi=1$ and $\lam>0$.
(b) The parameter space of the quantities $f_0/\Phi_0^2$, $\mu$ and $\xi$, for
    $\lam=-1/5$ or $\frac{2\lam}{1+4\lam}=-2$ and $\mu<-1$. The graphs depict the functions
    $F_{eff}(\lam, \mu,\xi)$, $F_{br}(\lam,\mu,\xi)$ and $F_{B}(\lam,\mu,\xi)$.}
   \label{quad-lam-pos-neg}
  %\end{center}
\end{figure}
%%%%%%%%%%%%%%%%%%%%%%%%%%%%%%%%%%%%%%%%%%%%%%

\item[\bf(ii)] For $\lam\in\left(-\frac{1}{4},0\right)$,  $\frac{2\lam}{1+4\lam}\neq n,\,n\in\mathbb{Z}^<$,
we have $\mu > 0$, and we obtain the same qualitative behaviour as in the previous case.
However, when $\frac{2\lam}{1+4\lam}=n$, we have $\mu\in(-\infty,-1)\cup(0,\infty)$.
In this case, the position of the surfaces $F_{eff}(\lam,\mu,\xi)$, $F_{br}(\lam,\mu,\xi)$ and $F_{B}(\lam,\mu,\xi)$
are different in the region of the parameter space where $\mu<-1$ and in the region where
$\mu>0$.  Specifically, in this case we find that 
\eq$\begin{array}{ll}
\displaystyle{F_{br}(\lam,\mu,\xi)>F_{B}(\lam,\mu,\xi)>F_{eff}(\lam,\mu,\xi)}, & \mu<-1\,,
\\[3mm]
\displaystyle{F_{eff}(\lam,\mu,\xi)>F_{B}(\lam,\mu,\xi)>F_{br}(\lam,\mu,\xi)}, & \mu>0\,.
\end{array}\nonum$
Again, there is no point in the parameter space at which we can satisfy simultaneously all inequalities.
For $\mu>0$, the situation is similar to the one of case (i) depicted in Fig. \myref{quad-lam-pos-neg}{quad-lam-pos}.
In this case, we may easily obtain a positive effective gravitational constant and a positive total
energy-density on the brane. For $\mu<-1$, though, as Fig. \myref{quad-lam-pos-neg}{quad-lam-neg-even1} also reveals,
we have the choice of supplementing the positivity of the effective gravitational constant by either a positive
total energy-density on the brane or by a bulk matter that satisfies the energy conditions close to
our brane.

\item[\bf(iii)] For $\lam=-1/4$, due to the different form of the solution, the functions
$F_{eff}(-1/4,\mu,\xi)$, $F_{br}(-1/4,\mu,\xi)$ and $F_{B}(-1/4,\mu,\xi)$ are given by different
expressions. Now, these are found to satisfy the relations 
\eq$\begin{array}{ll}
\displaystyle{F_{eff}(-1/4,\mu,\xi)>F_{B}(-1/4,\mu,\xi)>F_{br}(-1/4,\mu,\xi)}, & \mu<0\,,
\\[3mm]
\displaystyle{F_{br}(-1/4,\mu,\xi)>F_{B}(-1/4,\mu,\xi)>F_{eff}(-1/4,\mu,\xi)}, & \mu>0\,.
\end{array}\nonum$
In this case, for $\mu<0$, we may obtain only the combination of a positive
effective gravitational scale and a positive total energy-density on the brane, in the region
of the parameter space in which the value of the ratio $f_0/\Phi_0^2$ is greater than
the value of the function $F_{eff}(-1/4,\mu,\xi)$; the relative positions of the different
surfaces are the same as in Fig. \myref{quad-lam-pos-neg}{quad-lam-pos}. On the other hand, for
$\mu >0$, we again have the choice of satisfying either Eqs. \eqref{quad-eff-con} and 
\eqref{quad-weak-con}, in the region where $F_{eff}(-1/4,\mu,\xi)<f_0/\Phi_0^2<F_{B}(-1/4,\mu,\xi)$,
or Eqs. \eqref{quad-eff-con} and \eqref{quad-brane-ene-con}, in the region where
$f_0/\Phi_0^2>F_{br}(-1/4,\mu,\xi)$. This situation is in turn similar to the one depicted
in Fig. \myref{quad-lam-pos-neg}{quad-lam-neg-even1}.  

%%%%%%%%%%%%%%%%%%%%%%%%%%%%%%%%%%%%%%%%%%%%%%%%%%%%%%
\begin{figure}[t]
    \centering
    \begin{subfigure}[b]{0.4\textwidth}
        \includegraphics[height=0.31\textheight]{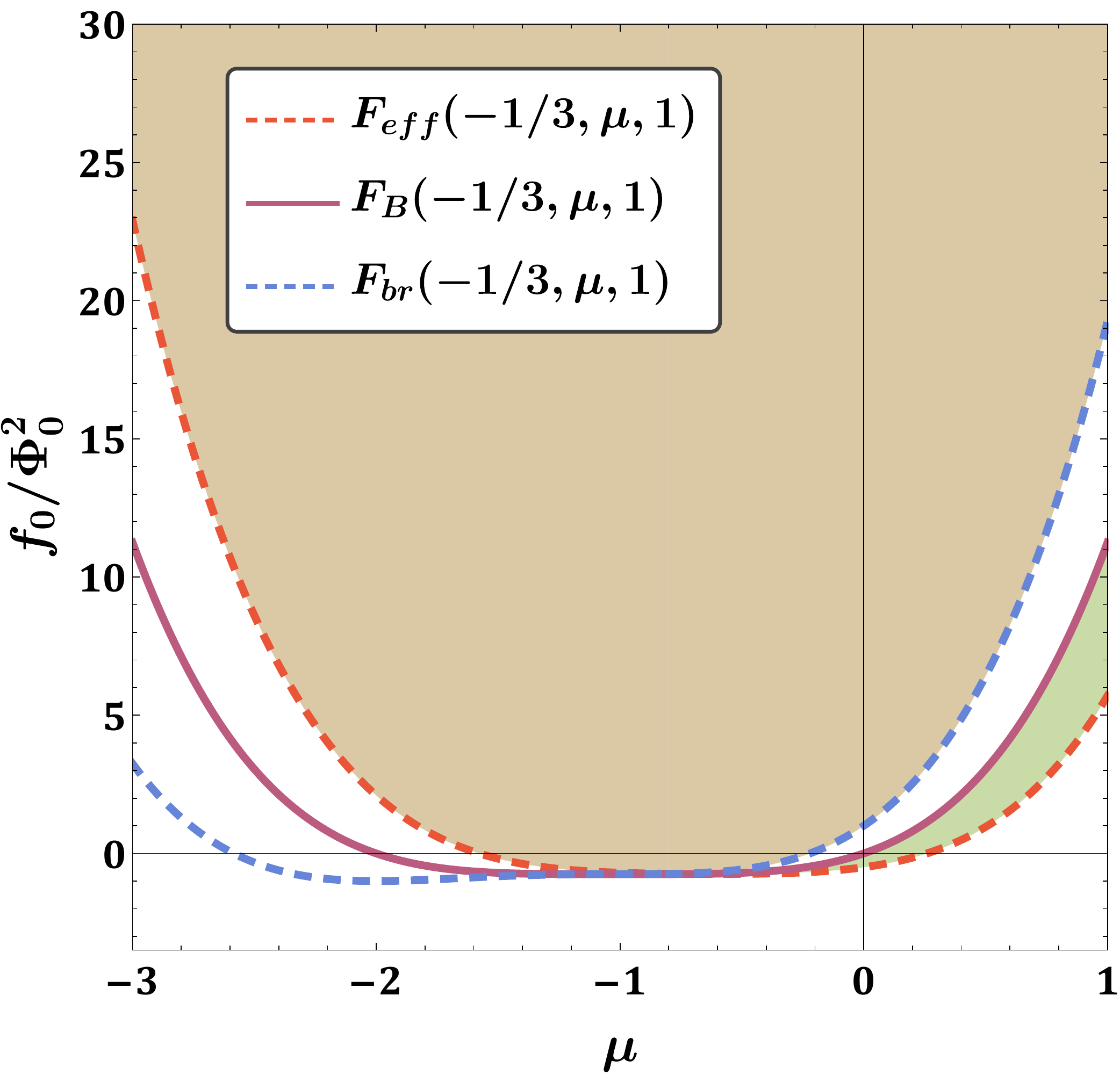}
        \caption{\hspace*{-5.9em}}
        \label{quad-lam-neglowereven}
    \end{subfigure}
    \hspace{3em}
    ~ %add desired spacing between images, e. g. ~, \quad, \qquad, \hfill etc. 
      %(or a blank line to force the subfigure onto a new line)
    \begin{subfigure}[b]{0.4\textwidth}   
        \includegraphics[height=0.3\textheight]{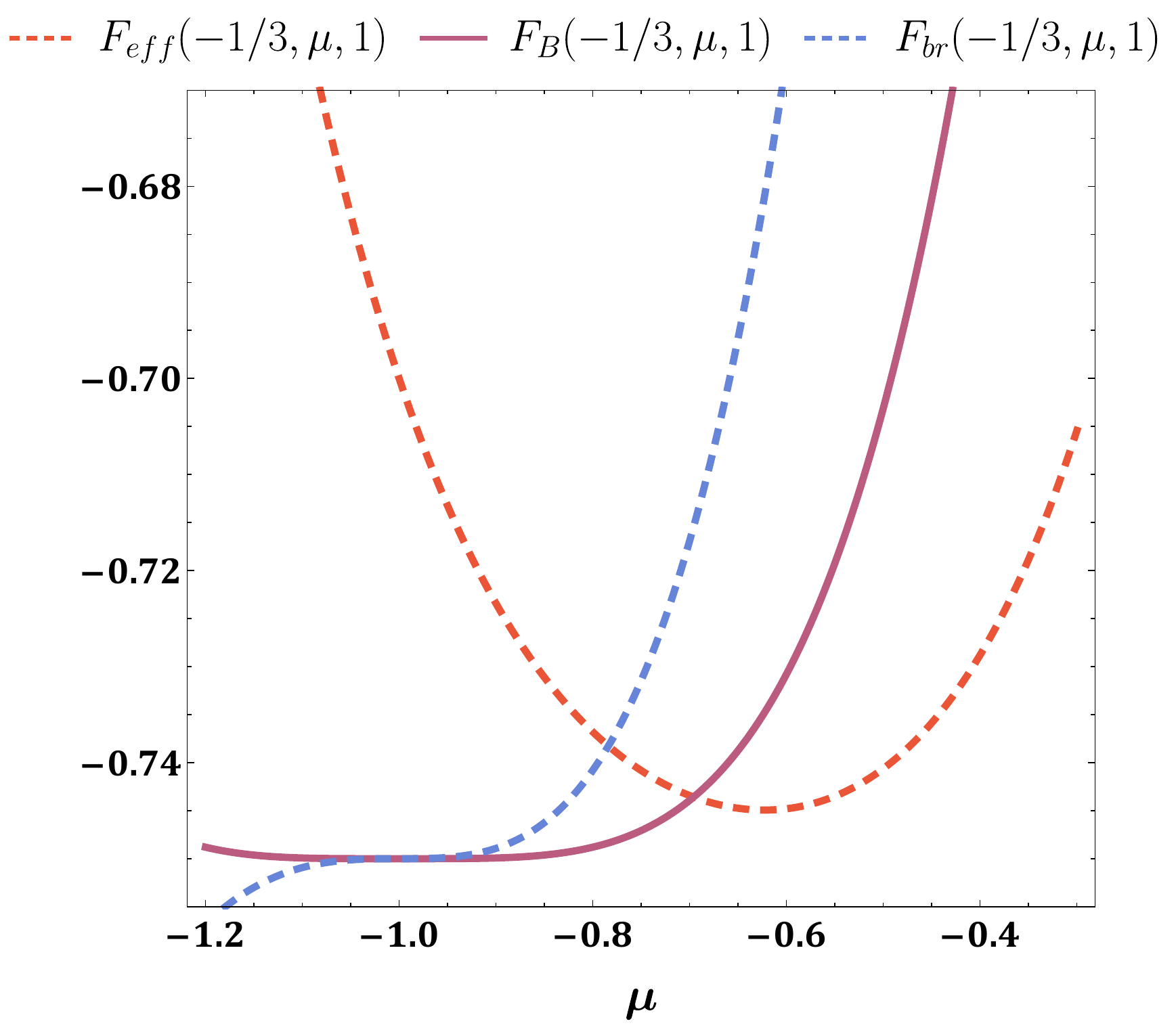}
        \vspace*{-.2em} \caption{\hspace*{-7.2em}}
        \label{quad-lam-neglowerevenzoom}
    \end{subfigure}
    ~ %add desired spacing between images, e. g. ~, \quad, \qquad, \hfill etc. 
    %(or a blank line to force the subfigure onto a new line)
    \vspace{-0.5em}    
    \caption{(a) The parameter space of the quantities $f_0/\Phi_0^2$ and $\mu$, for $\xi=1$,
    $\lam=-1/3$, and (b) a magnification of a particular region of the 
    previous figure in order to get a clear picture of the behaviour of the functions $F_{eff}(
    -1/3,\mu,1)$, $F_{br}(-1/3,\mu,1)$ and $F_{B}(-1/3,\mu,1)$ close to $\mu=-0.8$.}
   \label{quad-lam-neglowereven-1-2}
  %\end{center}
\end{figure}
%%%%%%%%%%%%%%%%%%%%%%%%%%%%%%%%%%%%%%%%%%%%%%%%%%%%%%%%%%%%%%%%%%

\item[\bf(iv)] For $\lam<-1/4$, $\frac{2\lam}{1+4\lam}\neq n,\,n\in\mathbb{Z}^>$, and
for every allowed value of the parameters $\mu \geq 0$ and $\xi\in\mathbb{R}\setminus\{0\}$,
we always have
$$F_{br}(\lam,\mu,\xi)>F_{B}(\lam,\mu,\xi)>F_{eff}(\lam,\mu,\xi)\,.$$
In this case, the situation is similar to the one depicted in Fig. \myref{quad-lam-pos-neg}{quad-lam-neg-even1}, 
and we have again the choice of combining a positive effective gravitational constant with
either a positive energy-density on the brane or a bulk matter that satisfies the weak
energy conditions close to and on our brane. 

\item[\bf(v)] For $\lam<-1/4$ and $\frac{2\lam}{1+4\lam}= n,\,n\in\mathbb{Z}^>$, the
parameter $\mu$ is free to take values in the whole set of real numbers. In Figs.
\myref{quad-lam-neglowereven-1-2}{quad-lam-neglowereven} and 
\myref{quad-lam-neglowereven-1-2}{quad-lam-neglowerevenzoom}, we depict
the parameter space of the ratio $f_0/\Phi_0^2$ and $\mu$ together with the curves of
the functions $F_{eff}(\lam,\mu,\xi)$, $F_{br}(\lam,\mu,\xi)$ and $F_{B}(\lam,\mu,\xi)$. 
Note that, for clarity of the graph, we have fixed the values of two parameters, i.e.
$\lam=-1/3$ and $\xi=1$, and thus present a two-dimensional graph. However, 
the situation remains the same for every other allowed value of the parameters 
$\lam$ and $\xi$. We observe that there always exist a region in the parameter space
in which we can have a positive value for the effective four-dimensional gravitation scale
and satisfy the weak energy conditions close to the brane (green region) and a region in
which both the four-dimensional gravitational constant and the total energy density on
the brane are positive (brown region). Since there is no overlapping between the green and
brown regions, as Fig. \myref{quad-lam-neglowereven-1-2}{quad-lam-neglowerevenzoom} reveals, there is no point
in the parameter space where all three conditions are satisfied. For comparison, we note
that the parameters in Fig. \myref{quad-plot-8}{quad-plot8b} have been chosen so that the depicted
solution falls into the green area of Fig. \myref{quad-lam-neglowereven-1-2}{quad-lam-neglowereven}.

\end{enumerate}

%%%%%%%%%%%%%%%%%%%%%%%%%%%%%%%%%%%%%%%%%%%%%%%%%%%%%%%%%%%%%%%%%%%%%%%%
%
%
%%%%%%%%%%%%%%%%%%%%%%%%%%%%%%%%%%%%%%%%%%%%%%%%%%%%%%%%%%%%%%%%%%%%%%%%%%

\section{An Inverse-power Coupling Function in terms of $y$}
\label{power}

\par In this and the following two sections, we will consider explicit forms of the coupling function
$f(y)$ in terms of the coordinate $y$. These forms cannot be easily expressed in terms of the scalar
field $\Phi$ in a closed form, they are however legitimate choices that satisfy the reality and
finiteness conditions imposed in Sec. \ref{th-frame}. We start with the following expression
\eq$\label{power-f}
f(y)=f_0+\frac{\Phi_0^2}{k^\lam(y+y_0)^\lam}\, ,$
where $(f_0, \Phi_0) \in \mathbb{R}\setminus\{0\}$ while ($\lam, y_0) \in(0,+\infty)$.
The factor $k^\lam$ in the denominator was introduced to make the product $k(y+y_0)$
dimensionless. %hence, it holds that $[f(y)]=[f_0]=[\Phi_0]^2$.

\subsection{The bulk solution}

\par Substituting the aforementioned coupling function in Eq. \eqref{grav-1-2} we obtain the
differential equation:
\eq$\label{power-der-Phi}
[\Phi'(y)]^2=\frac{\lam\,\Phi_0^2}{k^\lam(y+y_0)^{\lam+2}}\left[k(y+y_0)-\lam-1\right]\, .$
The r.h.s. of the above equation should be always positive; evaluating at $y=0$, the above yields 
the following constraint on the parameters of the model
\eq$\label{power-con1}
\frac{ky_0}{\lam+1}>1.$
%%%%%%%%%%%%%%%%%%%%%
The function $[\Phi'(y)]^2$ could, in principle, be zero at the point where $\Phi(y)$ has
an extremum. However, from Eq. \eqref{power-der-Phi} this may happen only at $y=y_0\left(\frac{
\lam+1}{ky_0}-1\right)$ which, upon using the constraint \eqref{power-con1}, turns out to be
negative. Therefore, the scalar field does not have an extremum in the whole domain $y\in[0,+
\infty)$, which also means that $\Phi(y)$ is an one-to-one function in the same regime. In 
addition, from Eq. \eqref{power-der-Phi} it is straightforward to deduce that, as $y\ra+\infty$, 
the physical constraint \eqref{con.2} is satisfied, thus, the scalar field does not diverge at 
infinity.
\par Let us now determine the explicit expression of the scalar field $\Phi(y)$ from Eq.
\eqref{power-der-Phi}. For simplicity and without loss of generality, we will assume that
$\Phi_0\in(0,+\infty)$. Then, after taking the square root of Eq. \eqref{power-der-Phi},
we have:
\eq$\label{power-phi1}
\Phi_{\pm}(y)=\pm \frac{\Phi_0\sqrt{\lam(\lam+1)}}{k^{\lam/2}}\int dy\ (y+y_0)^{-\frac{\lam}{2}-1}
\left[\frac{k(y+y_0)}{\lam+1}-1\right]^{\frac{1}{2}}\, .$
%%%%%%%%%%%%%%%%
Setting $u=\frac{k(y+y_0)}{\lam+1}$ and then $w=1-\frac{1}{u}$, the above integral takes the form
%%%%%%%%%%%%%%%%%%%%%
\bal$\int dy\ (y+y_0)^{-\frac{\lam}{2}-1}\left[\frac{k(y+y_0)}{\lam+1}-1\right]^{\frac{1}{2}}&
=\left(\frac{k}{\lam+1}\right)^{\frac{\lam}{2}}\int dw\ (1-w)^{\frac{\lam}{2}-\frac{3}{2}}\
w^{\frac{1}{2}}\nonum\\[2mm]
&=\left(\frac{k}{\lam+1}\right)^{\frac{\lam}{2}}\int_0^w dt\ t^{\frac{1}{2}}(1-t)^{\frac{\lam}{2}
-\frac{3}{2}}+C_1\nonum\\[2mm]
\label{power-int-sol}
&=\left(\frac{k}{\lam+1}\right)^{\frac{\lam}{2}}w^{\frac{3}{2}}\int_0^1 dt'\ {t'}^{\frac{1}{2}}
(1-wt')^{\frac{\lam}{2}-\frac{3}{2}}+C_1\,,$ 
%%%%%%%%%%%%%%%%%%%%%
where in the last line we have made the change of variable $t'=\frac{t}{w}$. Using the integral
representation of the hypergeometric function \eqref{int-rep-hyper}, 
Eq. \eqref{power-phi1} leads to the following expression for the scalar field $\Phi(y)$. 
%%%%%%%%%%%%%%
\eq$\label{power-Phi}
\Phi_{\pm}(y)=\pm \frac{2\Phi_0}{3}\sqrt{\frac{\lam}{(\lam+1)^{\lam-1}}}\left[1-\frac{\lam+1}{k(y+y_0)}
\right]^{\frac{3}{2}}\,_2F_1\left(\frac{3}{2}-\frac{\lam}{2},\frac{3}{2};\frac{5}{2};1-\frac{\lam+1}
{k(y+y_0)}\right)\, .$
In the above, we have also used the translational symmetry of the gravitational field equations
with respect to the scalar field, discussed also in Sec. \ref{linear-bulk}, to set $C_1=0$.

A solution for the scalar field similar to Eq. (\ref{power-Phi}) was derived in the context of
our previous analysis \cite{KNP2} for an exponential coupling function $f(y)$ and an
anti-de Sitter brane ($\Lambda<0$). The mathematical properties of the solution were studied there
in detail, therefore, here, we adapt those results in the present case and present our solutions 
for the scalar field without repeating the analysis---we refer the interested reader
to our previous work for further information. 

Trying to simplify Eq. (\ref{power-Phi}), we first note that, for every value of the coordinate $y$,
the argument $1-\frac{\lam+1}{k(y+y_0)}$ of the hypergeometric function is positive and smaller
than unity. Therefore, one can expand the hypergeometric function in power series as
\cite{Abramowitz, KNP2}
%%%%%%%%%%%%%%%%%%%%%%
\eq$\label{hyper-expr}
\,_2F_1\left(\frac{3}{2}-\frac{\lam}{2},\frac{3}{2};\frac{5}{2};1-\frac{\lam+1}
{k(y+y_0)}\right)=\sum_{n=0}^\infty \frac{\Gamma\left(\frac{3}{2}-\frac{\lam}{2}+n\right)}
{\Gamma\left(\frac{3}{2}-\frac{\lam}{2}\right)}\frac{3}{(2n+3)n!}\left[1-\frac{\lam+1}
{k(y+y_0)}\right]^n\,.$
%%%%%%%%%%%%%%%%%
There are two interesting categories of values for the parameter $\lam$ which lead to even simpler
and more elegant expressions for the hypergeometric function and subsequently for the scalar field.
These are:
%%%%%%%%%%%%%%%%%%%%%%%%%%
\begin{enumerate}
\item[\bf{(i)}] If $\lam=1+2q$ with $q \in {\mathbb{Z}}^{>}$, then, from Eq. \eqref{hyper-expr},
we have:
\bal$
_2F_1  \left(\frac{3-\lam}{2},\frac{3}{2};\frac{5}{2};1-\frac{\lam+1}
{k(y+y_0)}\right)
&=\left\{\begin{array}{cr}
1\,, &  q=1\\[3mm]
1+\sum_{n=1}^{q-1}\frac{3(-q+1)(-q+2)\cdots(-q+n)}{(2n+3)n!}\left[1-\frac{\lam+1}
{k(y+y_0)}\right]^n, &  q>1
\end{array}\right\} .
\label{F-series2}$
%%%%%%%%%%%%%%%

\begin{SCfigure}[][h!]
    \centering
    \includegraphics[width=0.5\textwidth]{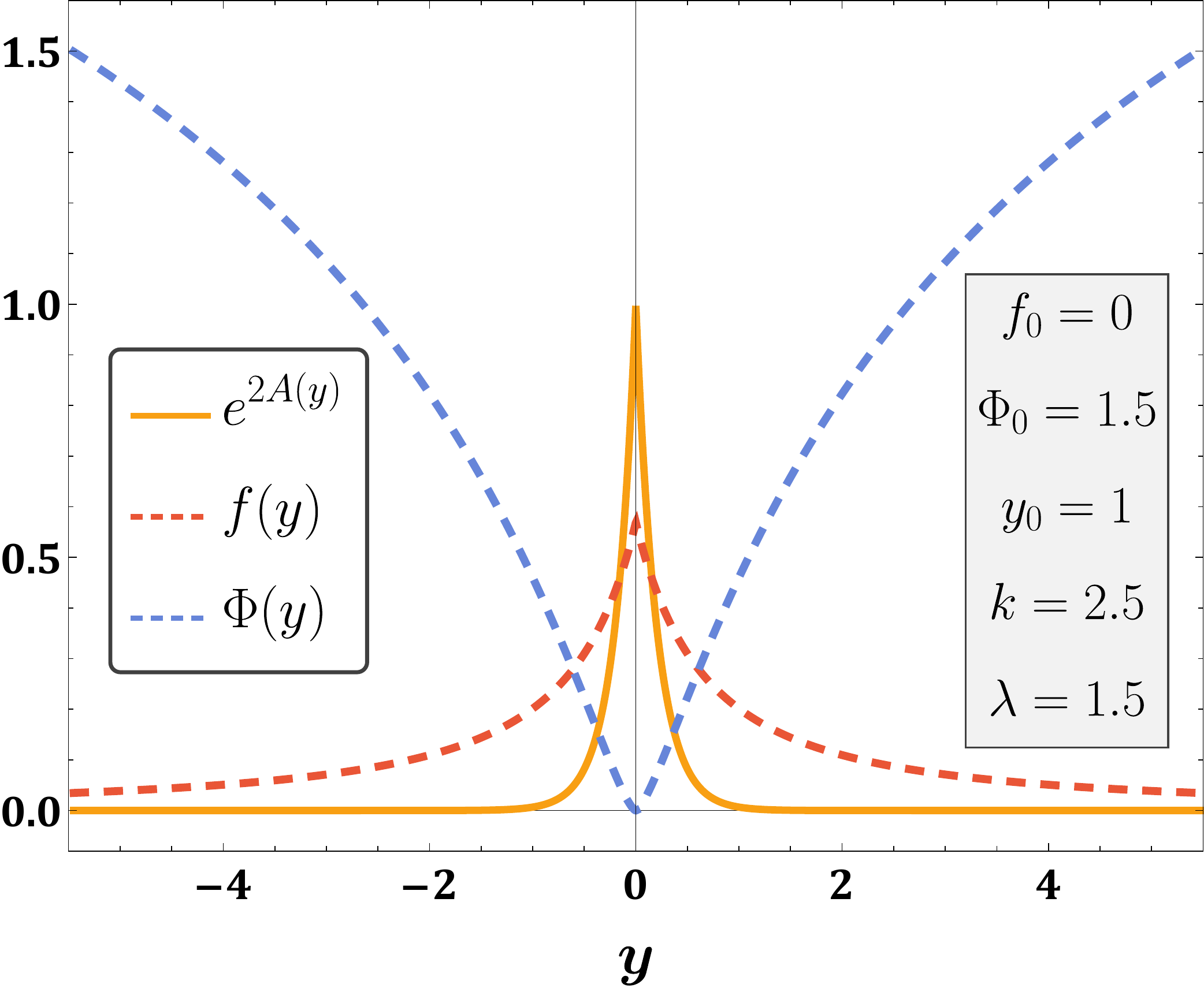}
    \caption{The warp factor $e^{2A(y)}=e^{-2k|y|}$, the coupling function $f(y)$ and the
    scalar field $\Phi(y)$ in terms of the coordinate $y$ for $f_0=0$, $\Phi_0=1.5$, $y_0=1$,
    $k=2.5$ and $\lam=1.5$.\\
    \vspace{1.3em}}
    \label{hyper-plot1}  
\end{SCfigure}

The solution for the scalar field then easily follows by using Eqs. (\ref{power-Phi}) and (\ref{F-series2})
and substituting the selected values for the parameter $\lam$ (or $q$).
As indicative cases, we present below the form of the scalar field for\footnote{For completeness,
we present here also the solution for the limiting case
with $\lam=1$ (i.e. for $q=0$); this has the form
%%%%%%%%%%%%%
$$\Phi_{\pm}(y)=\pm\frac{2\Phi_0}{3}
\left[\,{\rm arctanh}\left(\sqrt{1-\frac{2}{k(y+y_0)}}\,\right)-\sqrt{1-\frac{2}{k(y+y_0)}}
\,\right].$$}
%%%%%%%%%%%%%%
$\lambda=3$ (i.e. $q=1$) 
%%%%%%%%%%%%%%%%
$$\Phi_{\pm}(y)=\pm\frac{\Phi_0}{2\sqrt{3}}\left[1-\frac{4}{k(y+y_0)}\right]^{3/2},$$
%%%%%%%%%%%%%%%
and $\lambda=5$ (i.e. $q=2$)
$$\Phi_{\pm}(y)=\pm\frac{\Phi_0\sqrt{5}}{54}\left[1-\frac{6}{k(y+y_0)}\right]^{3/2}
\left\{1-\frac{3}{5}\left[1-\frac{6}{k(y+y_0)}\right]\right\}.$$
%%%%%%%%%%%

\item[\bf{(ii)}] If $\lam=2q$ with $q \in {\mathbb{Z}}^{>}$, we can always express the 
hypergeometric function in Eq. (\ref{power-Phi}) in terms of elementary functions, namely 
$\arcsin$, square roots and powers of its argument. For $\lam=2$ (i.e. $q=1$), it is
\eq$\label{hyper-lam-2}
\,_2F_1\left(\frac{1}{2},\frac{3}{2};\frac{5}{2};u^2\right)=\frac{3}{2}\frac{1}{u^2}\left(
\frac{\arcsin\, u}{u}-\sqrt{1-u^2}\right)\,.$
Therefore, from Eq. (\ref{power-Phi}), the scalar field for $\lam=2$ can be written in the form
$$\Phi_{\pm}(y)=\pm\frac{\Phi_0\sqrt{2}}{3}\left[1-\frac{3}{k(y+y_0)}\right]^{\frac{1}{2}}
\left[\left(1-\frac{3}{k(y+y_0)}\right)^{-\frac{1}{2}}\arcsin\left(\sqrt{1-\frac{3}{k(y+y_0)}}
\right)-\sqrt{\frac{3}{k(y+y_0)}}\ \right].$$
%%%%%%%%%%%%%%%%%
For larger values of $\lambda$ (i.e. for $q=1+\ell$, with $\ell \in {\mathbb{Z}}^{>}$),
the following relation holds 
\bal$\,_2F_1\left(\frac{1}{2}-\ell,\frac{3}{2};\frac{5}{2};u^2\right)u^2=&
\alpha\left(\frac{\arcsin u}{u}-\sqrt{1-u^2} \right)\nonum\\[2mm]
&+\sqrt{1-u^2}\left(\beta_1\,u^2+\beta_2\,u^4+\dots+\beta_{\ell-1}\,u^{2(\ell-1)}+
\beta_\ell\,u^{2\ell}\right),
\label{hyper-lam>2}$
where $\alpha, \beta_1, \dots, \beta_\ell$ are constant coefficients, which satisfy 
a system of $\ell+1$ linear algebraic equations \cite{KNP2}---the solution
of this system readily determines the unknown coefficients $\alpha, \beta_1, \dots, \beta_\ell$.
For example, for $\ell=1$ (i.e. for $q=2$, or equivalently $\lam=4$), this set of equations
gives $\alpha=3/8$ and $\beta_1=3/4$. Upon substituting these in \eqref{hyper-lam>2}, the
solution for the scalar field follows from Eq. \eqref{power-Phi} and has the form
%%%%%%%%%%%%%%%%%%%%%
\gat$\Phi_{\pm}(y)=\pm\frac{\Phi_0}{5\sqrt{5}}\left[1-\frac{5}{k(y+y_0)}\right]^{\frac{1}{2}}
\left[\frac{1}{2}\left(1-\frac{5}{k(y+y_0)}\right)^{-\frac{1}{2}}\arcsin\left(\sqrt{1-\frac{5}
{k(y+y_0)}}\right)\right.\nonum\\[3mm]
\hspace{15em}+\left.\sqrt{\frac{5}{k(y+y_0)}}\left(\frac{1}{2}-\frac{5}{k(y+y_0)}\right)\right].$
%%%%%%%%%%%%%%%%%%%%%%
\end{enumerate}

\par In Fig. \ref{hyper-plot1}, we depict the coupling function $f(y)$ and the scalar field $\Phi(y)$
for the indicative set of parameters $f_0=0$, $\Phi_0=1.5$, $y_0=1$, $k=2.5$ and $\lam=1.5$. For
comparison, we also display the exponentially decreasing warp factor. The coupling function remains
localised near the brane and asymptotically decreases to the constant value $f_0$, which here has
been taken to be zero. The scalar field starts from a constant value at the location of the brane,
which for this set of parameters turns out to be zero, and goes asymptotically to a constant value
that depends on the values of $\Phi_0$ and $\lambda$. Although this is not very clear from 
Fig. \ref{hyper-plot1}, it easily follows from Eq. (\ref{power-Phi}) with the asymptotic value of
the scalar field, as $y \rightarrow \pm\infty$, coming out to be
%%%%%%%%%%%%%
\eq$\lim_{y\ra\pm\infty}\Phi_{\pm}(y)=\pm \frac{\sqrt{\pi}\Phi_0}{2}\sqrt{\frac{\lam}{(\lam+1)^{\lam-1}}}
\ \frac{\Gamma\left(\frac{\lam}{2}-\frac{1}{2}\right)}{\Gamma\left(\frac{\lam}{2}+1\right)}\,.$
%%%%%%%%%%%%%%%%
It is worth noting that the profiles of both $f(y)$ and $\Phi(y)$ do not change with the variation
of the values of the parameters. 

\begin{SCfigure}[][t]
    \centering
    \includegraphics[width=0.55\textwidth]{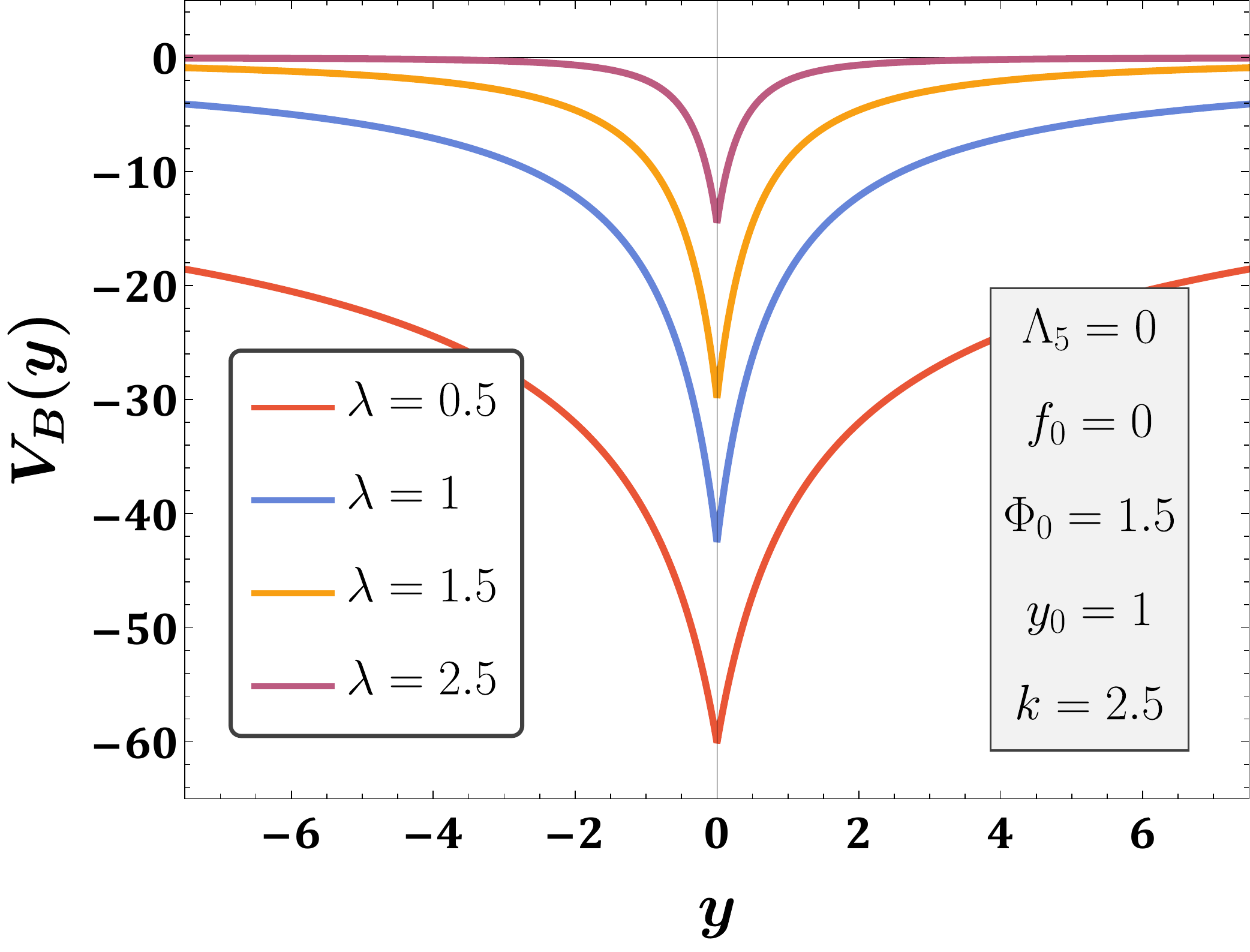}
    \caption{The scalar potential $V_B$ in terms of the extra dimension $y$ for 
    $\Lambda_5=0$, $f_0=0$, $\Phi_0=1.5$, $y_0=1$, $k=2.5$ and $\lam=0.5,\, 1,\, 1.5,\, 2.5$ 
    (from bottom to top).\\
    \vspace{2em}}
    \label{hyper-plot-pot}  
\end{SCfigure}

\par The potential of the scalar field $V_B(y)$ in the bulk can be determined from Eq. \eqref{V-B} 
using the expression of the coupling function $f(y)$. Thus, we find
\eq$\label{hyper-V-y}
V_B(y)=-\Lambda_5-6k^2f_0-\frac{\Phi_0^2}{2k^\lam (y+y_0)^{\lam+2}}\left[12k^2(y+y_0)^2+7\lam k(y+
y_0)+\lam(\lam+1)\right]\, .$
Since $\lam>0$, the last term in the above expression is negative-definite; it also vanishes as
$y \rightarrow +\infty$ leaving the parameters $\Lambda_5$ and $f_0$ to determine its asymptotic
value. Depending on the values of the parameters, the potential may be either positive or negative
at the location of the brane or asymptotically far away. In Fig. \ref{hyper-plot-pot}, one
can observe the aforementioned behaviour of the scalar potential $V_B(y)$. The values of the fixed
parameters $\Lambda_5,\, f_0,\, \Phi_0,\, y_0,\, k$ are the same as in 
Fig. \ref{hyper-plot1}, while the parameter $\lam$ varies.

\par The energy density $\rho(y)$ and pressure $p(y)=p^i(y)=p^y(y)$ may be finally computed by
employing  Eqs. \eqref{linear-rho} and \eqref{linear-p}. Then, we are led to the result
\eq$\label{power-rho}
\rho(y)=-p(y)=-6k^2f(y)=-6k^2\left[f_0+\frac{\Phi_0^2}{k^\lam(y+y_0)^\lam}\right].$
If we wish to satisfy the weak energy conditions close and on the brane, we should have $\rho(0)>0$,
which in turn means $f(0)<0$; in that case, the parameters of the model should satisfy the
following inequality: 
\eq$\label{power-weak-con}
\frac{f_0}{\Phi_0^2}<-\frac{1}{(ky_0)^\lam}\,.$

%%%%%%%%%%%%%%%%%%%%%%%%%%%%%%%%%%%%%%%%%%%%%%%%

\subsection{Junction conditions and the effective theory}

\par From the field equations (\ref{grav-1}) and (\ref{phi-eq}), we obtain the following
junction conditions for the matter on the brane:
\gat$\label{jun-con1}
3f(y)[A']=-[f']-(\sig+V_b)\,, \\[3mm]
\label{jun-con2}
[\Phi']=4[A']\pa_\Phi f+\pa_\Phi V_b\,,$
where all quantities are again evaluated at $y=0$. The only difference in this case is that we
have used the derivatives of the coupling function with respect to the coordinate $y$ rather
than the one with respect to the scalar field. This is due to the fact that the explicit expression
of the function $f(\Phi)$ is not know---although $\Phi(y)$ is a one-to-one function, it 
cannot in general be inverted. Taking advantage of the $\mathbf{Z}_2$ symmetry in the bulk,
we can easily evaluate the total energy density on the brane by Eq. \eqref{jun-con2}, which
is given by
\eq$\label{power-jc1}
\sig+V_b(\Phi)\Big|_{y=0}=6kf_0+\frac{2\Phi_0^2}{k^\lam\,y_0^{\lam+1}}(3ky_0+\lam)=6k\Phi_0^2
\left[\frac{f_0}{\Phi_0^2}+\frac{3ky_0+\lam}{3(ky_0)^{\lam+1}}\right]\,.$
If we demand the total energy density on the brane to be positive, namely $\sig+V_b(\Phi)|_{y=0}>0$,
then we straightforwardly deduce the constraint
\eq$\label{power-brane-ene-con}
\frac{f_0}{\Phi_0^2}>-\frac{3ky_0+\lam}{3(ky_0)^{\lam+1}}\,.$

In order to evaluate the first jump condition (\ref{jun-con1}), we write:
$\partial_\Phi f= \partial_y f/\Phi'$ and $\partial_\Phi V_b= \partial_y V_b/\Phi'$.
We are allowed to do this since, as we mentioned previously, the function $\Phi(y)$
does not possess any extrema in the bulk, therefore $\Phi'(y)$ never vanishes. Then, 
multiplying both sides of Eq. (\ref{jun-con1}) by $\Phi'$ and using Eq. 
(\ref{power-der-Phi}), we obtain the condition
\eq$\label{power-jc2}
\pa_y V_b\Big|_{y=0}=-\frac{2\lam\Phi_0^2}{k^\lam\, y_0^{\lam+2}}(3ky_0+\lam+1)\,.$
Due to the fact that $\lam>0$, $k>0$ and $y_0>0$, the r.h.s. of the above equation
never vanishes, which means that $V_b\neq const.$

\par Let us now focus on the effective four-dimensional theory on the brane. Using
Eq. \eqref{action_eff} and the  expression for the coupling function, from
Eq. \eqref{power-f} we obtain:
\bal$
\frac{1}{\kappa_4^2}&=
\frac{f_0}{k}+\frac{2\Phi_0^2}{k^\lam}\int_0^\infty dy \frac{e^{-2ky}}{(y+y_0)^\lam}
=\frac{f_0}{k}+\frac{2\Phi_0^2\,e^{2ky_0}}{k^\lam}\int_0^\infty dy\frac{e^{-2k(y+y_0)}}
{(y+y_0)^\lam}\, .$
Setting $t=2k(y+y_0)$, the above relation takes the form
\eq$\label{power-kappa1}
\frac{1}{\kappa_4^2}=\frac{f_0}{k}+\frac{2\Phi_0^2\, e^{2ky_0}}{k^\lam}(2k)^{\lam-1}
\int_{2ky_0}^{\infty}dt\ t^{-\lam} \,e^{-t}=\frac{f_0}{k}+\frac{2^\lam\Phi_0^2\, e^{2ky_0}}
{k}\,\Gamma(1-\lam,2ky_0)\,.$
Above, we have used the upper incomplete gamma function $\Gamma(s,x)$, defined
as follows
\eq$\label{upper-gamma}
\Gamma(s,x)\equiv\int_x^\infty dt\ t^{s-1}\, e^{-t}\,.$
The properties of the incomplete gamma function as well as the expressions giving its
numerical values are discussed in Appendix \ref{incom-gamma}. With the use of Eq.
\eqref{power-kappa1} and the relation $1/\kappa_4^2=M_{Pl}^2/(8\pi)$, we 
finally obtain
\eq$\label{power-planck}
M_{Pl}^2=\frac{8\pi\Phi_0^2}{k}\left\{\frac{f_0}{\Phi_0^2}+2^\lam\,e^{2ky_0}\,\Gamma(1-\lam,2ky_0)
\right\}\, .$
%%%%%%%%%%%%%%%%%%%
Demanding the positivity of the effective four-dimensional gravitational scale $M_{Pl}^2$, we
are led to the additional constraint
\eq$\label{power-eff-con}
\frac{f_0}{\Phi_0^2}>-2^\lam\,e^{2ky_0}\,\Gamma(1-\lam,2ky_0)\,.$
\par Finally, substituting the total energy density on the brane from Eq. \eqref{power-jc1} and the
expressions of the functions $f(y)$, $\Phi(y)$\footnote{For the calculation of the effective
four-dimensional cosmological constant on the brane $\Lambda_4$, it is more convenient to
use the relation \eqref{power-der-Phi} instead of the explicit form of the scalar field $\Phi(y)$
as given by Eq. \eqref{power-Phi}.} and $V_B(y)$ in Eq. \eqref{cosm_eff}, we can verify 
that the effective four-dimensional cosmological constant on the brane is zero, as expected.

%%%%%%%%%%%%%%%%%%%%%%%%%%%%%%%%%%%%%%%%%%%%%%%%%%%%%%%%%%%%%%%%%%%%%%%%%

\subsection{Energy conditions and the parameter space}

In this subsection, we will study the parameter space of the ratio $f_0/\Phi_0^2\,$
and the dimensionless parameter $ky_0$. The value of the parameter $\lam$ may be also varied,
however, once fixed, it determines the allowed values of the parameter $ky_0$ through the
constraint \eqref{power-con1}. As usually, we will investigate the parameter regimes where
the inequalities \eqref{power-weak-con}, \eqref{power-brane-ene-con} and \eqref{power-eff-con} 
are satisfied.

%%%%%%%%%%%%%%%%%%%%
\begin{SCfigure}[][t]
    \centering
    \includegraphics[width=0.5\textwidth]{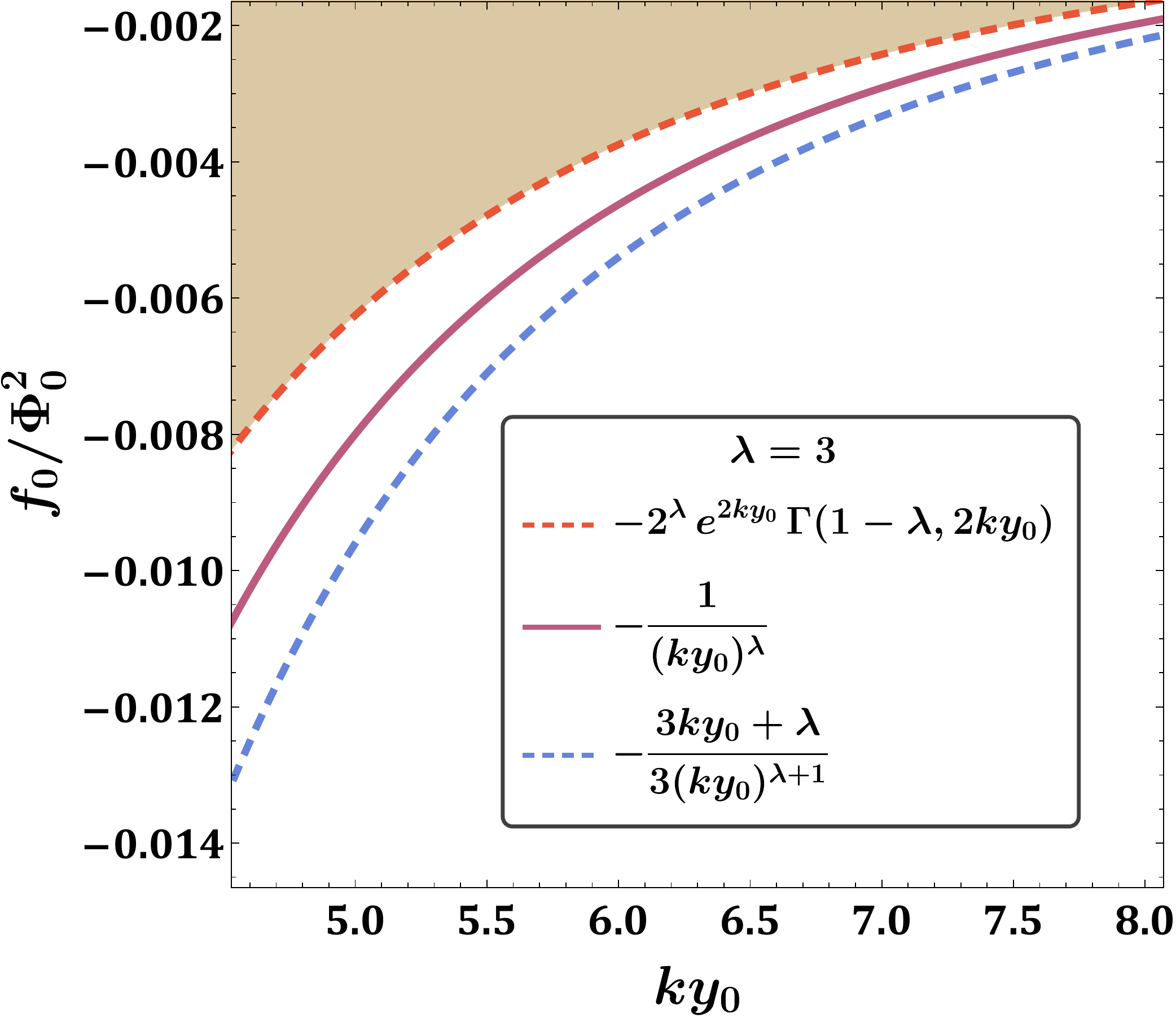}
    \caption{The parameter space between the ratio $f_0/\Phi_0^2$
    and the parameter $ky_0$, for $\lam=3$. The figure depicts also the plots of the 
    expressions appearing on the r.h.s.'s of the inequalities \eqref{power-weak-con},
    \eqref{power-brane-ene-con} and \eqref{power-eff-con}.\\
    \vspace{2em}}
    \label{hyper-plot-par}  
\end{SCfigure}
%%%%%%%%%%%%%%%%%%

In Fig. \ref{hyper-plot-par}, we depict the aforementioned parameter space for the 
value $\lam=3$. We also depict the curves of the expressions on the r.h.s.'s of the inequalities
\eqref{power-weak-con}, \eqref{power-brane-ene-con} and \eqref{power-eff-con}. 
Although the corresponding  curves have been drawn for a particular value of $\lam$, it turns
out that their relative position remains the same for any allowed value of the parameters
$\lam$ and $ky_0$, namely we always have:
$$-2^\lam\,e^{2ky_0}\,\Gamma(1-\lam,2ky_0)>-\frac{1}{(ky_0)^\lam}>-\frac{3ky_0+\lam}{3(k
y_0)^{\lam+1}}\,.$$
Clearly, this means that only the inequalities \eqref{power-brane-ene-con} and \eqref{power-eff-con}
can be simultaneously satisfied. Therefore, we may easily obtain a model with a positive four-dimensional 
gravitational constant and a positive total energy density on the brane. However, in that case, 
we will not be able to satisfy the weak energy conditions by the bulk matter close and on the brane. 
This means that the energy density $\rho$ will be negative at the location of the brane with
the pressure having the exact opposite value.

%%%%%%%%%%%%%%%%%%%%%%%%%%%%%%%%%%%%%%%%%%%%%%%%%%%%%%%%%%%%%%%%%%%
%
%
%%%%%%%%%%%%%%%%%%%%%%%%%%%%%%%%%%%%%%%%%%%%%%%%%%%%%%%%%%%%%%%%%%%

\section{A Linear-Exponential Coupling Function in terms of $y$}

In this case, we consider the following coupling function $f(y)$ in terms
of the coordinate $y$:
\eq$\label{exp-f}
f(y)=f_0+f_1\, ky\,e^{-\lam ky}\,.$
We also assume that $f_1\in \mathbb{R}\setminus\{0\}$ and $\lam\in(0,+\infty)$ in order
for $f(y)$ to  satisfy the physical constraints discussed at the end of Sec. \ref{th-frame}. 

Let us start by deriving first the bulk solution. Substituting the aforementioned coupling
function in Eq. \eqref{grav-1-2}, we obtain:
\eq$\label{exp-der-Phi}
[\Phi'(y)]^2=f_1\,k^2\,e^{-\lam ky}\left[2\lam-1-\lam(\lam-1)ky\right]\,.$
%%%%%%%%%%%%%%%%%%%%%%
Since the scalar field $\Phi(y)$ should be a real-valued function, it is obvious
that $[\Phi'(y)]^2\geq 0$ for all values of $y$ which are greater or equal to zero.
Let us first assume that $f_1<0$; then, demanding that $[\Phi'(0)]^2 \geq 0$, we obtain
the constraint $\lam\leq 1/2$. On the other hand, for large values of the $y$-coordinate
(i.e. at $y=y_0\gg 1$), demanding that $[\Phi'(y_0)]^2 \geq 0$ leads to $\lam\geq 1$\footnote{
Here, we have used the fact that, for large values of $y$, only the term proportional to $ky$ mainly
contributes to the value of $[\Phi'(y)]^2$.}. However, these two constraints are incompatible,
which leads us to deduce that the parameter $f_1$ should be strictly positive. In that
case, a similar argument as above leads to the allowed regime $\lam\in\left[\frac{1}{2},1\right]$. 
Moreover, since $f_1$ is positive, we may set $f_1=\Phi_0^2$, and assume for simplicity
that $\Phi_0\in(0,+\infty)$.

For $\lam=1$, we can easily integrate Eq. \eqref{exp-der-Phi} with respect to $y$, and
determine the expression of the function of the scalar field $\Phi(y)$. Then, we obtain
%%%%%%%%%%%%%%%%%%%%%%
\eq$\label{exp-phi1}
\Phi_{\pm}(y)=\pm\, 2\Phi_0\, e^{-ky/2}\,.$
Above, we have used again the translational symmetry of the gravitational field equations 
with respect to the value of the scalar field in order to eliminate an additive integration
constant. By inverting the above function, we can express the coupling function in terms
of the scalar field $\Phi$, namely
\eq$\label{exp-f-phi}
f(\Phi)=f_0-\frac{\ \Phi^2}{2}\ln\left(\frac{\Phi}{2\Phi_0}\right)\,.$

Equation \eqref{exp-der-Phi} is more difficult to solve in the remaining $\lam$-parameter
regime, i.e. for $\lam\in\left[\frac{1}{2},1\right)$. In that case, Eq. \eqref{exp-der-Phi} 
leads to 
%%%%%%%%%%%%%%%%%%%%%%%%
\bal$\Phi_{\pm}(y)&=\pm\,\Phi_0\,k\int dy\, e^{-\lam ky/2}\sqrt{2\lam-1-\lam(\lam-1)
ky}\nonum\\[3mm]
&=\pm\,\frac{2\Phi_0}{\lam}\left[-e^{-\lam ky/2}\sqrt{2\lam-1-\lam(\lam-1)ky}
+\int dy\, e^{-\lam ky/2}\,\frac{d}{dy}\left(\sqrt{2\lam-1-\lam(\lam-1)ky}\right)
\right].
\label{exp-phi2-1}$
%%%%%%%%%%%%%%%%%%%%%%%
Focusing on the second term of the r.h.s. of the above relation, and due to the
fact that $\lam\in\left[\frac{1}{2},1\right)$, we can write
%%%%%%%%%%%%%%%%%%%%%%
\bal$\int dy\,e^{-\frac{\lam ky}{2}}\,\frac{d}{dy}\left(\text{\small{$\sqrt{2\lam
-1-\lam(\lam-1)ky}$}}\right)&=e^{\frac{2\lam-1}{2(1-\lam)}}\sqrt{2(1-\lam)}\int dy\,
e^{-\frac{2\lam-1-\lam
(\lam-1)ky}{2(1-\lam)}} \frac{d}{dy}\text{\small{$\sqrt{\frac{2\lam-1-
\lam(\lam-1)ky}{2(1-\lam)}}$}}\nonum\\[3mm]
&=\sqrt{\frac{\pi(1-\lam)}{2}}\,e^{\frac{2\lam-1}{2(1-\lam)}}\,\text{erf}\left(
\text{\small{$\sqrt{\frac{2\lam-1-\lam(\lam-1)ky}{2(1-\lam)}}$}}\,\right)\,,
\label{exp-phi2-2}$
where we have used the error function, defined as
$$\text{erf}(x)=\frac{2}{\sqrt\pi}\int_0^x dt\, e^{-t^2}\,,$$
and its property
$$\frac{d}{dx}\, \text{erf}(g(x))=\frac{2}{\sqrt\pi}e^{-g(x)^2}\, \frac{dg(x)}{dx}\,.$$
Combining Eqs. \eqref{exp-phi2-1} and \eqref{exp-phi2-2} we obtain
%%%%%%%%%%%%%%%%%%%%%%%
\eq$\label{exp-phi2}
\Phi_{\pm}(y)=\pm\,\frac{2\Phi_0}{\lam}\left[-e^{-\lam ky/2}\text{\footnotesize{$\sqrt{
2\lam-1-\lam(\lam-1)ky}$}}+\sqrt{\frac{\pi(1-\lam)}{2}}\,e^{\frac{2\lam-1}{2(1-
\lam)}}\,\text{erf}\left(
\text{\footnotesize{$\sqrt{\frac{2\lam-1-\lam(\lam-1)ky}{2(1-\lam)}}$}}\right)\right].$
%%%%%%%%%%%%%%%%%%%%%%%%%%
In this case, it is not possible to invert the function $\Phi(y)$ in order to find the
form of the coupling function $f(\Phi)$. However, from Eq. \eqref{exp-der-Phi} and for
$\lam\in\left[\frac{1}{2},1\right)$, it is straightforward to deduce that $\Phi'(y)\neq 0$
for all $y>0$; this, again, means that $\Phi(y)$ does not have any extremum, 
and is therefore a one-to-one function. This property will be of use in the evaluation
of the junction conditions on the brane.

%%%%%%%%%%%%%%%%%%%%%%%%%%%%%%%%%%%%%%%%%
\begin{figure}[t]
    \centering
    \begin{subfigure}[b]{0.495\textwidth}
        \includegraphics[width=\textwidth]{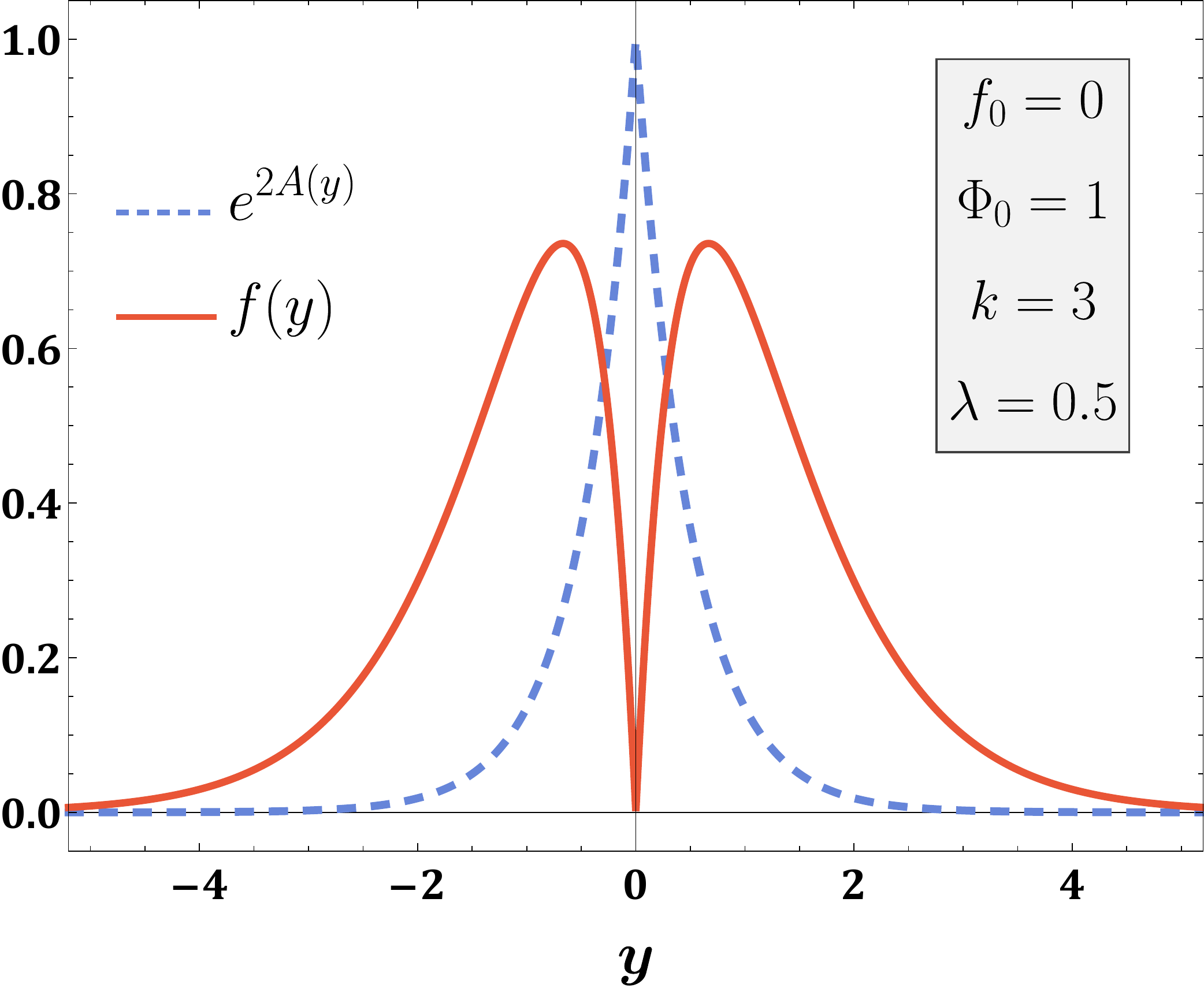}
        \caption{\hspace*{-1.7em}}
        \label{erf-plot1}
    \end{subfigure}
   % \hfill
    ~ %add desired spacing between images, e. g. ~, \quad, \qquad, \hfill etc. 
      %(or a blank line to force the subfigure onto a new line)
    \begin{subfigure}[b]{0.44\textwidth}
        \includegraphics[width=\textwidth]{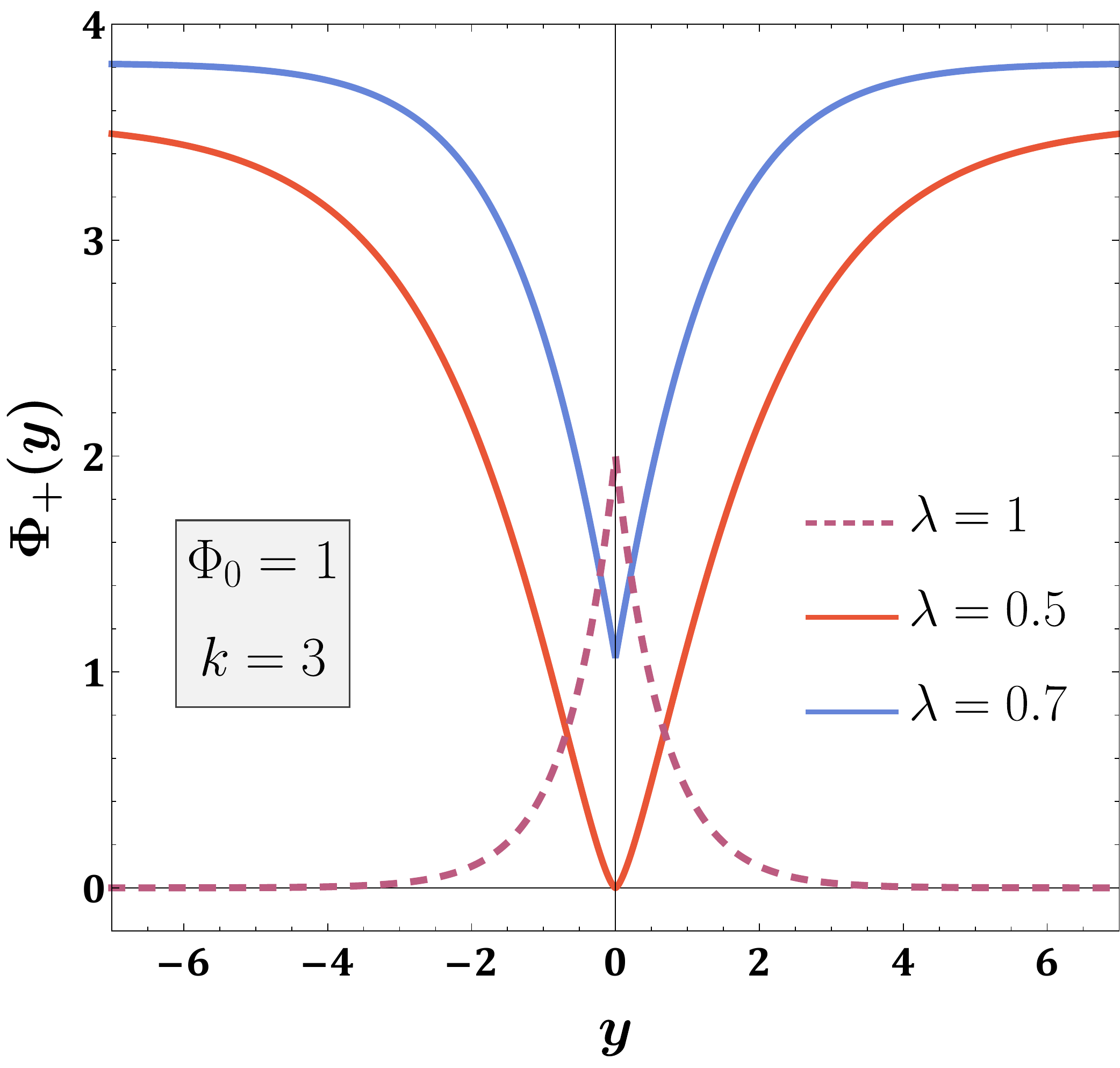}
        \caption{\hspace*{-2.6em}}
        \label{erf-plot2}
    \end{subfigure}
    
    \vspace{-0.5em}
    \caption{(a) The coupling function $f(y)$ and the warp factor $e^{-2A(y)}$ in terms
    of the $y$-coordinate for $f_0=0$, $\Phi_0=1$, $k=3$ and $\lam=0.5$, and (b)
    the scalar field $\Phi_+(y)$ for $\Phi_0=1$, $k=3$ and $\lam=1,\,0.5,\,0.7$.}
    \label{erf-plot-1-2}
\end{figure}
%%%%%%%%%%%%%%%%%%%%%%%%%%%%%

In Figs. \myref{erf-plot-1-2}{erf-plot1} and \myref{erf-plot-1-2}{erf-plot2}, we present
the warp factor, the coupling function and the scalar field for particular choices of values
for the parameters of the model. The coupling function $f(y)$ adopts the same constant
value $f_0$ at the location of our brane and at asymptotic infinity while reaching a
maximum value at some intermediate distance off our brane, as depicted in 
Fig. \myref{erf-plot-1-2}{erf-plot1}. In Fig. \myref{erf-plot-1-2}{erf-plot2}, the
scalar field presents two distinct profiles, for $\lam=1$ and $\lam \in\left[\frac{1}{2}
,1\right)$ due to the two different solutions given by Eqs. \eqref{exp-phi1} and
\eqref{exp-phi2}, respectively. In all cases, though, $\Phi_\pm(y)$ remains everywhere finite
approaching a constant value at asymptotic infinity: for $\lam=1$ this constant is zero,
while for $\lam \in\left[\frac{1}{2},1\right)$ this is given by the expression
%%%%%%%%%%%%%%
\eq$\label{exp-phi-inf}
\lim_{y\ra\pm\infty}\Phi_{\pm}(y)=\pm\frac{2\Phi_0}{\lam}\sqrt{\frac{\pi(1-\lam)}{2}}\,
e^{\frac{2\lam-1}{2(1-\lam)}},\hspace{1em}\lam\in\Bigl[\frac{1}{2},1\Bigr)\,.$
%%%%%%%%%%%%%%%
In the above, we have used the fact that the limit of the error function appearing in
Eq. \eqref{exp-phi2}, as $y \rightarrow +\infty$, is unity. Due to the $\mathbf{Z}_2$
symmetry imposed on our model, the same limit will hold for the scalar field also
for $y\ra-\infty$.

\par
\begin{SCfigure}
    \centering
    \includegraphics[width=0.48\textwidth]{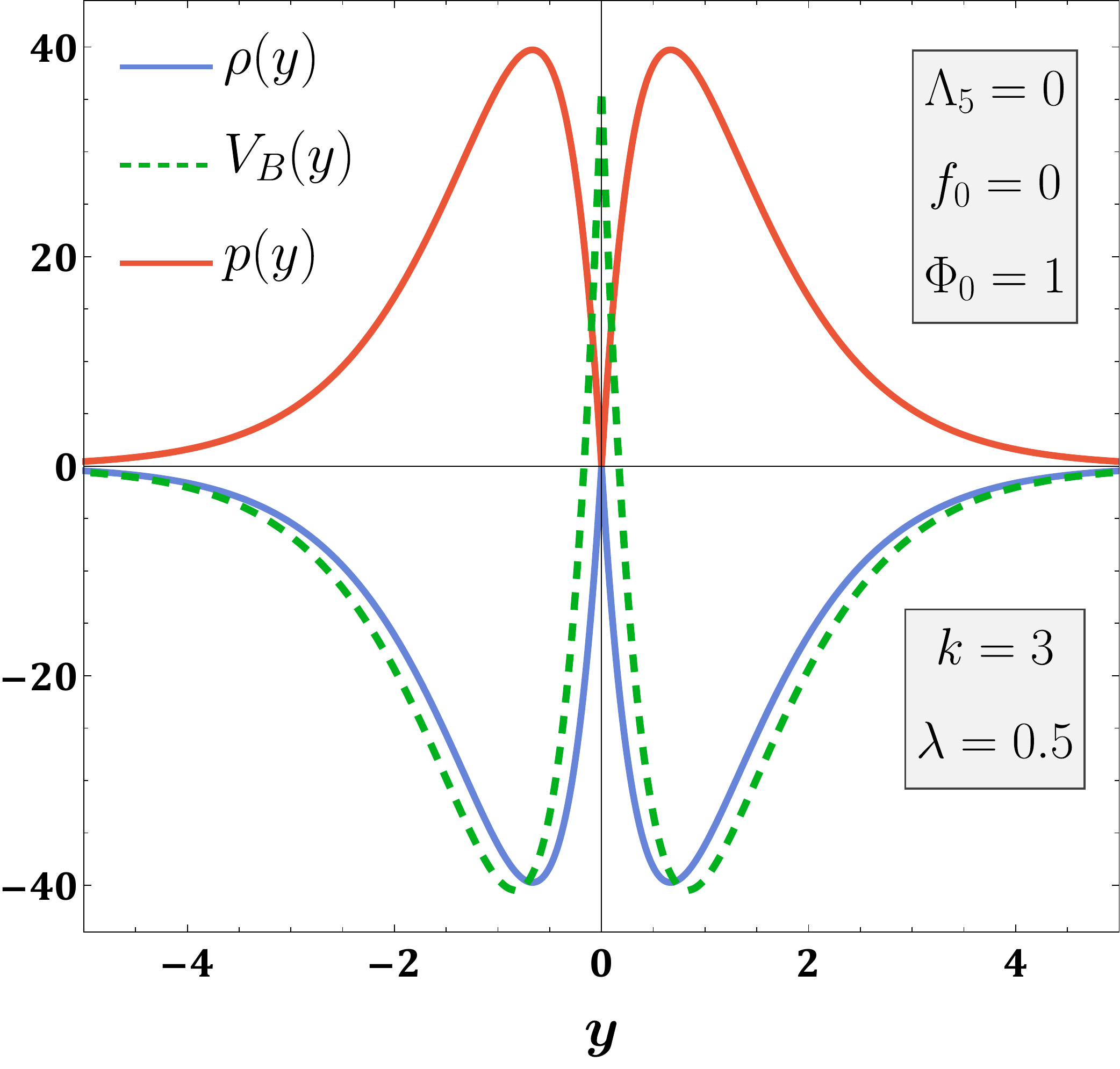}
    \caption{The energy density $\rho$ and pressure $p$ of the system together with
    the scalar potential $V_B$ in terms of the coordinate $y$ for $\Lambda_5=0$, $f_0=0$,
    $\Phi_0=1$, $k=3$, and $\lam=0.5$.\\
    \vspace{1.8em}}
    \label{erf-plot3}  
\end{SCfigure}

 From Eq. \eqref{V-B} , we may now determine the potential of the scalar field $V_B(y)$
in the bulk by using the expression of the coupling function $f(y)$. Then, we find
\eq$\label{exp-V-y}
V_B(y)=-\Lambda_5-6k^2f_0+\Phi_0^2\,k^2\,e^{-\lam ky}\left[\frac{7}{2}+\lam-ky\left(\frac{
\ \lam^2}{2}+\frac{7\lam}{2}+6\right)\right],\hspace{1em}\lam\in\left[\frac{1}{2},1\right]\, .$
%%%%%%%%%%%%%%%%%%%%
On the other hand, the energy density $\rho(y)$ and pressure $p(y)=p^i(y)=p^y(y)$ may be
computed by employing Eqs. \eqref{linear-rho} and \eqref{linear-p}; then, we obtain
\eq$\label{exp-rho}
\rho(y)=-p(y)=-6k^2f(y)=-6k^2\left(f_0+\Phi_0^2\, ky\,e^{-\lam ky}\right),$
In order to satisfy the weak energy conditions close and on the brane, we should have again
$\rho(0) \geq 0$, or equivalently $f(0) \leq 0$; hence, we are led to the following inequality: 
\eq$\label{exp-weak-con}\frac{f_0}{\Phi_0^2} \leq 0\,.$
%%%%%%%%%%%%%%%%
In Fig. \ref{erf-plot3}, we present the energy-density and pressure as well as the profile
of the bulk potential for the same values of parameters as in Fig. \ref{erf-plot-1-2}
for easy comparison. We observe that both components and the bulk potential are
everywhere finite, reach their maximum values at a finite distance from our brane and
reduce to a constant value (which here is taken to be zero) at large distances.

%%%%%%%%%%%%%%%%%%%%%%%%%%%%
%%%%%%%%%%%%%%%%%%%%%%%%%%%%%%%

Let us now turn to the junction conditions introduced in the theory at the location of the brane.
From Eqs. \eqref{jun-con1} and \eqref{jun-con2}, we obtain in a similar way the conditions:
\gat$\label{exp-jc1}
\sig+V_b(\Phi)\Big|_{y=0}=2k(3f_0-\Phi_0^2),\\[2mm]
\label{exp-jc2}
\pa_yV_b\Big|_{y=0}=8k^2\,\Phi_0^2\left(\lam^2-\lam+\frac{5}{4}\right),$
for $\lam\in\left[\frac{1}{2},1\right]$. The total energy density on the brane will be 
positive if and only if $\sig+V_b(\Phi)|_{y=0}>0$, which results to
\eq$\label{exp-brane-ene-con}
\frac{f_0}{\Phi_0^2}>\frac{1}{3}\, .$

Next, we are going to evaluate the effective four-dimensional gravitational constant on the brane.
Using Eq. \eqref{action_eff}, we obtain:
\gat$\frac{1}{\kappa_4^2}=\frac{M_{Pl}^2}{8\pi}=\frac{\Phi_0^2}{k}\left[\frac{f_0}{\Phi_0^2}+
\frac{2}{(2+\lam)^2}\right].$
For a robust effective theory on the brane, it is imperative to have a positive four-dimensional
gravitational constant, thus, we must satisfy the following constraint:
\eq$\label{exp-eff-con}
\frac{f_0}{\Phi_0^2}>-\frac{2}{(2+\lam)^2}\,.$
Once again, as expected, the effective four-dimensional cosmological constant on the brane
may be found to be zero with the use of Eq. \eqref{cosm_eff}.
 
It is straightforward to study whether the inequalities \eqref{exp-weak-con}, \eqref{exp-brane-ene-con}
and \eqref{exp-eff-con} can be simultaneously satisfied. By merely observing the first two of
them, it is easy to deduce that they are incompatible since the value of $f_0/\Phi_0^2$ can be
either positive or negative. Additionally, as we already mentioned, the parameter $\lam$
takes values in the range $\left[\frac{1}{2},1\right]$. In this case, it holds that 
\eq$\label{exp-eff-val}
-\frac{8}{25}\leq -\frac{2}{(2+\lam)^2}\leq -\frac{2}{9}\,.$
Hence, we can simultaneously satisfy either the inequalities \eqref{exp-weak-con} and
\eqref{exp-eff-con}, or \eqref{exp-brane-ene-con} and \eqref{exp-eff-con}. In particular,
a positive four-dimensional gravitational scale $M_{Pl}^2$ can be combined
with the bulk matter satisfying the weak energy conditions close to the brane, for
$$-\frac{2}{9}\leq \frac{f_0}{\Phi_0^2} \leq 0\,,$$
or with a positive total energy-density on the brane, for
$$\frac{f_0}{\Phi_0^2}>\frac{1}{3}\,.$$
The particular solution depicted in Fig. \ref{erf-plot3} corresponds to the value $f_0=0$;
therefore, it is characterised by a negative energy density inside the bulk, which
violates the energy conditions. Note, however, that at the location of our brane,
both the energy density and pressure are zero while the bulk potential is positive.

%\newpage
%%%%%%%%%%%%%%%%%%%%%%%%%%%%%%%%%%%%%%%%%%%%%%%%%%%%%%%%%%%%%%%%%%%%%%%%
%
%
%%%%%%%%%%%%%%%%%%%%%%%%%%%%%%%%%%%%%%%%%%%%%%%%%%%%%%%%%%%%%%%%%%%%%%%%

\section{A Double-Exponential Scalar Field in terms of $y$}

\begin{figure}[t]
    \centering
    \begin{subfigure}[b]{0.47\textwidth}
        \includegraphics[width=\textwidth]{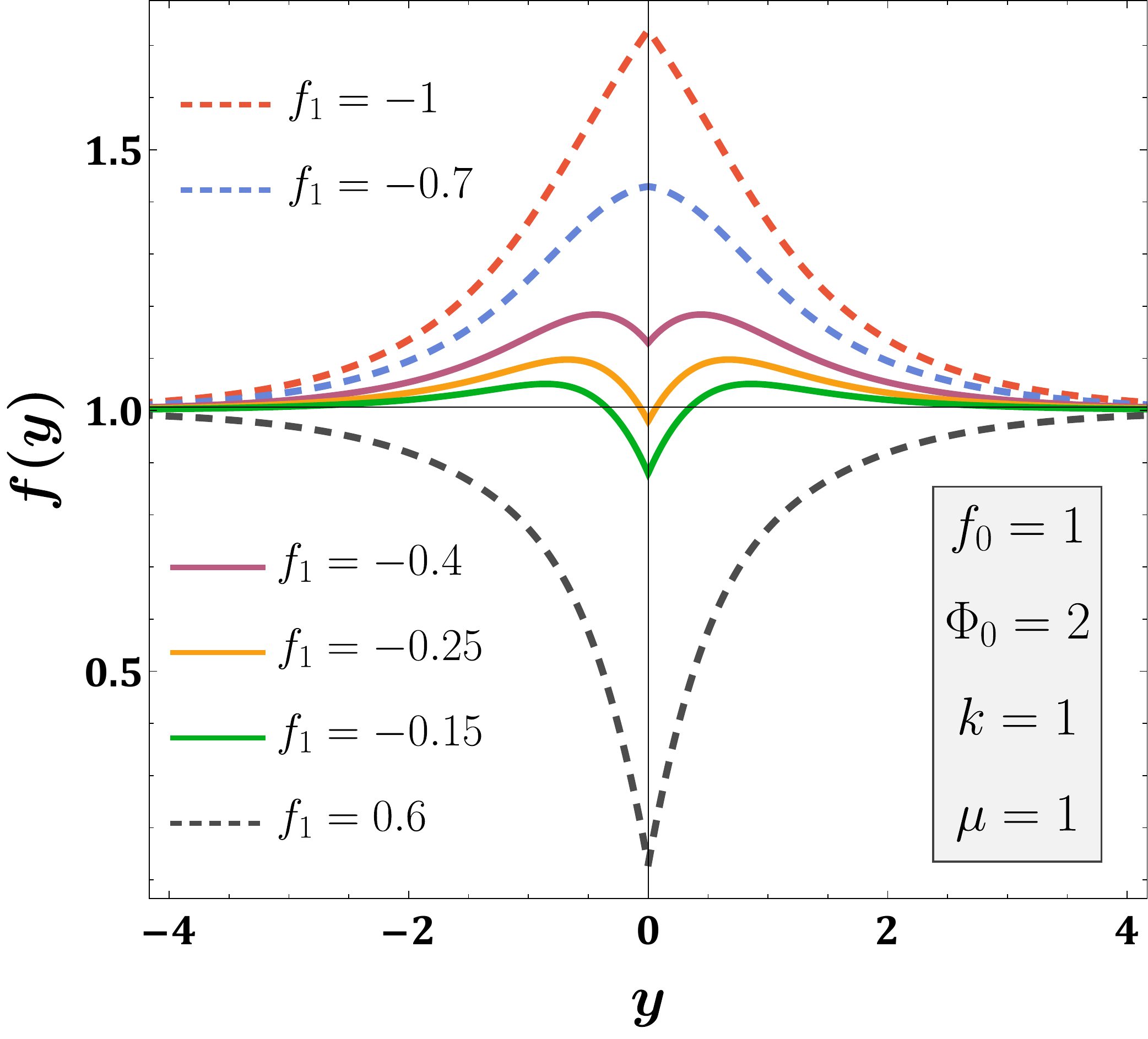}
        \caption{\hspace*{-3.4em}}
        \label{double-plot1}
    \end{subfigure}
    \hspace*{-0.1cm}
    ~ %add desired spacing between images, e. g. ~, \quad, \qquad, \hfill etc. 
      %(or a blank line to force the subfigure onto a new line)
    \begin{subfigure}[b]{0.47\textwidth}
        \includegraphics[width=\textwidth]{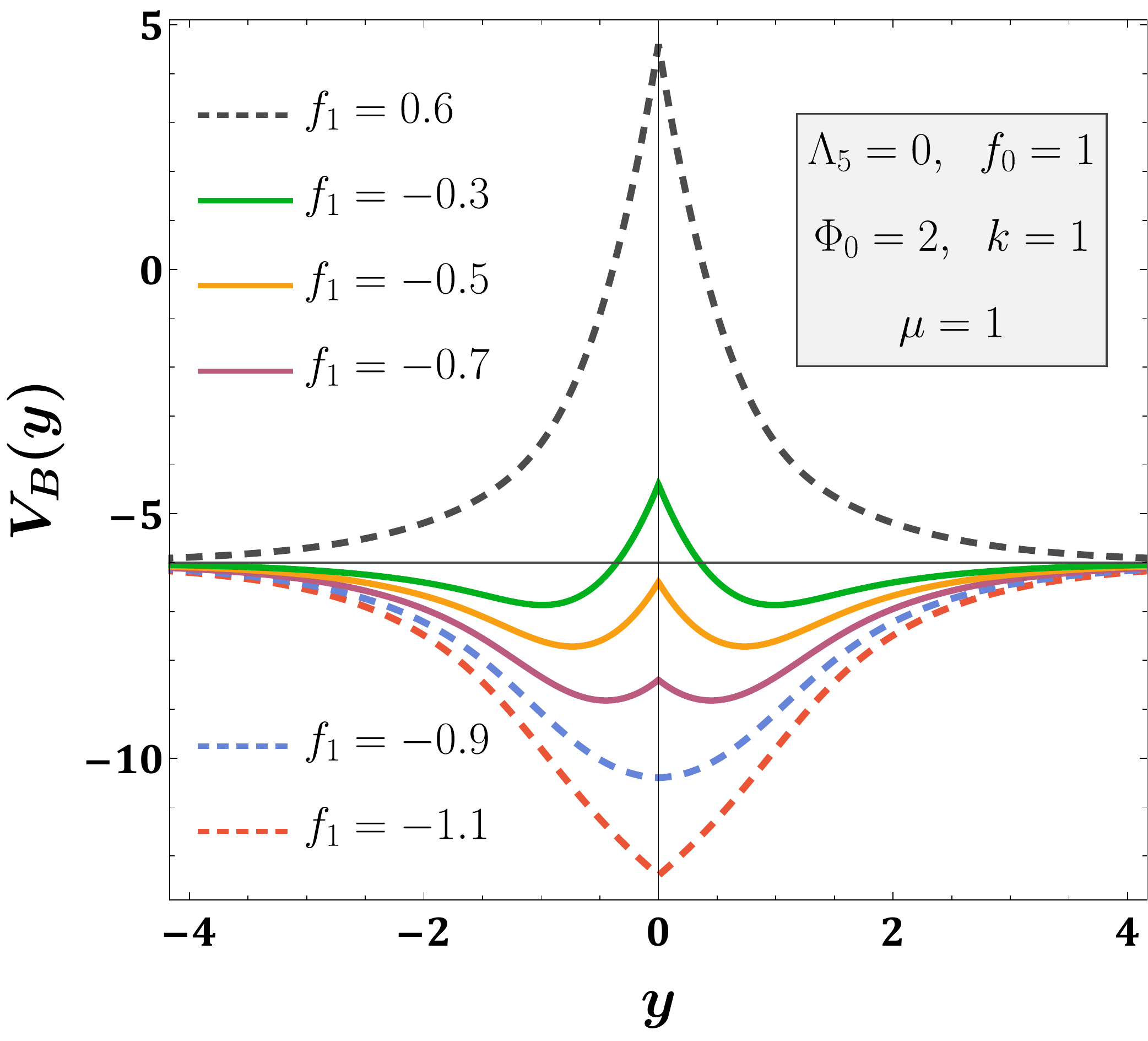}
        \caption{\hspace*{-3.7em}}
        \label{double-plot2}
    \end{subfigure}
    
    \vspace{-0.5em}
    \caption{(a) The coupling function in terms of the $y$-coordinate for $f_0=1$, $\Phi_0=2$,
    $k=1$, $\mu=1$ and $f_1=-1,\,-0.7,\,-0.4,\,-0.25,\,-0.15,\,0.6$ (from top to bottom), and
    (b) the scalar potential $V_B$ in terms of the coordinate $y$ for $\Lambda_5=0$,
    $f_0=1$, $\Phi_0=2$, $k=1$, $\mu=1$ and $f_1=-1.1,\,-0.9,\,-0.7,\,-0.5,\,-0.3,\,0.6$ (from
    bottom to top).}
    \label{double-plot-1-2}
\end{figure}

In this section, we follow an alternative approach and consider the following expression for the scalar
field in terms of the coordinate $y$:
\eq$\label{double-phi}
\Phi(y)=\Phi_0\, e^{-\mu^2 e^{ky}}\,.$
Although this expression seems similar to the sub-case of the quadratic coupling function
with $\lam=-1/4$, it differs significantly as it will become clear from the
expressions of the coupling function $f(\Phi)$ and the scalar potential $V_B(\Phi)$.
Moreover, it is obvious that both parameters $\Phi_0$ and $\mu$ can now take values in the 
entire set of real numbers except zero. 
With the form of the scalar field already known, it is straightforward to derive the corresponding
forms of the coupling function, bulk potential and components of the energy-momentum tensor.
Starting with the coupling function, upon substituting the aforementioned expression of the scalar
field in Eq. \eqref{grav-1-2}, we readily obtain
%%%%%%%%%%%%%%%%%
\eq$\label{double-f-y}
f(y)=f_0-f_1\,e^{-ky}-\frac{\Phi_0^2}{4\mu^2}\,e^{-2\mu^2e^{ky}}(\mu^2+e^{-ky})\,.$
%%%%%%%%%%%%%%%%%%
In the above result, the parameter $f_1$ is allowed to take values in the whole set of real
numbers, while the allowed values for the parameter $f_0$ will be examined shortly. Inverting 
the function $\Phi(y)$, the expression of the coupling function in terms of the scalar field reads
\eq$\label{double-f-phi}
f(\Phi)=f_0+\frac{f_1\, \mu^2}{\ln(\Phi/\Phi_0)}-\frac{\ \Phi^2}{4}\left(1-\frac{1}{\ln(
\Phi/\Phi_0)}\right)\,.$
The scalar potential $V_B(y)$ can then be evaluated employing Eqs. \eqref{V-B} and
\eqref{double-f-y}. Then, we find
\eq$\label{double-V-y}
V_B(y)=-\Lambda_5-6k^2f_0+10k^2f_1\,e^{-ky}+\frac{\Phi_0^2\,k^2}{2\mu^2}\,e^{-2\mu^2e^{ky}}
\left(5e^{-ky}+7\mu^2+4\mu^4\,e^{ky}+\mu^6\,e^{2ky}\right)\,,$
in terms of the $y$-coordinate, or
\eq$\label{double-V-phi}
V_B(\Phi)=-\Lambda_5-6k^2f_0-\frac{10k^2f_1\,\mu^2}{\ln(\Phi/\Phi_0)}+\frac{\Phi_0^2\,k^2}{
2}\left\{-\frac{5}{\ln(\Phi/\Phi_0)}+7-4\ln\left(\frac{\Phi}{\Phi_0}\right)+\left[\ln\left(
\frac{\Phi}{\Phi_0}\right)\right]^2\right\},$
%%%%%%%%%%%%%%%%%%%%%%
in terms of the scalar field. In Figs. \myref{double-plot-1-2}{double-plot1} and
\myref{double-plot-1-2}{double-plot2}, we display the profiles of the coupling function and
scalar potential in terms of the $y$ coordinate, for particular values of the parameters of
the model. The varying parameter here is $f_1$, which is clearly the decisive one for the
form of both functions. 
We observe that for positive $f_1$, the coupling function takes its lowest value at the
location of the brane while, as $f_1$ gradually takes larger negative values, the coupling function
eventually exhibits a peak at the location of the brane. The bulk potential has almost the exact
opposite profile of the coupling function: it acquires a maximum, positive value at the location
of our brane for $f_1>0$ while it turns to globally negative values for $f_1<0$. For every set of
values of the parameters, though, both functions are everywhere finite and reduce to a constant
value at large distances---this value, when $\Lambda_5=0$, is determined by $f_0$.

%%%%%%%%%%%%%%%%%
%%%%%%%%%%%%%%%%%%%%%%%%%%%%

Finally, the energy density $\rho(y)$ and pressure $p(y)=p^i(y)=p^y(y)$ components
may be computed as usually by employing Eqs. \eqref{linear-rho} and \eqref{linear-p}, in which case
we are led to the results 
%%%%%%%%%%%%%%%%%%%%
\eq$\label{double-rho}
\rho(y)=-p(y)=-6k^2f(y)=-6k^2\left[f_0-f_1\,e^{-ky}-\frac{\Phi_0^2}{4\mu^2}\,
e^{-2\mu^2e^{ky}}(\mu^2+e^{-ky})\right].$
In order to satisfy the weak energy conditions close and on the brane, we demand again that
$f(0)<0$; hence, we obtain the following inequality
\gat$\label{double-weak-con}
\frac{f_0}{\Phi_0^2}<\frac{f_1}{\Phi_0^2}+\frac{\mu^2+1}{4\mu^2}\,e^{-2\mu^2}\,.$

Turning now to the junction conditions, from Eqs. \eqref{jun-con1}, \eqref{jun-con2} and using
also the relations \eqref{double-phi}-\eqref{double-f-phi}, we obtain:
%%%%%%%%%%%%%%%%%%%%
\gat$\label{double-jc1}
\sig+V_b(\Phi)\Big|_{y=0}=6kf_0-8kf_1-\frac{\Phi_0^2\,k\,e^{-2\mu^2}}{2\mu^2}(4+5\mu^2+2\mu^4)
\,,\\[3mm]
\label{double-jc2}
\pa_yV_b\Big|_{y=0}=8k^2f_1+\frac{2\Phi_0^2\,	k^2}{\mu^2}\,e^{-2\mu^2}(1+2\mu^2+2\mu^4+\mu^6)\,.$

%%%%%%%%%%%%%%%%%%%%%%%%%%%%%%%%%
\begin{figure}[t]
    \centering
    \begin{subfigure}[b]{0.51\textwidth}
        \includegraphics[width=\textwidth]{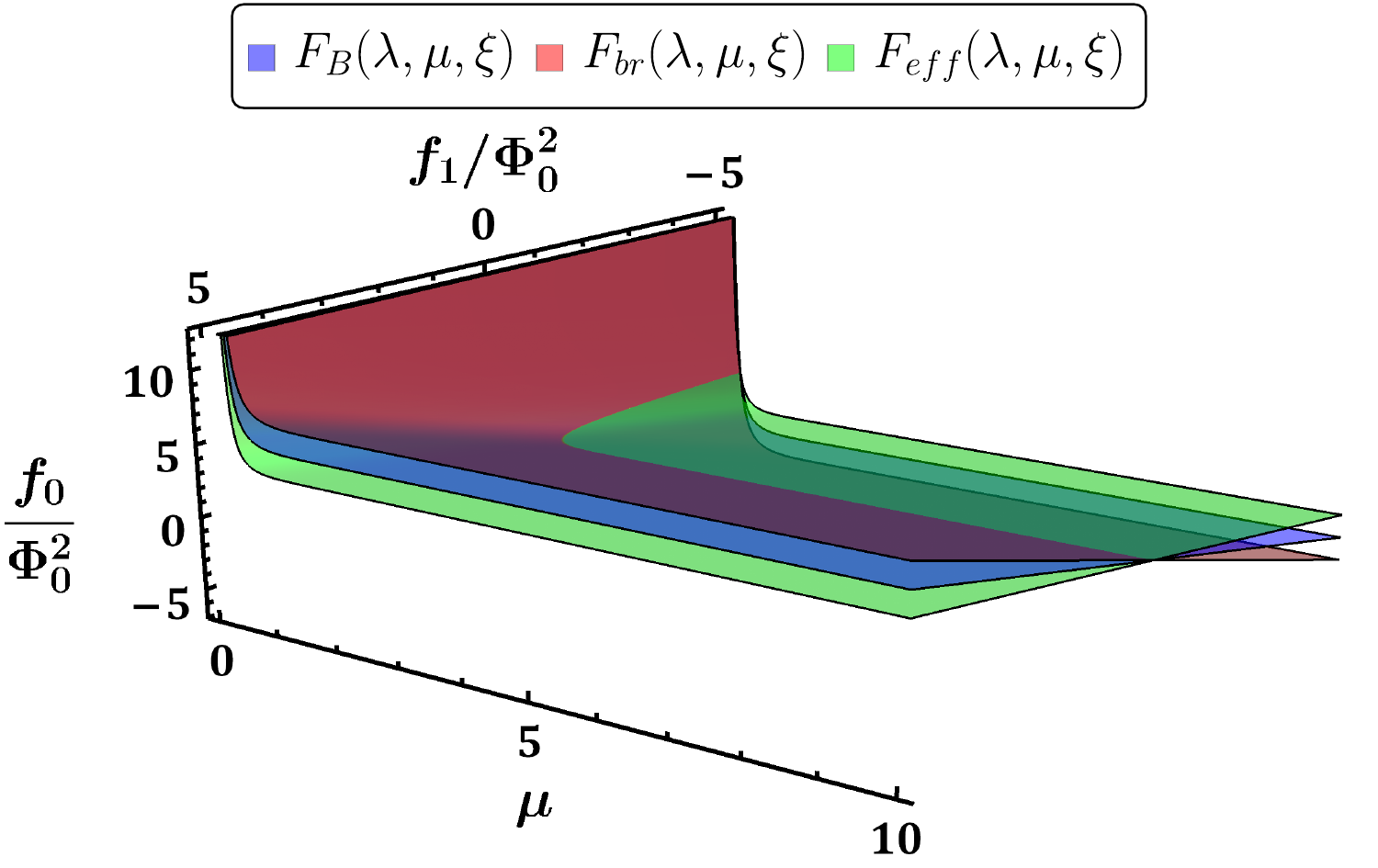}
        \caption{\hspace*{5.4em}}
        \label{double-plot-par1}
    \end{subfigure}
    ~ %add desired spacing between images, e. g. ~, \quad, \qquad, \hfill etc. 
      %(or a blank line to force the subfigure onto a new line)
    \begin{subfigure}[b]{0.42\textwidth}
        \includegraphics[width=\textwidth]{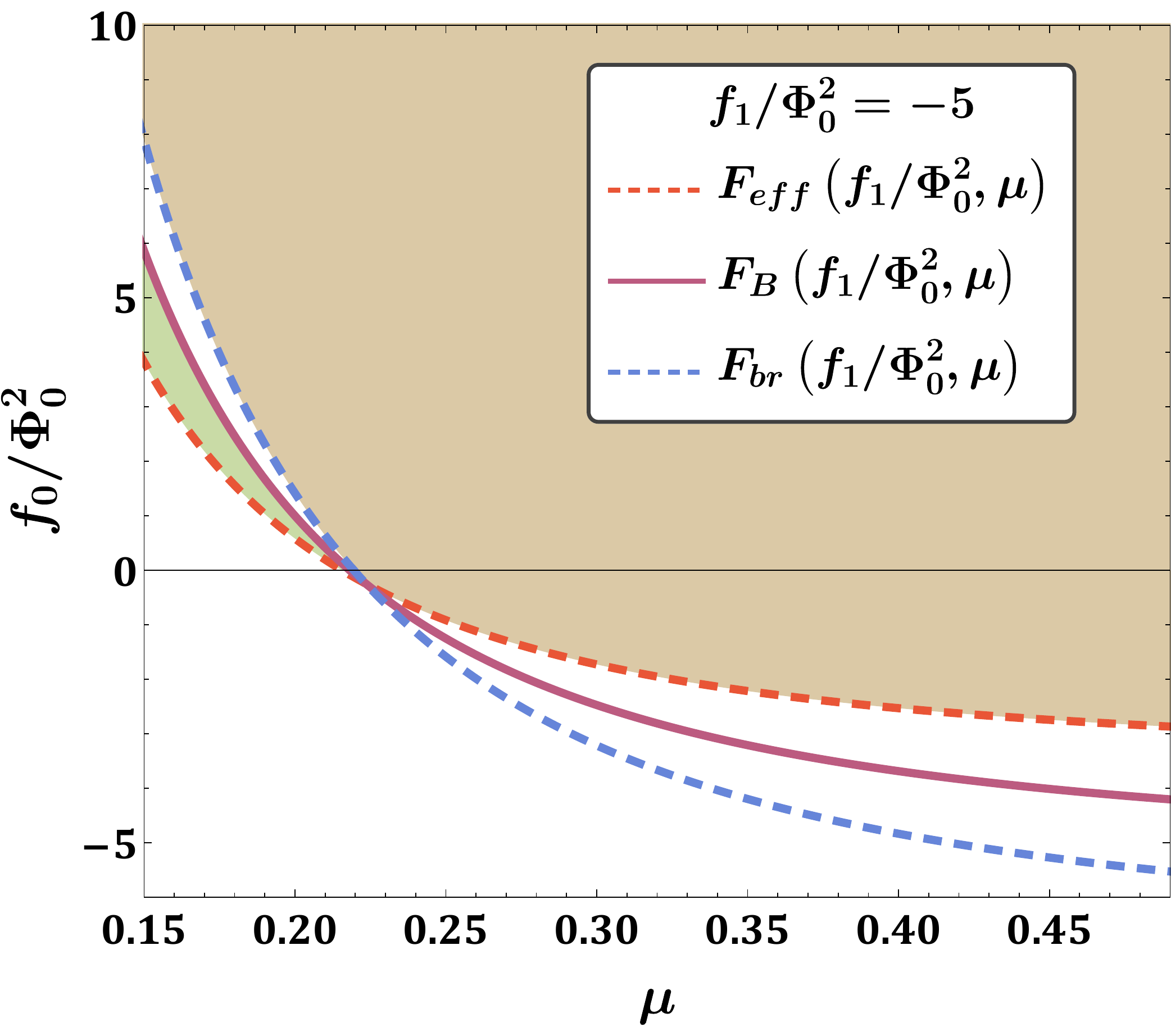}
        \caption{\hspace*{-2.8em}}
        \label{double-plot-par2}
    \end{subfigure}\vspace*{1em}
    \vspace{-0.5em}
    \caption{(a) The 3D parameter space of the quantities $f_0/\Phi_0^2$, $f_1/\Phi_0^2$ and $\mu$. 
   (b) The 2D parameter space of the quantities $f_0/\Phi_0^2$ and $\mu$ for $f_1/\Phi_0^2=-5$.
    The figures also present the corresponding surfaces or curves of $F_{B}\left(f_1/\Phi_0^2,\mu\right)$, 
    $F_{br}\left(f_1/\Phi_0^2,\mu\right)$ and $F_{eff}\left(f_1/\Phi_0^2,\mu\right)$.}
    \label{double-plot-par}
\end{figure}
%%%%%%%%%%%%%%%%%%%%%%%%%%%%%%%%%%%%%%

In the second of the above equations, we have used the relation $\pa_y V_b=\Phi'\,\pa_\Phi V_b$.
In order to have a positive total energy-density on the brane, we should have
\eq$\label{double-brane-ene-con}
\frac{f_0}{\Phi_0^2}>\frac{4}{3}\frac{f_1}{\Phi_0^2}+\frac{e^{-2\mu^2}}{12\mu^2}(4+5\mu^2+2\mu^4)\,.$
In the context of the effective theory on the brane, we may evaluate the four-dimensional gravitational
scale using Eqs. \eqref{action_eff} and \eqref{double-f-y}. Then, 
%%%%%%%%%%%%%%%%%%%%%
\bal$ \frac{1}{\kappa_4^2}&=
\frac{f_0}{k}-\frac{2f_1}{3k}-\frac{2\Phi_0^2}{4\mu^2}\int_{0}^{\infty} dy\, 
e^{-2\mu^2e^{ky}-2ky}(\mu^2+e^{-ky}) \nonumber \\[2mm]
& = \frac{f_0}{k}-\frac{2f_1}{3k}-\frac{2\Phi_0^2\, \mu^2}{k}
\left(\frac{e^{-2\mu^2}}{8\mu^4}-\mu^2\int_{2\mu^2}^\infty dt\ t^{-4}\,e^{-t}
\right)\,,
\label{linear_effG}$
where, in the second line, we first set $e^{-ky}=w$ and then $\frac{2\mu^2}{w}=t$.  The integral in
the above expression is the upper incomplete gamma function $\Gamma\left(-3,2\mu^2\right)$ as
one may easily conclude from Eq. \eqref{upper-gamma}. The latter quantity, through 
Eq. \eqref{upper-gamma-explicit}, may be written as
%%%%%%%%%%%%%%%%%%%%%%
\gat$\label{double-upper-gamma1}
\Gamma\left(-3,2\mu^2\right)=\frac{1}{6}\left[\frac{e^{-2\mu^2}}{4\mu^6}\left(
1-\mu^2+2\mu^4\right)-\Gamma\left(0,2\mu^2\right)\right],$
where
\gat$\label{double-upper-gamma2}
\Gamma\left(0,2\mu^2\right)=-\gamma-\ln\left(2\mu^2\right)-\sum_{m=1}^\infty \frac{
(-1)^m\,2^m\,\mu^{2m}}{m(m!)}\,.$
%%%%%%%%%%%%%%%%%%%%%
Hence, we finally obtain
\eq$\label{double-eff2}
\frac{1}{\kappa_4^2}=\frac{M_{Pl}^2}{8\pi}=\frac{f_0}{k}-\frac{2f_1}{3k}-\frac{\Phi_0^2
\,e^{-2\mu^2}}{12k\,\mu^2}(2+\mu^2-2\mu^4)-\frac{\Phi_0^2\,\mu^4}{3k}\,\Gamma\left(0,2
\mu^2\right).$
Demanding as usually a positive four-dimensional gravitational constant,  we
find that the following inequality must be satisfied:
\eq$\label{double-eff-con}
\frac{f_0}{\Phi_0^2}>\frac{2}{3}\frac{f_1}{\Phi_0^2}+\frac{e^{-2\mu^2}}{12\mu^2}(2+\mu^2
-2\mu^4)+\frac{\ \mu^4}{3}\,\Gamma\left(0,2\mu^2\right).$
As before, the evaluation of the effective four-dimensional cosmological constant gives $\Lambda_4=0$.

\begin{SCfigure}[][t]
    \centering
    \includegraphics[width=0.46\textwidth]{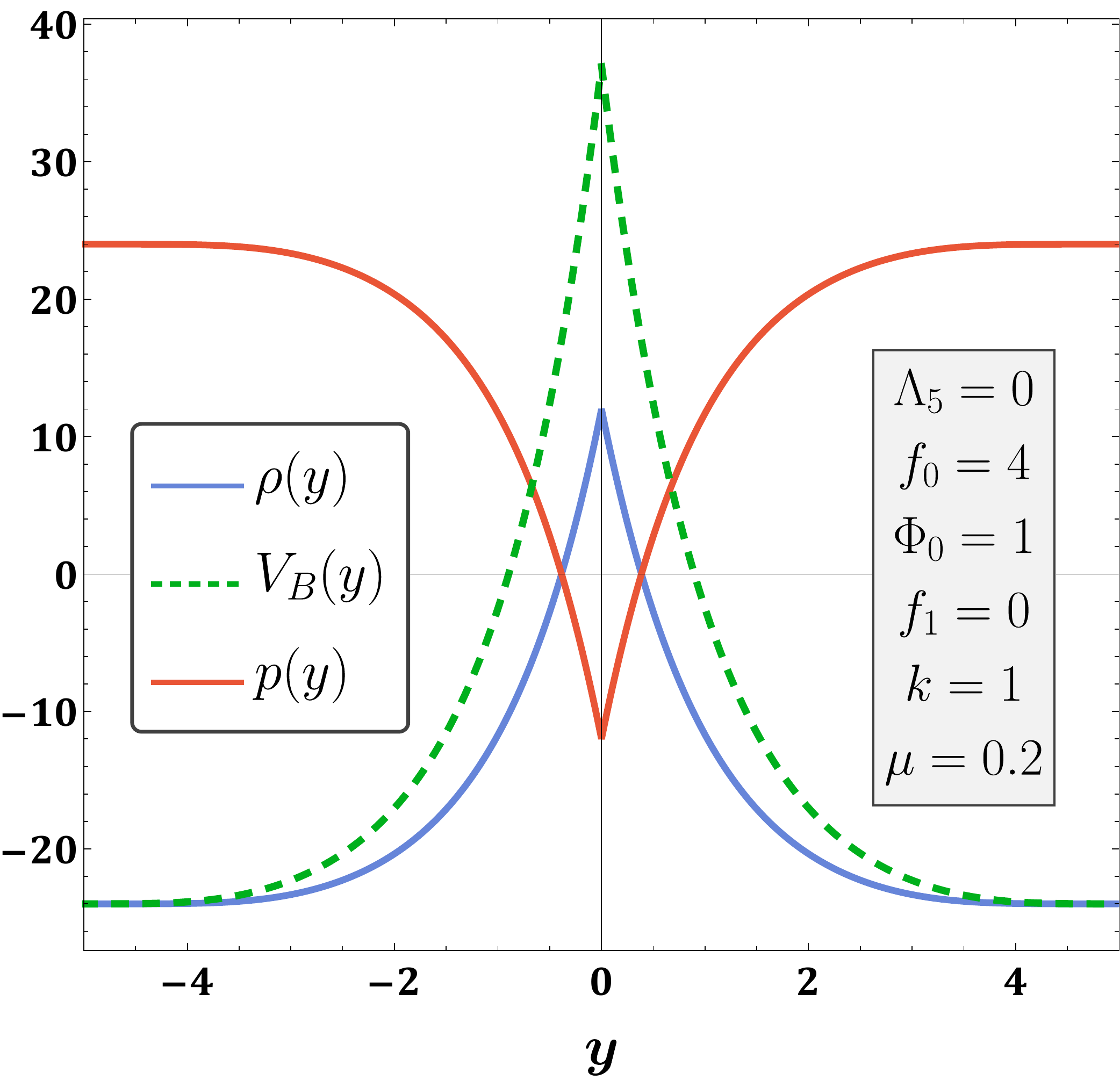}
    \caption{The energy density $\rho$ and pressure $p$ of the system together with
    the scalar potential $V_B$ in terms of the coordinate $y$ for $\Lambda_5=0$, $f_0=4$,
    $\Phi_0=1$, $f_1=0$, $k=1$, and $\mu=0.2$.\\ 
    \vspace{1.5em}}
    \label{double-plot3}  
\end{SCfigure}

Finally, we investigate the parameter space of the quantities
$f_0/\Phi_0^2$, $f_1/\Phi_0^2$ and $\mu$, in an attempt to simultaneously satisfy
the inequalities \eqref{double-weak-con}, \eqref{double-brane-ene-con} and
\eqref{double-eff-con}. In Fig. \myref{double-plot-par}{double-plot-par1}, we depict 
the aforementioned parameter
space as well as the surfaces which correspond to the r.h.s.'s of these inequalities.
We observe that there is no point in the parameter space at which all three inequalities can
be satisfied. It is possible though to satisfy two out of these three inequalities simultaneously;
which two are satisfied depends on the values of the parameters. For $f_1/\Phi_0^2=5$, 
for example, the situation is simple as the relative position of the three surfaces remains
the same independently of the value of $\mu$: thus, we may have a positive effective
cosmological constant and a positive total energy-density on the brane for low values of
$f_0/\Phi_0^2$ whereas for large values of $f_0/\Phi_0^2$ we have a positive $M_{Pl}^2$
and the weak energy conditions are satisfied close to our brane. For $f_1/\Phi_0^2=-5$,
the situation changes and the pair of conditions satisfied depends on the values of all
three parameters---the exact situation is depicted in Fig. \myref{double-plot-par}{double-plot-par2}
where the green region corresponds to the area where the inequalities \eqref{double-weak-con}
and \eqref{double-eff-con} are satisfied, and the brown region to the area
where \eqref{double-brane-ene-con} and \eqref{double-eff-con} are satisfied.

In Fig. \ref{double-plot3}, we present the graphs of the energy density $\rho(y)$ and pressure 
$p(y)$ as well as the potential of the scalar field $V_B(y)$ in terms of the $y$-coordinate for
$\Lambda_5=0$, $f_0=4$, $\Phi_0=1$, $f_1=0$, $k=1$, and $\mu=0.2$. As we can see in the
figure, the values of the parameters are appropriately chosen to satisfy the weak 
energy conditions close to the brane.

%%%%%%%%%%%%%%%%%%%%%%%%%%%%%%%%%%%%%%%%%%%%%%%%%%%%%%%%%%%%%%%%%%%%%
%
%
%%%%%%%%%%%%%%%%%%%%%%%%%%%%%%%%%%%%%%%%%%%%%%%%%%%%%%%%%%%%%%%%%%%%%

\section{A Hyperbolic-Tangent Scalar Field in terms of $y$}
\label{hyper-tang}

Following the same line of thinking as in the previous section, we now consider the following
expression for the scalar field in terms of the coordinate $y$:
\eq$\label{tanh-phi}
\Phi(y)=\Phi_0\, \tanh(ky)\,,$
where $\Phi_0\in\mathbb{R}\setminus\{0\}$. Substituting the above expression of the scalar 
field in Eq. \eqref{grav-1-2}, we obtain the form of the coupling function
\eq$\label{tanh-f-y}
f(y)=f_0-f_1\,e^{-ky}+\Phi_0^2\ e^{-ky}\,\arctan(e^{ky})-\frac{\Phi_0^2\ e^{2ky}(e^{2ky}-1)}{
3(e^{2ky}+1)^2}\,.$
Again, the parameter $f_1$ is allowed to take values in the whole set of real numbers, while the
allowed values for the parameter $f_0$ will be examined shortly. By inverting the function
$\Phi(y)$, we may express the coupling function in terms of the scalar field to get
\eq$\label{tanh-f-phi}
f(\Phi)=f_0-f_1\sqrt{\frac{\Phi_0-\Phi}{\Phi_0+\Phi}}+\Phi_0^2\sqrt{\frac{\Phi_0-\Phi}{\Phi_0+
\Phi}}\,\arctan\left(\sqrt{\frac{\Phi_0+\Phi}{\Phi_0-\Phi}}\right)-\frac{\Phi(\Phi+\Phi_0)}{6}\,.$ 

Similarly, the scalar potential $V_B(y)$ can be evaluated from Eq. \eqref{V-B} with the use of 
\eqref{tanh-f-y}; then, we find
\bal$\label{tanh-V-y}
V_B(y)=&-\Lambda_5-6k^2f_0+10k^2\,e^{-ky}\left[f_1-\Phi_0^2\,\arctan(e^{ky})\right]\nonum\\[2mm]
&+\frac{2k^2\Phi_0^2}{3}\frac{6+19e^{2ky}+19e^{4ky}-3e^{6ky}+3e^{8ky}}{(e^{2ky}+1)^4}\,.$
In terms of the scalar field, the scalar potential is alternatively written as
\bal$\label{tanh-V-phi}
V_B(\Phi)=&-\Lambda_5-6k^2f_0+10k^2\sqrt{\frac{\Phi_0-\Phi}{\Phi_0+\Phi}}\left[f_1-\Phi_0^2\,
\arctan\left(\sqrt{\frac{\Phi_0+\Phi}{\Phi_0-\Phi}}\right)\right]\nonum\\[2mm]
&+\frac{k^2}{6\Phi_0^2}\left(3\Phi^4+8\Phi^3\Phi_0+4\Phi^2\Phi_0^2-14\Phi\Phi_0^3+11\Phi_0^4\right).$
The profiles of the coupling function and scalar potential in this case are qualitatively the
same as the ones in the double-exponential case of the previous section depicted in
Figs. \myref{double-plot-1-2}{double-plot1} and \myref{double-plot-1-2}{double-plot2}. Again, as the 
parameter $f_1$ changes from positive to negative
values, the coupling function acquires an increasingly larger positive value at the location of our
brane; with the same variation, the scalar potential changes from globally positive-definite to
globally negative-definite values. As before, both functions remain finite everywhere in the bulk
and adopt constant values at large distances. 

\par The energy density $\rho(y)$ and pressure $p(y)=p^i(y)=p^y(y)$ may be computed by 
employing Eqs. \eqref{linear-rho} and \eqref{linear-p}, and we are led to the result
\eq$\label{tanh-rho}
\rho(y)=-p(y)=-6k^2f(y)=-6k^2\left[f_0-f_1\,e^{-ky}+\Phi_0^2\ e^{-ky}\,\arctan(e^{ky})-
\frac{\Phi_0^2\ e^{2ky}(e^{2ky}-1)}{3(e^{2ky}+1)^2}\right].$

In order to satisfy the weak energy conditions close and on the brane, we impose the condition
that $\rho(0)>0$, or $f(0)<0$; hence, we obtain the following inequality: 
\eq$\label{tanh-weak-con}
\frac{f_0}{\Phi_0^2}<\frac{f_1}{\Phi_0^2}-\frac{\pi}{4}\,.$

The junction conditions \eqref{jun-con1}, \eqref{jun-con2}, employing the relations 
\eqref{tanh-phi}-\eqref{tanh-f-phi}, now yield:
\gat$\label{tanh-jc1}
\sig+V_b(\Phi)\Big|_{y=0}=6kf_0-8kf_1+2k\Phi_0^2\left(\pi-\frac{1}{3}\right)\,,\\[2mm]
\label{tanh-jc2}
\pa_yV_b\Big|_{y=0}=8k^2f_1+2k^2\Phi_0^2\left(\frac{7}{3}-\pi\right)\,.$
Therefore, in order to have a positive total energy density on the brane, we demand the condition
\eq$\label{tanh-brane-ene-con}
\frac{f_0}{\Phi_0^2}>\frac{4}{3}\frac{f_1}{\Phi_0^2}+\frac{1}{9}-\frac{\pi}{3}\,.$
Let us also evaluate the effective four-dimensional gravitational constant on the brane.
Using Eq. \eqref{action_eff}, we obtain:
\bal$\frac{1}{\kappa_4^2}&=2\int_0^\infty dy\, e^{-2ky}\left[
f_0-f_1\,e^{-ky}+\Phi_0^2\ e^{-ky}\,\arctan(e^{ky})-\frac{\Phi_0^2\ e^{2ky}(e^{2ky}-
1)}{3(e^{2ky}+1)^2}\right].$
Evaluating the above integral, we obtain the result
\eq$\label{tanh-eff}
\frac{1}{\kappa_4^2}=\frac{M_{Pl}^2}{8\pi}=\frac{f_0}{k}-\frac{2}{3}\frac{f_1}{k}+\frac{\pi}{6}
\frac{\Phi_0^2}{k}\,.$
Since it is imperative to have a positive four-dimensional gravitational constant, 
we must satisfy also the following constraint:
\eq$\label{tanh-eff-con}
\frac{f_0}{\Phi_0^2}>\frac{2}{3}\frac{f_1}{\Phi_0^2}-\frac{\pi}{6}\,.$

%%%%%%%%%%%%%%%%%%%%%%
\begin{SCfigure}[][t!]
    \centering
    \includegraphics[width=0.48\textwidth]{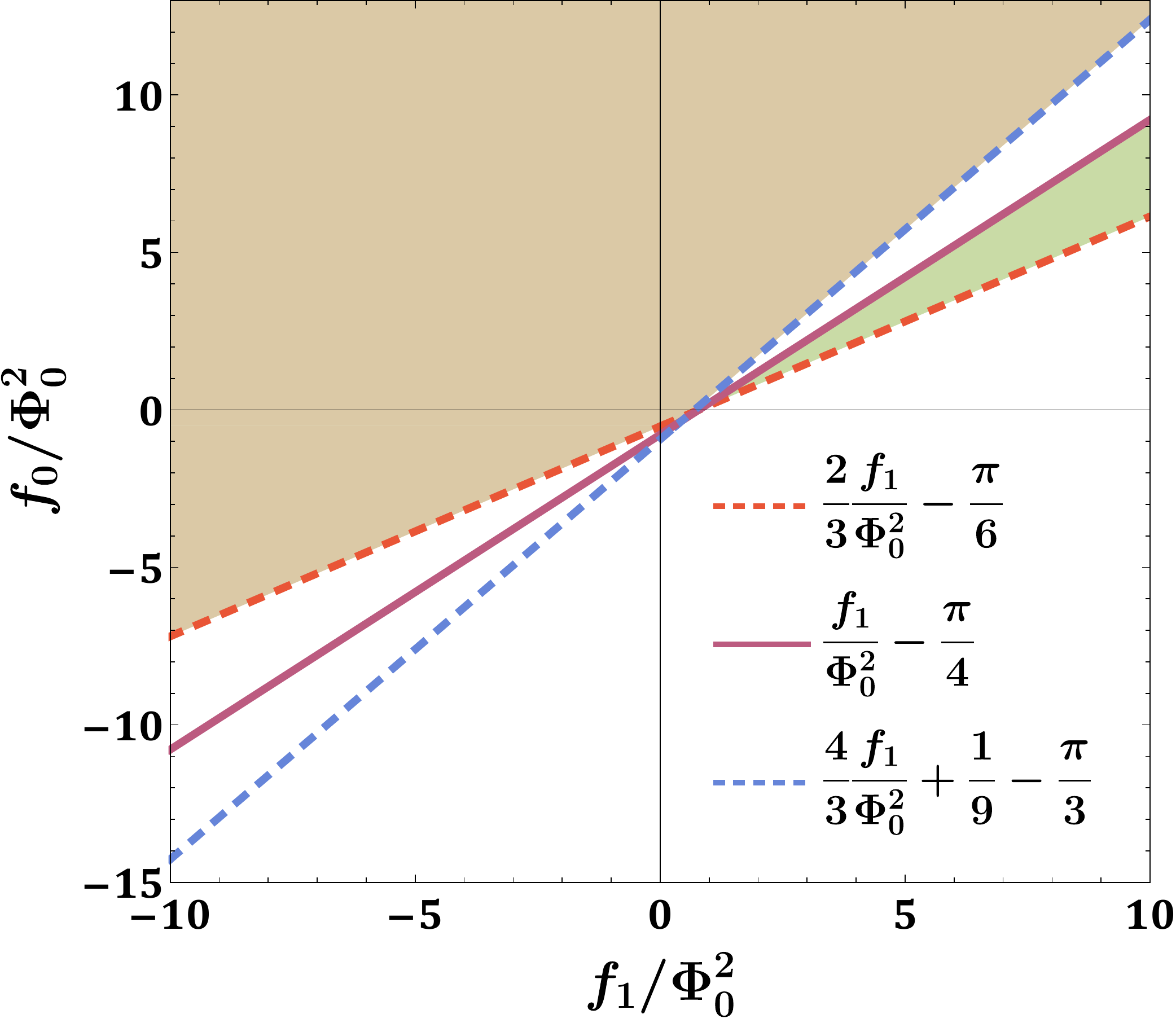}
    \caption{The parameter space of the quantities $f_0/\Phi_0^2$ and $f_1/\Phi_0^2$.
    The curves corresponding to the expressions on the r.h.s.'s of the inequalities
    \eqref{tanh-weak-con}, \eqref{tanh-brane-ene-con} and \eqref{tanh-eff-con}
    are depicted as well.\\ 
    \vspace{2em}}
    \label{tanh-plot-par}  
\end{SCfigure}
%%%%%%%%%%%%%%%%%%%%

In Fig. \ref{tanh-plot-par}, we present the parameter space of the quantities
$f_0/\Phi_0^2$ and $f_1/\Phi_0^2$, in an attempt to satisfy simultaneously
the inequalities \eqref{tanh-weak-con}, \eqref{tanh-brane-ene-con} and
\eqref{tanh-eff-con}. As it is clear, there is again no point where all three
inequalities can be satisfied. The green area defines the part of the parameter
space where $M_{Pl}^2$ is positive and the weak energy conditions are satisfied
by the bulk matter close to our brane, while the brown area defines the part
of the parameter space where both the effective gravitational constant and the
total energy-density of the brane are positive.

%%%%%%%%%%%%%%%%%%%%%%%%%%%%%%%%%%%%%%%%%%%%%%%%%%%%%%%%%%%%%%%%%%

\section{Discussion and Conclusions}
\label{conc}

In the context of this work, we have investigated the emergence of brane-world solutions
in the framework of a general scalar-tensor theory where the scalar field is non-minimally
coupled to the five-dimensional Ricci scalar. In the bulk, these solutions are characterised
by a Randall-Sundrum-type, exponentially decaying warp factor and a $y$-dependent
scalar field with a bulk potential. On the brane, the space-time takes in general the form
of a Schwarzschild-(anti-)de Sitter solution. The present work completes our previous two
analyses \cite{KNP, KNP2}, where the cases of a de Sitter and an anti-de Sitter brane
were considered, and focuses on the case of a flat, Minkowski brane with $\Lambda=0$.
The complete five-dimensional solution for the gravitational background in this case may
describe either a non-homogeneous black string, when the metric parameter $M$ is
non-zero, or a regular anti-de Sitter space-time, when $M=0$.  

The above features characterise our solutions irrespectively of the form of the coupling
function between the bulk scalar field and the five-dimensional Ricci scalar. In this work,
we have performed a comprehensive study of the types of brane-world solutions that
emerge in the context of this theory by considering a plethora of forms of the coupling
function, all supported by physical arguments regarding the reality and finiteness of
its value everywhere in the bulk. We have thus considered the cases of a linear and
quadratic coupling function in terms of the scalar field $\Phi$, but also an inverse-power
and a linear-exponential form in terms of the $y$-coordinate. From a different perspective,
we also considered given forms for the scalar field which again satisfied the finiteness condition,
namely a double-exponential and a hyperbolic-tangent form in terms of the $y$-coordinate,
and determined subsequently the form of the coupling function. In all cases, the profile of
the coupling function remains finite along the fifth coordinate as expected, reducing either
to zero or to a constant value far away from our brane---in both cases, the coupling
between the scalar field and the bulk Ricci scalar becomes trivial and as a result the
scalar-tensor theory naturally reduces to a purely gravitational theory at large distances.
Gravity by itself is also localised due to the exponentially decaying warp factor. 

In each case, we have also determined in an analytical way the corresponding solutions
for the profiles of the scalar field and scalar bulk potential. These also remain finite
over the entire bulk for every solution found, and their behaviour resembles the one of
the coupling function reducing either to zero or to a constant value away from our brane.
Depending on the values of the parameters of the solutions, the bulk 
potential in particular could adopt a variety of forms being non trivial close to our brane
and reducing to a constant, positive or negative, value at asymptotic infinity. What was
of particular importance is the fact that the Randall-Sundrum-type, exponentially
decaying warp factor is supported independently of the presence of the negative bulk
cosmological constant $\Lambda_5$, which is usually introduced in a ad hoc way in
brane-world models.

The case of a zero effective cosmological constant, studied in the context of this work,
allowed for the maximum flexibility regarding the form and characteristics of the
coupling function, when compared to the cases of a positive or negative effective
cosmological constant \cite{KNP, KNP2}. For $\Lambda>0$, the coupling function had
to be negative-definite at large distances from our brane, while, for $\Lambda <0$,
a fast localised profile was necessary in order to avoid an ill-defined behaviour for
the scalar field at the bulk boundaries. For $\Lambda=0$, though, no such requirements
are necessary. In order, however, to derive physically-acceptable brane-world solutions,
we have imposed three additional conditions: the positivity of the effective gravitational 
constant $\kappa_4^2$ on our brane defined as
%%%%%%%%%%
\beq
\frac{1}{\kappa_4^2}\equiv 2\,\int_{0}^{\infty} dy\, e^{-2 k y}\,f(y)\,,
\label{Geff_gen}
\eeq
%%%%%%%%%%
the positivity of the total energy density of our brane, which follows
from the junction condition (\ref{jun_con1}) and may be rewritten as
%%%%%%%%%%%
\beq 
\sigma+V_b=6 k f(0) - 2 f'(0)\,,
\label{jun_gen}
\eeq
%%%%%%%%%%%%%%
and the validity of the weak energy conditions by the bulk matter in the vicinity of our brane;
the latter, using Eqs. (\ref{linear-rho})-(\ref{linear-py}), may be expressed as
%%%%%%%%%%%%%
\beq
f(0)<0\,.
\label{weak-gen}
\eeq
%%%%%%%%%%%
In each solution found, we have thus performed a careful study of the effective theory
on the brane, the junction conditions introduced by the presence of the brane and the 
profiles of the energy-density and pressure of the bulk matter. Subsequently, we conducted
a thorough investigation of the corresponding parameter space in order to deduce whether it
is possible to simultaneously satisfy all three aforementioned conditions. 

We have found that, for all solutions presented, this is not possible. The aforementioned
three constraints are not a priori incompatible: Eq. (\ref{weak-gen}) constrains the value
of the coupling function $f(y)$ at the location of the brane, Eq. (\ref{jun_gen}) dictates
that its first derivative must be also negative and decreasing fast at the same point,
while Eq. (\ref{Geff_gen}) imposes a constraint on its integral over the entire bulk.
Note that if we had demanded the validity of the weak energy condition everywhere in the
bulk, i.e. $\rho(y)>0$, that would imply $f(y)<0$, for $\forall y$. This would be in
obvious contradiction with the positivity of the effective gravitational constant through
Eq. (\ref{Geff_gen}). Demanding the validity of the weak energy condition only at the
vicinity of our brane, as in Eq. (\ref{weak-gen}), allows the coupling function to be
negative close to our brane and become positive at some distance off it, so that the
integral in Eq. (\ref{Geff_gen}) turns out to be positive-definite. That was indeed realised
for some of our solutions but the parameter space corresponding to those solutions
was always severely restricted. Imposing the third constraint (\ref{jun_gen}) on the
value of $f'(0)$, on top of the previous two constraints, in an attempt to make the
energy-density of the brane also positive, we were led to contradictions for all the
analytical solutions we have found. These contradictions are translated to the absence
of a single point in the parameter space in which all the above constraints can be
simultaneously satisfied. In contrast, relaxing
the weak energy condition, which involves bulk quantities, and demanding instead the 
validity of Eqs. (\ref{Geff_gen}) and (\ref{jun_gen}), which are relevant for the
4-dimensional observer on the brane, has led to a plethora of analytic solutions with
an extended parameter space. The question of whether a solution satisfying all three
constraints could be constructed, either analytically or numerically, naturally emerges,
and could be pursued in a future work. That solution, however, would have to be not only
a mathematically consistent solution of the set of field equations satisfying the
constraints  (\ref{Geff_gen})-(\ref{weak-gen}) but to be also characterised by a
physically acceptable behaviour throughout the bulk---the analytical solutions
presented in this work were carefully constructed in order to have a physically
acceptable behaviour regarding the profiles of the scalar field, its coupling
function and potential throughout the bulk.

The question of the stability of the solutions found in this work is also an important
one. The presence of the scalar field, which is non-minimally coupled to gravity in
the context of our theory, considerably complicates the stability analysis of the solutions
found. Such an analysis will inevitably involve a coupled system of gravitational and
scalar-field equations with the particular form of the coupling fuction, characterising
each solution, playing perhaps a desicive role in the outcome of the analysis. 
We note that, in all solutions presented in this work, the coupling function becomes
trivial at large distances from the brane and the scalar field acquires a constant value.
As a result, the non-minimally coupled scalar-tensor theory reduces there to a pure
gravitational theory. Thus, at large distances from the brane, our solutions reduce to
the black-string solutions derived in \cite{CHR} and shown to be unstable in \cite{RuthGL}.
However, the non-trivial configurations of both the coupling function and the scalar
field as we approach the brane may significantly alter the stability behaviour of our
solutions compared to that of the black string of \cite{CHR}. It is quite likely that each
of the obtained solutions has its own stability behaviour under perturbations, and their
future study may provide valuable restrictions on the exact form of the coupling function,
bulk potential and profile of the scalar field itself.

In conclusion, the well-known generalised gravitational theory of a scalar field 
non-minimally coupled to the Ricci scalar admits, upon embedded in
a five-dimensional brane-world context, a variety of solutions with a number
of attractive features, such as the support of an exponentially decaying warp factor,
and thus of graviton localisation, without the need for a negative bulk cosmological
constant. In the particular case of $\Lambda=0$ studied here, this is always supplemented
by a regular scalar field, a finite coupling function, which becomes naturally trivial at
the outskirts of the bulk, a physically-acceptable brane with a positive total energy-density
and a robust effective four-dimensional theory on our brane.

\vspace*{1em}

{\bf Acknowledgements.} The research of T.N. was co-financed by Greece and the
European Union (European Social Fund- ESF) through the Operational Programme
``Human Resources Development, Education and Lifelong Learning'' in the context
of the project ``Strengthening Human Resources Research Potential via Doctorate
Research – 2nd Cycle'' (MIS-5000432), implemented by the State Scholarships Foundation (IKY).
The research of N. P. was implemented under the “Strengthening Post-Doctoral Research” 
scholarship program (Grant No. 2016-050-0503-7626) by the Hellenic State Scholarships 
Foundation as part of the Operational Program “Human Resources Development Program, 
Education and Lifelong Learning,” cofinanced by the European Social Fund-ESF and the Greek 
government.

\appendix

\numberwithin{equation}{section}	

\section{Restrictions on the allowed values of the parameter $\mu$ in quadratic case}
\label{App-mu}

We shall now determine the range of values for the parameter $\mu$ in the case
of the quadratic coupling function \eqref{quad-f}. The allowed values of $\mu$ will be
obtained by demanding that the scalar field \eqref{quad-Phi} remains real and finite,
and depend primarily on the assumed value of the parameter $\lam$. In what
follows, we will consider in detail every possible case:
%%%%%%%%%%%%%%%%%%%%%%5
\begin{enumerate}
\item[\bf(i)] \begin{flushleft}
\underline{$\lam>0$:}
\end{flushleft}
\par Using Eq. \eqref{quad-Phi} we get
$$\lim_{y\ra +\infty}\Phi(y)=\frac{\Phi_0}{2\lam}\left(\xi\ \mu^\frac{2\lam}{1+4\lam}-1
\right)\,.$$
Thus, demanding the functions $\Phi(y)$, $f(y)$ to be real-valued in their whole domain,
it is necessary to have $\mu\geq 0$. \\

\item[\bf(ii)] \begin{flushleft}
\underline{$\lam\in\left(-\frac{1}{4},0\right)\ \land\ \frac{2\lam}{1+4\lam}\neq n, \ n\in\mathbb{Z}^<$:}
\end{flushleft}
In this case $\frac{2\lam}{1+4\lam}$ is a negative rational number. Hence, one may write
\eq$\lim_{y\ra +\infty}\Phi(y)=\frac{\Phi_0}{2\lam}\left(\xi\ \mu^\frac{2\lam}{1+4\lam}-1
\right)=-\frac{\Phi_0}{2|\lam|}\left[1-\xi\left(\frac{1}{\mu}\right)^{\left|\frac{2\lam}
{1+4\lam}\right|}\right]\,.\nonum$
Therefore, in order to avoid having a complex scalar field we should demand $\mu>0$.\\

\item[\bf(iii)]\begin{flushleft}
\underline{$\lam\in\left(-\frac{1}{4},0\right)\ \land\ \frac{2\lam}{1+4\lam}=n,\ n\in
\mathbb{Z}^<$:}
\end{flushleft}
In this case we have $\frac{2\lam}{1+4\lam}=n$ or $\lam=\frac{n}{2(1-2n)}$.
Thus, one may write
$$\Phi(y)=\frac{\Phi_0}{2\lam}\left[\xi(\mu+e^{-ky})^{\frac{2\lam}{1+4\lam}}-1\right]=
\frac{\Phi_0(1-2n)}{n}
\left[\xi\left(\frac{1}{\mu+e^{-ky}}\right)^{|n|}-1\right]\,.$$
It is clear that the parameter $\mu$ is allowed to take negative values. However, we 
should not allow values in the range $[-1,0]$, because then at $y_0=-\frac{1}{k}\ln
(-\mu)>0$ we would encounter infinities regarding both the scalar field and the coupling 
function in a finite distance away from the brane. Thus, $\mu\in(-\infty,-1)\cup(0,\infty)$.\\

\item[\bf(iv)]\begin{flushleft}
\underline{$\lam=-\frac{1}{4}$:}
\end{flushleft}
In this particular case it is obvious from Eqs. \eqref{quad-Phi} and \eqref{quad-f-y} that 
the parameter $\mu$ is allowed to take any value in the set of the real numbers except 
zero. Thus, $\mu\in(-\infty,0)\cup(0,+\infty)$.\\

\item[\bf(v)]\begin{flushleft}
\underline{$\lam<-\frac{1}{4}\ \land\ \frac{2\lam}{1+4\lam}\neq n,\ n\in\mathbb{Z}^>$:}
\end{flushleft}
In this case $\frac{2\lam}{1+4\lam}$ is a positive rational number. Thus, we have
$$\lim_{y\ra + \infty}\Phi(y)=\frac{\Phi_0}{2\lam}\left(\xi\ \mu^\frac{2\lam}{1+4\lam}-1
\right)\,.$$
Therefore, $\mu\geq 0$ to avoid a complex-valued scalar field.\\

\item[\bf(vi)]\begin{flushleft}
\underline{$\lam<-\frac{1}{4}\ \land\ \frac{2\lam}{1+4\lam}=n,\ n\in\mathbb{Z}^>$:}
\end{flushleft}
In this case, it is $\frac{2\lam}{1+4\lam}=n$ and $\lam=\frac{n}{2(1-2n)}$. Thus, from Eq. 
\eqref{quad-Phi} we have
$$\Phi(y)=\frac{\Phi_0}{2\lam}\left[\xi(\mu+e^{-ky})^{\frac{2\lam}{1+4\lam}}-1\right]=
\frac{\Phi_0(1-2n)}{n}\left[\xi \left(\mu+e^{-ky}\right)^{n}-1\right],$$
which allows $\mu$ to take values in the whole set of the real number: $\mu\in\mathbb{R}$.\\
\end{enumerate}

The aforementioned results are summarised in Table \ref{quad-par-val}.

%%%%%%%%%%%%%%%%%%%%%%%%%%%%%%%%%%%%%%%%%%%%%%%%%%%%%%%%%%%%%%%%%%%

\section{The upper and lower incomplete gamma functions}
\label{incom-gamma}

The upper incomplete gamma function $\Gamma(s,x)$ is defined as follows
\eq$\label{upper-gamma}
\Gamma(s,x)\equiv\int_x^\infty dt\ t^{s-1}\, e^{-t}=\Gamma(s)-\gamma(s,x)\, ,$
where
\eq$\label{lower-gamma}
\gamma(s,x)\equiv \int_0^x dt\ t^{s-1}\, e^{-t}\, ,$
is the lower incomplete gamma function. Both upper and lower incomplete gamma functions, as
defined above, are valid for real and positive $s$ and $x$. However, both functions can
be extended for almost all combinations of complex $s$ and $x$. One can show that, for
all complex $s$ and $z$, the lower incomplete gamma function can be expanded in the following
power series
\eq$\label{lower-gamma-exp}
\gamma(s,z)=z^s\,\Gamma(s)\,e^{-z}\sum_{k=0}^\infty \frac{z^k}{\Gamma(s+k+1)}\, .$
Locally, the sum in the r.h.s. of the previous relation converges uniformly for all $s\in\mathbb{C}$ 
and $z\in\mathbb{C}$. Using the relation $\Gamma(s,z)=\Gamma(s)-\gamma(s,z)$ we obtain the
values of the upper incomplete gamma function for complex $s$ and $z$, but only for the
points $(s,z)$ in which the r.h.s. exists. The numerical value of the upper incomplete gamma
function can be given by the following expressions:
\eq$\label{upper-gamma-explicit}
\Gamma(s,x)=\left\{\begin{array}{ll}
\displaystyle{\Gamma(s)-x^s\,\Gamma(s)\,e^{-x}\sum_{k=0}^\infty \frac{x^k}{\Gamma(s+k+1)}}\, , &
 s\neq-n,\ n\in\mathbb{Z}^>\\[7mm]
\displaystyle{-\gamma-\ln(x)-\sum_{k=1}^\infty \frac{(-x)^k}{k(k!)}}\, , & s=0\\[7mm]
\displaystyle{\frac{1}{n!}\left[\frac{e^{-x}}{x^n}\sum_{k=0}^{n-1}(-x)^k(n-k-1)!+(-1)^n\,\Gamma(0,x)
\right]} , & s=-n,\ n\in\mathbb{Z}^>
\end{array}\right\}\, ,$
where $\gamma$ is the Euler-Mascheroni constant. In our case, for $s=1-\lam$, namely
$s\in(-\infty,1)$ and $x=2ky_0>0$, we obtain the expressions
\eq$\label{gamma-inc}
\Gamma(1-\lam,2ky_0)=\left\{\begin{array}{ll}
\displaystyle{\Gamma(1-\lam)\left[1-(2ky_0)^{1-\lam}\,e^{2ky_0}\sum_{m=0}^\infty \frac{(2ky_0)^m}{
\Gamma(2+m-\lam)}\right]}, & \lam\neq n+1,\\[-0.8em]
&\ \ n\in\mathbb{Z}^>\\[8mm]
\displaystyle{-\gamma-\ln(2ky_0)-\sum_{m=1}^\infty \frac{(-2ky_0)^m}{m(m!)}}, & \lam=1\\[8mm]
\displaystyle{\frac{1}{n!}\left[\frac{e^{-2ky_0}}{(2ky_0)^n}\sum_{m=0}^{n-1}(-2ky_0)^m(n-m-1)!+(-1)^n
\,\Gamma(0,2ky_0)\right]},& \lam= n+1,\\[-0.8em]
&\ \ n\in\mathbb{Z}^>
\end{array}\right\}.$

\clearpage
\bibliography{Bibliography}{}
\bibliographystyle{utphys}

\end{document}